%% file: main.tex
\documentclass[journal]{IEEEtran}
\IEEEoverridecommandlockouts
\usepackage{cite}
\usepackage{multirow}
\usepackage{tabularray}
\usepackage{amsmath,amssymb,amsfonts}
\usepackage{algorithmic}
\usepackage{graphicx}
\usepackage{svg}
\usepackage{textcomp}
\usepackage{pifont}
\usepackage{tcolorbox}
\usepackage{xcolor}
\usepackage{tikz}
\usepackage{pgfplots}
\usepgfplotslibrary{groupplots}
\pgfplotsset{compat=1.12}
\usepackage{subfig}
\usepackage{algorithmic}
\usepackage{algorithm}
\usepackage{longtable}
\usepackage{svg}
\usepackage{xcolor,colortbl}
\usepackage{hyperref}
 
\newcommand*\halfcirc[1][1ex]{%
  \begin{tikzpicture}
  \draw[fill] (0,0)-- (90:#1) arc (90:270:#1) -- cycle ;
  \draw (0,0) circle (#1);
  \end{tikzpicture}}

\def\BibTeX{{\rm B\kern-.05em{\sc i\kern-.025em b}\kern-.08em
    T\kern-.1667em\lower.7ex\hbox{E}\kern-.125emX}}

\begin{document}

\title{Generative AI in Cybersecurity: A Comprehensive Review of LLM Applications and Vulnerabilities}

\author{
\IEEEauthorblockN{Mohamed~Amine~Ferrag\IEEEauthorrefmark{1}\IEEEauthorrefmark{6}, Fatima Alwahedi\IEEEauthorrefmark{2}, Ammar Battah\IEEEauthorrefmark{2}, Bilel Cherif\IEEEauthorrefmark{2}, Abdechakour Mechri\IEEEauthorrefmark{3}, \\ Norbert Tihanyi\IEEEauthorrefmark{2}, Tamas Bisztray\IEEEauthorrefmark{4}, and  Merouane Debbah\IEEEauthorrefmark{5}}
 \\\IEEEauthorblockA{\IEEEauthorrefmark{1}Department of Computer Science, Guelma University, Algeria}
 \\\IEEEauthorblockA{\IEEEauthorrefmark{2}Technology Innovation Institute, UAE}
 \\\IEEEauthorblockA{\IEEEauthorrefmark{3}Concordia University, Canada}
\\\IEEEauthorblockA{\IEEEauthorrefmark{4}University of Oslo, Norway}
\\\IEEEauthorblockA{\IEEEauthorrefmark{5}Khalifa University of Science and Technology, UAE}
\\\IEEEauthorblockA{\IEEEauthorrefmark{6}Corresponding author: ferrag.mohamedamine@univ-guelma.dz}
 }





\maketitle
\thispagestyle{plain}
\pagestyle{plain}
\begin{abstract}
This paper provides a comprehensive review of the future of cybersecurity through Generative AI and Large Language Models (LLMs). We explore LLM applications across various domains, including hardware design security, intrusion detection, software engineering, design verification, cyber threat intelligence, malware detection, and phishing detection. We present an overview of LLM evolution and its current state, focusing on advancements in models such as GPT-4, GPT-3.5, Mixtral-8x7B, BERT, Falcon2, and LLaMA. Our analysis extends to LLM vulnerabilities, such as prompt injection, insecure output handling, data poisoning, DDoS attacks, and adversarial instructions. We delve into mitigation strategies to protect these models, providing a comprehensive look at potential attack scenarios and prevention techniques. Furthermore, we evaluate the performance of 42 LLM models in cybersecurity knowledge and hardware security, highlighting their strengths and weaknesses. We thoroughly evaluate cybersecurity datasets for LLM training and testing, covering the lifecycle from data creation to usage and identifying gaps for future research. In addition, we review new strategies for leveraging LLMs, including techniques like Half-Quadratic Quantization (HQQ), Reinforcement Learning with Human Feedback (RLHF), Direct Preference Optimization (DPO), Quantized Low-Rank Adapters (QLoRA), and Retrieval-Augmented Generation (RAG). These insights aim to enhance real-time cybersecurity defenses and improve the sophistication of LLM applications in threat detection and response. Our paper provides a foundational understanding and strategic direction for integrating LLMs into future cybersecurity frameworks, emphasizing innovation and robust model deployment to safeguard against evolving cyber threats.

\end{abstract}


\section*{List of Abbreviations}
\begin{tabular}{ll}
AI & Artificial Intelligence \\
AIGC & Artificial Intelligence Generated Content \\
APT & Advanced Persistent Threat \\
CNN & Convolutional Neural Network \\
CTG & Controllable Text Generation \\
CVE & Common Vulnerabilities and Exposures \\
CWE & Common Weakness Enumeration \\
FNN & Feed-Forward Neural Network \\
FRR & False Refusal Rate \\
GPT & Generative Pre-trained Transformers \\
GRU & Gated Recurrent Units \\
\end{tabular}
\begin{tabular}{ll}
GQA & Grouped-Query Attention \\
HPC & High-Performance Computing \\
HLS & High-Level Synthesis Design Verification \\
HQQ &  Half-Quadratic Quantization \\
IDS & Intrusion Detection System \\
LLM & Large Language Model \\
LoRA & Low-rank Adapters \\
LSTM & Long Short-Term Memory \\
ML & Machine Learning \\
MLP & Multi-Layer Perceptron \\
MQA & Multi-Query Attention \\
NIST & National Institute of Standards and Technology \\
NLP & Natural Language Processing \\
NLU & Natural Language Understanding \\
ORPO & Odds Ratio Preference Optimization  \\
PEFT & Parameter Efficient Fine-Tuning \\
PLM & Pre-trained Language Model \\
PPO & Proximal Policy Optimization \\
RAG & Retrieval Augmentation Generation \\
RLHF & Reinforcement Learning from Human Feedback \\
RNN & Recurrent Neural Networks \\
RTL & Register-Transfer Level \\
SARD & Software Assurance Reference Dataset \\
SFT & Supervised Fine-Tuning \\
SVM & Support Vector Machine \\
TRPO & Trust Region Policy Optimization \\
\end{tabular}

\section{Introduction}

The history of Natural Language Processing (NLP) dates back to the 1950s when the Turing test was developed. However, NLP has seen significant advancements in recent decades with the introduction of Recurrent Neural Networks (RNN) \cite{pascanu2013construct}, Long Short-Term Memory (LSTM) \cite{hochreiter1997long}, Gated Recurrent Units (GRU) \cite{dey2017gate}, and Transformer methods \cite{vaswani2017attention}. RNN was first introduced in the 1990s to model data sequences. LSTM, a variant of RNN, was introduced in 1997, which addressed the vanishing gradient problem and allowed for longer-term memory in NLP models. GRU, another variant of RNN, was introduced in 2014, which reduced the number of parameters and improved computational efficiency \cite{yan2024protecting}. The latest breakthrough in NLP was the introduction of Transformers in 2017, enabling parallel processing of sequential data and revolutionizing tasks like machine translation. These methods have significantly improved various NLP tasks, including sentiment analysis, language generation, and translation~\cite{vaswani2017attention,myers2024foundation,tonmoy2024comprehensive,ferrag2023generative}.

Cybersecurity is an ever-evolving field, with threats becoming increasingly sophisticated and complex. As organizations and individuals rely on digital technologies for communication, commerce, and critical infrastructure, the need for robust cybersecurity measures has never been greater \cite{sarker2024multi}. The scale and diversity of cyber threats make it a daunting challenge for security professionals to effectively identify, detect, and defend against them. In this context, Large Language Models (LLMs) have emerged as a game-changing technology with the potential to enhance cybersecurity practices significantly \cite{yao2024survey, yan2024depending,sladic2023llm,tann2023using,g2024harnessing}.  These models, powered by advanced NLP and Machine Learning (ML) techniques, offer a new frontier in the fight against cyber threats~\cite{ebert2023artificial,wang2024software}. This article explores the motivations and applications of LLMs in cybersecurity.

\begin{figure}[h]
    \centering
       \includegraphics[width=0.45\textwidth]{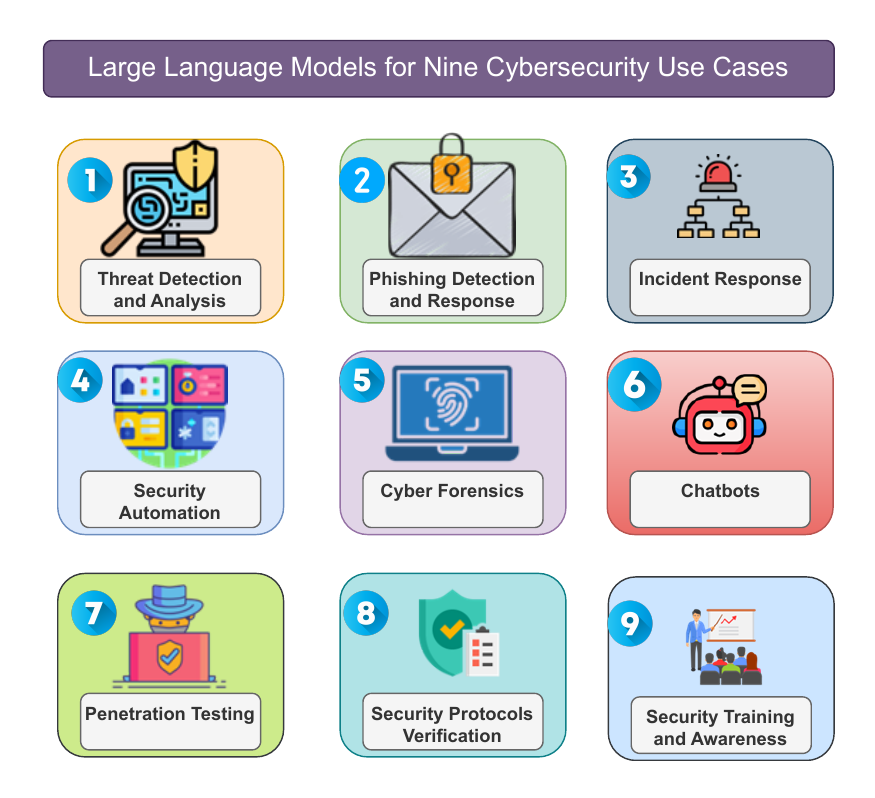}
    \caption{LLM Use Cases And Applications for Cybersecurity.}
    \label{fig:usescases}
\end{figure}

Cybersecurity professionals often need to sift through a vast amount of textual data, including security alerts, incident reports, threat feeds, and research papers, to stay ahead of evolving threats. LLMs, like Falcon 180b \cite{almazrouei2023falcon}, possess natural language understanding capabilities that enable them to parse, summarize, and contextualize this information efficiently \cite{zhou2024large,lai2024survey,tonmoy2024comprehensive}. They can assist in rapidly identifying relevant threat intelligence, allowing analysts to make more informed decisions and prioritize responses \cite{ferrag2024revolutionizing}. LLMs can excel in various domains within cybersecurity \cite{tihanyi2024dynamic,tihanyi2024cybermetricnew}. Figure~\ref{fig:usescases} highlights the top nine use cases and applications for LLMs in this field \cite{10527236,friha2024llm}.

\begin{enumerate}
    \item \textbf{Threat Detection and Analysis:} LLMs can analyze vast network data in real-time to detect anomalies and potential threats. They can recognize patterns indicative of cyber attacks, such as malware, phishing attempts, and unusual network traffic \cite{ferrag2024revolutionizing}.
    \item \textcolor{black}{\textbf{Phishing Detection and Response:} LLMs can identify phishing emails by analyzing the text for malicious intent and comparing it to known phishing examples. They can also generate alerts and recommend preventive actions \cite{jamal2024improved}.}
    \item \textcolor{black}{\textbf{Incident Response:} During a cybersecurity incident, LLMs can assist by providing rapid analysis of the situation, suggesting mitigation strategies, and automating responses where applicable \cite{zhao2023survey}.}
    \item \textbf{Security Automation:} LLMs can facilitate the automation of routine security tasks such as patch management, vulnerability assessments, and compliance checks. This reduces the workload on cybersecurity teams and allows them to focus on more complex tasks \cite{yao2024survey}.
    \item \textbf{Cyber Forensics:} LLMs can help in forensic analysis by parsing through logs and data to determine the cause and method of attack, thus aiding in the recovery process and future prevention strategies \cite{10445123}.
    \item \textcolor{black}{\textbf{Chatbots:} LLMs significantly enhance the capabilities of chatbots in cybersecurity environments by providing User Interaction, Incident Reporting and Handling, Real-time Assistance, Training and Simulations, and FAQ Automation \cite{10433480}.}
    \item \textbf{Penetration Testing:} LLMs can help generate scripts or modify existing ones to automate certain parts of the penetration testing process. This includes scripts for vulnerability scanning, network mapping, and exploiting known vulnerabilities \cite{fang2024llm}.
    \item \textbf{Security Protocols Verification:} LLMs can help verify the security of protocols such as TLS/SSL, IPSec, …etc. 
    \item \textbf{Security Training and Awareness:} LLMs can generate training materials tailored to an organization's needs. They can also simulate phishing attacks and other security scenarios to train employees to recognize and respond to security threats \cite{chang2023survey}.
\end{enumerate}

\begin{figure*}[h]
    \centering
       \includegraphics[width=1\textwidth]{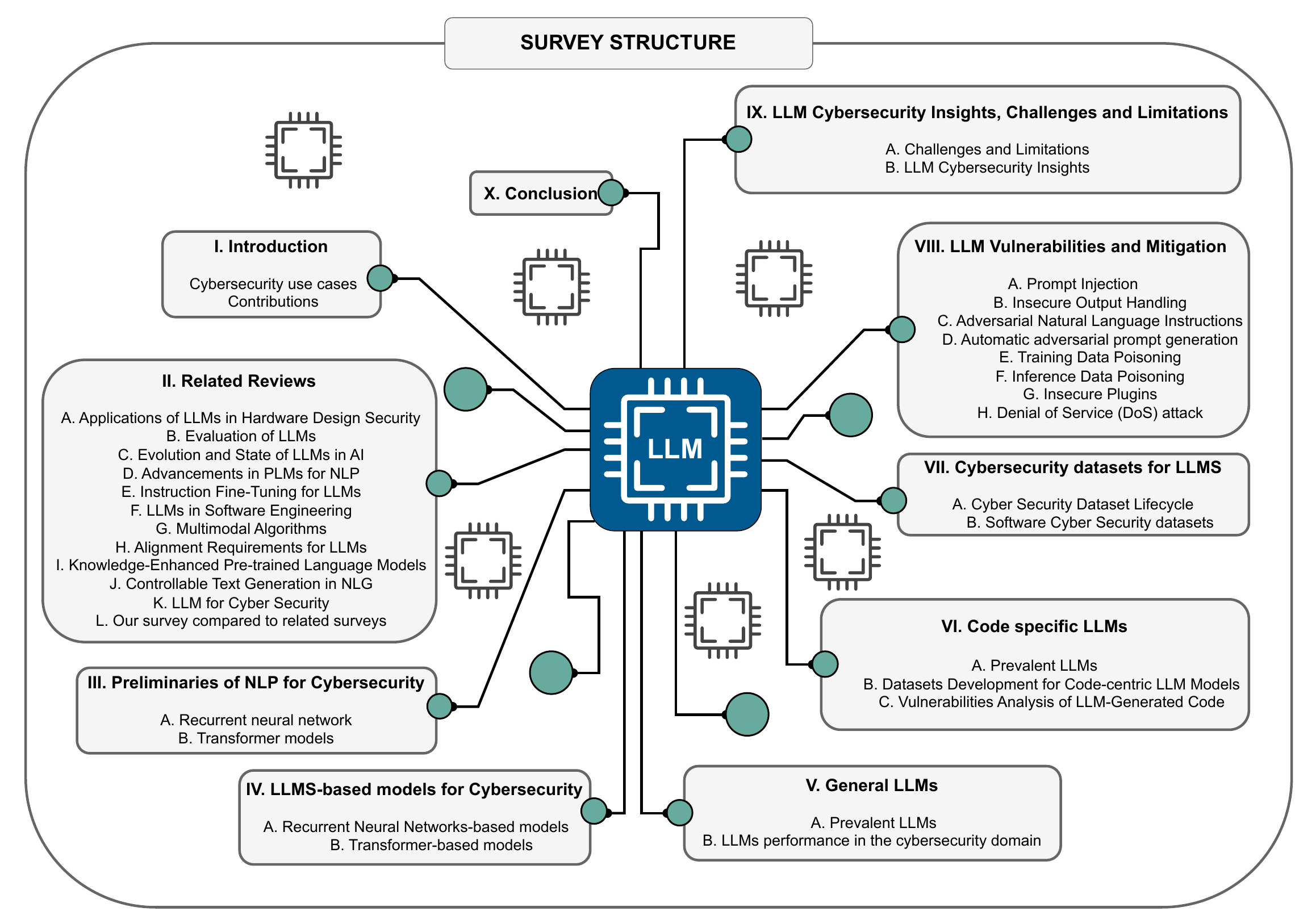}
    \caption{Survey Structure (From Section I. to Section X.)}
    \label{fig:structure}
\end{figure*}
The primary aim of this paper is to provide an in-depth and comprehensive review of the future of cybersecurity using Generative AI and LLMs, covering all relevant topics in the cyber domain. The contributions of this study are summarized below:

\begin{itemize}
    \item We review LLMs' applications for cybersecurity use cases, such as hardware design security, intrusion detection, software engineering, design verification, cyber threat intelligence, malware detection, phishing, and spam detection,  etc., providing a nuanced understanding of LLM capabilities across different cybersecurity domains;
    \item We present a comprehensive overview of LLMs in cybersecurity, detailing their evolution and current state, including advancements in 42 specific models, such as GPT-4o, GPT-4, BERT, Falcon, and LLaMA models;
    \item We analyze the vulnerabilities associated with LLMs, including prompt injection, insecure output handling, training data poisoning, inference data poisoning, DDoS attacks, and adversarial natural language instructions. We also examine the mitigation strategies to safeguard these models from such vulnerabilities, providing a comprehensive look at potential attack scenarios and prevention techniques;
    \item We evaluated the performance of 42 LLM models in different datasets in the cybersecurity domain.
    \item We thoroughly evaluate cybersecurity datasets tailored for LLM training and testing. This includes a lifecycle analysis from dataset creation to usage, covering various stages such as data cleaning, preprocessing, annotation, and labeling. We also compare cybersecurity datasets to identify gaps and opportunities for future research;
    \item We provide the challenges and limitations of employing LLMs in cybersecurity settings, such as dealing with adversarial attacks and ensuring robustness. We also discuss the implications of these challenges for future LLM deployments and the development of secure, optimized models;
    \item We discuss novel insights and strategies for leveraging LLMs in cybersecurity, including advanced techniques such as Half-Quadratic Quantization (HQQ), Reinforcement Learning with Human Feedback (RLHF), Direct Preference Optimization (DPO), Odds Ratio Preference Optimization (ORPO), GPT-Generated Unified Format (GGUF),  Quantized Low-Rank Adapters (QLoRA), and Retrieval-Augmented Generation (RAG). These insights aim to enhance real-time cybersecurity defenses and improve the sophistication of LLM applications in threat detection and response.
\end{itemize}

The rest of this paper is organized as follows. Section \ref{sec:2} presents an in-depth analysis of related reviews in the field, charting the evolution and state of LLMs in artificial intelligence. Section \ref{sec:3} delves into the preliminaries of NLP applications for cybersecurity, covering foundational models and their advancements. Section \ref{sec:4} discusses LLM-based solutions specific to cybersecurity. Section \ref{sec:5} reviews general LLM models. Section \ref{sec:6} reviews Code-specific LLMs models. Section \ref{sec:7} explores various cybersecurity datasets designed for LLM training and evaluation, detailing their development lifecycle and specific attributes. Section \ref{sec:8} focuses on the vulnerabilities associated with LLMs and the strategies for their mitigation, introducing a classification of potential threats and defense mechanisms. Section \ref{sec:9} offers comprehensive insights into the challenges and limitations of integrating LLMs into cybersecurity frameworks, including practical considerations and theoretical constraints. Finally, Section \ref{sec:10} concludes the paper by summarizing the key findings and proposing directions for future research in LLMs and cybersecurity. A brief overview of the paper's structure is illustrated in Figure~\ref{fig:structure}.

\begin{table*}[h]
\centering
\caption{Summary of Related Reviews on Large Language Models}
\scriptsize
\begin{tabular}{|p{2.7cm}|p{0.5cm}|p{1cm}|p{6.5cm}|p{0.6cm}|p{0.6cm}|p{0.6cm}|p{0.6cm}|p{0.6cm}|}
\hline
\textbf{Focused Area of Study} & \textbf{Year} & \textbf{Authors} & \textbf{Key Points} & \textbf{Data.} & \textbf{Vuln.} & \textbf{Comp.} & \textbf{Optim.} & \textbf{Hardw.} \\ 
\hline

LLMs in Enhancing Hardware Design Security & 2023 & Saha \textit{et al.} \cite{saha2023llm} & Discuss applications of LLMs in hardware design security, including vulnerability introduction, assessment, verification, and countermeasure development. &  \large \ding{56} & \large \ding{56} & \large \ding{56} & \large \ding{56}& \large \ding{56} \\ 
\hline
Comprehensive Evaluation Methodologies for LLMs & 2023 & Chang \textit{et al.} \cite{chang2023survey} & Provides an analysis of LLM evaluations focusing on criteria, context, methodologies, and future challenges. &  \large \ding{56} & \large \ding{56} & \large \ding{56} & \large \ding{56}& \large \ding{56} \\ 
\hline
The Evolutionary Path of LLMs in AI & 2023 & Zhao \textit{et al.} \cite{zhao2023survey} & Surveys the evolution of LLMs in AI, focusing on pre-training, adaptation tuning, utilization, and capacity evaluation. &  \large \ding{56} & \large \ding{56} & \large \ding{56} & \large \ding{56}& \large \ding{56} \\ 
\hline
Recent Advancements in PLMs for NLP & 2023 & Min \textit{et al.} \cite{min2023recent} & Reviews advancements in PLMs for NLP, covering paradigms like Pre-train then Fine-tune, Prompt-based Learning, and NLP as Text Generation. &  \large \ding{56} & \large \ding{56} & \large \ding{56} & \large \ding{56}& \large \ding{56} \\  
\hline
Exploring Instruction Fine-Tuning in LLMs & 2023 & Zhang \textit{et al.} \cite{zhang2023instruction} & Explores instruction fine-tuning for LLMs, covering methodologies, datasets, models, and multi-modality techniques.  &  \large \ding{56} & \large \ding{56} & \large \ding{56} & \large \ding{56}& \large \ding{56} \\ 
\hline
Applying LLMs in Software Engineering & 2023 & Fan \textit{et al.} \cite{fan2023large} & Survey the use of LLMs in Software Engineering, discussing applications, challenges, and hybrid approaches.  &  \large \ding{56} & \large \ding{56} & \large \ding{56} & \large \ding{56}& \large \ding{56} \\ 
\hline
Understanding Multimodal Algorithms & 2023 & Wu \textit{et al.} \cite{wu2023multimodal} & Provides an overview of multimodal algorithms, covering definition, evolution, technical aspects, and challenges. &  \large \ding{56} & \large \ding{56} & \large \ding{56} & \large \ding{56}& \large \ding{56} \\ 
\hline
Defining Alignment Requirements for LLMs & 2023 & Liu \textit{et al.} \cite{liu2023trustworthy} & Proposes a taxonomy of alignment requirements for LLMs and discusses harmful content concepts.  &  \large \ding{56} & \large \ding{56} & \large \ding{56} & \large \ding{56}& \large \ding{56} \\ 
\hline
Incorporating External Knowledge in PLMs & 2023 & Hu \textit{et al.} \cite{hu2023survey} & Reviews KE-PLMs, focusing on incorporating different types of knowledge into PLMs for NLP.  &  \large \ding{56} & \large \ding{56} & \large \ding{56} & \large \ding{56}& \large \ding{56} \\ 
\hline
Advances in Controllable Text Generation & 2023 & Zhang \textit{et al.} \cite{zhang2023survey} & Reviews CTG in NLG, focusing on Transformer-based PLMs and challenges in controllability.  &  \large \ding{56} & \large \ding{56} & \large \ding{56} & \large \ding{56}& \large \ding{56} \\ 
\hline
LLM for Blockchain Security & 2024 & He \textit{et al.} \cite{he2024large} & Analyze existing research to understand how LLMs can improve blockchain systems' security. & \halfcirc & \halfcirc & \large \ding{56} & \large \ding{56} & \large \ding{56}\\ 
\hline
LLM for Critical Infrastructure Protection & 2024 & Yigit \textit{et al.} \cite{yigit2024critical} & Proposing advanced strategies using Generative AI and Large Language Models to enhance resilience and security. & \large \ding{56} & \large \ding{56}& \large \ding{56} & \large \ding{56} & \large \ding{56} \\   
\hline
Software Testing with Large Language Models & 2024 & Wang \textit{et al.} \cite{10440574} & Explore how Large Language Models (LLMs) can enhance software testing, examining tasks, techniques, and future research directions.  &  \large \ding{56} & \large \ding{56} & \large \ding{56} & \large \ding{56} & \large \ding{56} \\ 
\hline
Malicious Insider Threat Detection Using Machine Learning Methods & 2024 & Alzaabi \textit{et al.} \cite{10445123} & Recommends advanced ML methods like deep learning and NLP for better detection and mitigation of insider threats in cybersecurity, emphasizing the need for integrating time-series techniques. & \halfcirc & \halfcirc & \large \ding{56} & \large \ding{56} & \large \ding{56}\\ 
\hline
Advancements in Large Language Models & 2024 & Raiaan \textit{et al.} \cite{10433480} & Reviews the evolution, architectures, applications, societal impacts, and challenges of LLMs, aiding practitioners, researchers, and experts in understanding their development and prospects. & \large \ding{56} & \large \ding{56} & \large \ding{56} & \large \ding{56} & \large \ding{56} \\ 
\hline
Applications of LLMs in cybersecurity tasks & 2024 & Xu \textit{et al.} \cite{xu2024large} & Highlights the diverse applications of LLMs in cybersecurity tasks such as vulnerability detection, malware analysis, and intrusion and phishing detection.  & \halfcirc & \halfcirc &  \large \ding{56} & \large \ding{56} &  \large \ding{56}\\ 
\hline
Retrieval-Augmented Generation for LLMs & 2024 & Zhao \textit{et al.} \cite{10433480} & Reviews how RAG has been integrated into various AIGC scenarios to overcome common challenges such as updating knowledge, handling long-tail data, mitigating data leakage, and managing costs associated with training and inference. & \large \ding{56} & \large \ding{56} & \large \ding{56} & \large \ding{56} & \large \ding{56} \\ 
\hline
Provides an overview of Parameter Efficient Fine-Tuning (PEFT) & 2024 & Han \textit{et al.} \cite{han2024parameter} & Reviews various PEFT algorithms, their effectiveness, and the computational overhead involved.  & \large \ding{56} & \large \ding{56} & \large \ding{56} & \large \ding{56} & \large \ding{56} \\ 
\hline
LLM for Cyber Security & 2024 & Zhang \textit{et al.} \cite{zhang2024llms} & The paper conducts a systematic literature review of over 180 works on applying LLMs in cybersecurity. & \halfcirc & \halfcirc & \large \ding{56} & \large \ding{56} & \large \ding{56} \\ 
\hline
LLM with security and privacy issues & 2024 & Yao \textit{et al.} \cite{yao2024survey} & Explores the dual impact of LLMs on security and privacy, highlighting their potential to enhance cybersecurity and data protection while also posing new risks and vulnerabilities. & \halfcirc & \halfcirc & \large \ding{56}  & \large \ding{56}  & \large \ding{56}  \\  
\hline
\textbf{THIS SURVEY} & 2024 & Ferrag \textit{et al.}  & This paper provides an in-depth review of using Generative AI and Large Language Models (LLMs) in cybersecurity.& \large \ding{52} & \large \ding{52} & \large \ding{52} & \large \ding{52} & \large \ding{52} \\
\hline
\end{tabular}\\
\ding{56} : Not covered; \halfcirc : Partially covered; \ding{52}: Covered;  Data.: Datasets used for training and fine-tuning LLMs for security use cases; Vuln.: LLM Vulnerabilities and Mitigation
; Comp.: Experimental Analysis of LLMs Models’ Performance in Cyber Security Knowledge;  Optim.: Optimization Strategies for Large Language Models in Cybersecurity; Hardw. : Experimental Analysis of LLMs Models’ Performance in Hardware Security.
\label{table:related-reviews}
\end{table*}

\section{Related reviews}
\label{sec:2}
This section delves into a curated collection of recent articles that significantly contribute to the evolving landscape of LLMs and their multifaceted applications. These reviews offer a comprehensive and insightful exploration into various dimensions of LLMs, including their innovative applications in hardware design security, evaluation methodologies, and evolving role in artificial intelligence. Further, they cover cutting-edge advancements in Pre-trained Language Models (PLMs) for NLP, delve into the intricacies of instruction fine-tuning for LLMs, and explore their impactful integration into software engineering. The section also encompasses an in-depth look at multimodal algorithms, examines the critical aspect of alignment requirements for LLMs, and discusses integrating external knowledge into PLMs to enhance NLP tasks. Lastly, it sheds light on the burgeoning field of Controllable Text Generation (CTG) in Natural Language Generation (NLG), highlighting the latest trends and challenges in this dynamic and rapidly advancing area of research \cite{cui2024survey,bai2024beyond,tian2024opportunities}. \textcolor{black}{ Table~\ref{table:related-reviews} presents a comprehensive summary of existing reviews on LLMS across various application domains.}

\subsection{Evaluation of LLMs}
Chang \textit{et al.} \cite{chang2023survey} offers a comprehensive analysis of LLM evaluations, addressing three key aspects: the criteria for evaluation (what to evaluate), the context (where to evaluate), and the methodologies (how to evaluate). It thoroughly reviews various tasks across different domains to understand the successes and failures of LLMs, contributing to future research directions. The paper also discusses current evaluation metrics, datasets, and benchmarks and introduces novel approaches, providing a deep understanding of the current evaluation landscape. Additionally, it highlights future challenges in LLM evaluation and supports the research community by open-sourcing related materials, fostering collaborative advancements in the field.

\subsection{Evolution and State of LLMs in AI}

Zhao \textit{et al.} \cite{zhao2023survey} provides an in-depth survey of LLMs' evolution and current state in artificial intelligence. It traces the progression from statistical language models to neural language models, specifically focusing on the recent emergence of pre-trained language models (PLMs) using Transformer models trained on extensive corpora. The paper emphasizes the significant advancements achieved by scaling up these models, noting that LLMs demonstrate remarkable performance improvements beyond a certain threshold and exhibit unique capabilities not found in smaller-scale models. The survey covers four critical aspects of LLMs: pre-training, adaptation tuning, utilization, and capacity evaluation, providing insights into both their technical evolution and the challenges they pose. Additionally, the paper discusses the resources available for LLM development and explores potential future research directions, underlining the transformative effect of LLMs on AI development and application.

\subsection{Advancements in PLMs for NLP}
Min \textit{et al.} \cite{min2023recent} surveys the latest advancements in leveraging PLMs for NLP, organizing the approaches into three main paradigms. Firstly, the "Pre-train then Fine-tune" method involves general pre-training on large unlabeled datasets followed by specific fine-tuning for targeted NLP tasks. Secondly, "Prompt-based Learning" uses tailored prompts to transform NLP tasks into formats akin to a PLM's pre-training, enhancing the model's performance, especially in few-shot learning scenarios. Lastly, the "NLP as Text Generation" paradigm reimagines NLP tasks as text generation problems, fully capitalizing on the strengths of generative models like GPT-2 and T5. These paradigms represent the cutting-edge methods in utilizing PLMs for various NLP applications.

\subsection{Instruction Fine-Tuning for LLMs}

Zhang \textit{et al.} \cite{zhang2023instruction} delves into the field of instruction fine-tuning for LLMs, offering a detailed exploration of various facets of this rapidly advancing area. It begins with an overview of the general methodologies used in instruction fine-tuning, then discusses the construction of commonly-used, representative datasets tailored for this approach. The survey highlights a range of instruction-fine-tuned models, showcasing their diversity and capabilities. It also examines multi-modality techniques and datasets, including those involving images, speech, and video, reflecting the broad applicability of instruction tuning. The adaptation of LLMs to different domains and applications using instruction tuning strategies is reviewed, demonstrating the versatility of this method. Additionally, the survey addresses efforts to enhance the efficiency of instruction fine-tuning, focusing on reducing computational and time costs. Finally, it evaluates these models, including performance analysis and critical perspectives, offering a holistic view of the current state and potential of instruction fine-tuning in LLMs.

\subsection{LLMs in Software Engineering}

Fan \textit{et al.} \cite{fan2023large} present a survey on using LLMs in Software Engineering (SE), highlighting their potential applications and open research challenges. LLMs, known for their emergent properties, offer novel and creative solutions across various Software Engineering activities, including coding, design, requirements analysis, bug fixing, refactoring, performance optimization, documentation, and analytics. Despite these advantages, the paper also acknowledges the significant technical challenges these emergent properties bring, such as the need for methods to eliminate incorrect solutions, notably hallucinations. The survey emphasizes the crucial role of hybrid approaches, which combine traditional Software Engineering techniques with LLMs, in developing and deploying reliable, efficient, and effective LLM-based solutions for Software Engineering. This approach suggests a promising pathway for integrating advanced AI models into practical software development processes.

\subsection{Multimodal Algorithms}
Wu \textit{et al.} \cite{wu2023multimodal} addresses a significant gap in understanding multimodal algorithms by providing a comprehensive overview of their definition, historical development, applications, and challenges. It begins by defining multimodal models and algorithms, then traces their historical evolution, offering insights into their progression and significance. The paper serves as a practical guide, covering various technical aspects essential to multimodal models, such as knowledge representation, selection of learning objectives, model construction, information fusion, and prompts. Additionally, it reviews current algorithms employed in multimodal models and discusses commonly used datasets, thus laying a foundation for future research and evaluation in this field. The paper concludes by exploring several applications of multimodal models and delving into key challenges that have emerged from their recent development, shedding light on both the potential and the limitations of these advanced computational tools.

\subsection{Alignment Requirements for LLMs}

Liu \textit{et al.} \cite{liu2023trustworthy} propose a taxonomy of alignment requirements for LLMs to aid practitioners in understanding and effectively implementing alignment dimensions and inform data collection efforts for developing robust alignment processes. The paper dissects the concept of "harmful" generated content into specific categories, such as harm to individuals (like emotional harm, offensiveness, and discrimination), societal harm (including instructions for violent or dangerous behaviors), and harm to stakeholders (such as misinformation impacting business decisions). Citing an imbalance in Anthropic’s alignment data, the paper points out the uneven representation of various harm categories, like the high frequency of "violence" versus the marginal appearance of "child abuse" and "self-harm." This observation supports the argument that alignment techniques heavily dependent on data cannot ensure that LLMs will uniformly align with human behaviors across all aspects. The authors' own measurement studies reveal that aligned models do not consistently show improvements across all harm categories despite the alignment efforts claimed by the model developers. Consequently, the paper advocates for a framework that allows a more transparent, multi-objective evaluation of LLM trustworthiness, emphasizing the need for a comprehensive and balanced approach to alignment in LLM development.

\subsection{Knowledge-Enhanced Pre-trained Language Models}
Hu \textit{et al.} \cite{hu2023survey} offers a comprehensive review of Knowledge-Enhanced Pre-trained Language Models (KE-PLMs), a burgeoning field aiming to address the limitations of standard PLMs in NLP. While PLMs trained on vast text corpora demonstrate impressive performance across various NLP tasks, they often fall short in areas like reasoning due to the absence of external knowledge. The paper focuses on how incorporating different types of knowledge into PLMs can overcome these shortcomings. It introduces distinct taxonomies for Natural Language Understanding (NLU) and Natural Language Generation (NLG) to distinguish between these two core areas of NLP. For NLU, the paper categorizes knowledge types into linguistic, text, knowledge graph (KG), and rule knowledge. In the context of NLG, KE-PLMs are classified into KG-based and retrieval-based methods. By outlining these classifications and exploring the current state of KE-PLMs, the paper provides not only clear insights into this evolving domain but also identifies promising future directions for the development and application of KE-PLMs, highlighting their potential to enhance the capabilities of PLMs in NLP tasks significantly.

\subsection{Controllable Text Generation in NLG}
Zhang~\cite{zhang2023survey} provides a critical and systematic review of Controllable Text Generation (CTG), a burgeoning field in NLG that is essential for developing advanced text generation technologies tailored to specific practical constraints. The paper focuses on using large-scale pre-trained language models (PLMs), particularly those based on transformer architecture, which have established a new paradigm in NLG due to their ability to generate more diverse and fluent text. However, the limited interpretability of deep neural networks poses challenges to the controllability of these methods, making transformer-based PLM-driven CTG a rapidly evolving and challenging research area. The paper surveys various approaches that have emerged in the last 3-4 years, each targeting different CTG tasks with varying controlled constraints. It provides a comprehensive overview of common tasks, main approaches, and evaluation methods in CTG and discusses the current challenges and potential future directions in the field. Claiming to be the first to summarize state-of-the-art CTG techniques from the perspective of Transformer-based PLMs, this paper aims to assist researchers and practitioners in keeping pace with the academic and technological developments in CTG, offering them an insightful landscape of the field and a guide for future research.

\subsection{LLM for Cyber Security}
Zhang \textit{et al.} \cite{zhang2024llms} examines the integration of LLMs within cybersecurity. Through an extensive literature review involving over  127 publications from leading security and software engineering venues, this paper aims to shed light on LLMs' multifaceted roles in enhancing cybersecurity measures. The survey pinpoints various applications for LLMs in detecting vulnerabilities, analyzing malware, and managing network intrusions and phishing threats. It highlights the current limitations regarding the datasets used, which often lack size and diversity, thereby underlining the necessity for more robust datasets tailored to these security tasks. The paper also identifies promising methodologies like fine-tuning and domain-specific pre-training, which could better harness the potential of LLMs in cybersecurity contexts.

Yao \textit{et al.} \cite{yao2024survey} explores the dual role of LLMs in security and privacy, highlighting their benefits in enhancing code security and data confidentiality and detailing potential risks and inherent vulnerabilities. The authors categorize the applications and challenges into "The Good," "The Bad," and "The Ugly," where they discuss LLMs' positive impacts, their use in offensive applications, and their susceptibility to specific attacks, respectively. The paper emphasizes the need for further research on threats like model and parameter extraction attacks and emerging techniques such as safe instruction tuning, underscoring the complex balance between leveraging LLMs for improved security and mitigating their risks.

\textcolor{black}{Saha \textit{et al.} \cite{saha2023llm} discussed several key applications of LLMs in the context of hardware design security. The paper illustrates how LLMs can intentionally introduce vulnerabilities and weaknesses into RTL (Register-Transfer Level) designs. This process is guided by well-crafted prompts in natural language, demonstrating the model's ability to understand and manipulate complex technical designs. The authors explore using LLMs to assess the security of hardware designs. The model is employed to identify vulnerabilities, weaknesses, and potential threats. It's also used to pinpoint simple coding issues that could evolve into significant security bugs, highlighting the model's ability to evaluate technical designs critically. In this application, LLMs verify whether a hardware design adheres to specific security rules or policies. The paper examines the model's proficiency in calculating security metrics, understanding security properties, and generating functional testbenches to detect weaknesses. This part of the study underscores the LLM's ability to conduct thorough and detailed verification processes. Finally, the paper investigates how effectively LLMs can be used to develop countermeasures against existing vulnerabilities in a design. This aspect focuses on the model's capability to solve problems and create solutions to enhance the security of hardware designs. Overall, the paper presents an in-depth analysis of how LLMs can be a powerful tool in various stages of hardware design security, from vulnerability introduction and assessment to verification and countermeasure development.}

\subsection{Our survey compared to related surveys}
Our paper presents a more specialized and technical exploration of generative artificial intelligence and large language models in cybersecurity than the previous literature review. Focusing on a broad array of cybersecurity domains such as hardware design security, intrusion detection systems, and software engineering, it targets a wider professional audience, including engineers, researchers, and industrial practitioners. This paper reviews 35 leading models like GPT-4, BERT, Falcon, and LLaMA, not only highlighting their applications but also their developmental trajectories, thereby providing a comprehensive insight into the current capabilities and future potentials of these models in cybersecurity. 

The paper also delves deeply into the vulnerabilities associated with LLMs, such as prompt injection, adversarial natural language instructions, and insecure output handling. It presents sophisticated attack scenarios and robust mitigation strategies, offering a detailed analysis crucial for understanding and protecting against potential threats. Additionally, the lifecycle of specialized cybersecurity datasets—covering creation, cleaning, preprocessing, annotation, and labeling—is scrutinized, providing essential insights into improving data integrity and utility for training and testing LLMs. This level of detail is vital for developing robust cybersecurity solutions that can effectively leverage the power of LLMs.

Lastly, the paper examines the challenges associated with deploying LLMs in cybersecurity contexts, emphasizing the necessity for model robustness and the implications of adversarial attacks. It introduces advanced methodologies such as Reinforcement Learning with Human Feedback (RLHF) and Retrieval-Augmented Generation (RAG) to enhance real-time cybersecurity operations. This focus not only delineates the current state of LLM applications in cybersecurity but also sets the direction for future research and practical applications, aiming to optimize and secure LLM deployments in an evolving threat landscape. This makes the paper an indispensable resource for anyone involved in cybersecurity and AI, bridging the gap between academic research and practical applications.



\section{Preliminaries of NLP for Cyber Security} \label{sec:3}

This section presents the preliminaries of NLP for cybersecurity, including recurrent neural networks (LSTMs and GRUs) and transformer models. 

\subsection{Recurrent neural networks}
Recurrent Neural Networks (RNNs) \cite{rumelhart1986learning} are artificial neural networks that handle data sequences such as time series or NLP tasks. The RNN model consists of two linked recurrent neural networks. The first RNN encodes sequences of symbols into a fixed-length vector, while the second decodes this vector into a new sequence. This architecture aims to maximize the conditional probability of a target sequence from a given source sequence. When applied to cybersecurity, this model could be instrumental in threat detection and response systems by analyzing and predicting network traffic or log data sequences that indicate malicious activity. Integrating the conditional probabilities generated by this model could enhance anomaly detection frameworks, improving the identification of subtle or novel cyber threats. The model's ability to learn meaningful representations of data sequences further supports its potential to recognize complex patterns and anomalies in cybersecurity environments \cite{kasongo2023deep,sohi2021rnnids}.

\subsubsection{Gated Recurrent Units}

GRUs are a recurrent neural network architecture designed to handle the vanishing gradient problem that can occur with standard recurrent networks. Introduced by Cho et al. in 2014 \cite{cho2014learning}, GRUs simplify the LSTM (Long Short-Term Memory) model while retaining its ability to model long-term dependencies in sequential data. GRUs achieve this through two main gates: the update gate, which controls how much a new state overwrites the old state, and the reset gate, which determines how much past information to forget. These gates effectively regulate the flow of information, making GRUs adept at tasks like time series prediction, speech recognition, and natural language processing. The main steps of GRUs are organized as follows:

\begin{itemize}

\item Update Gate: The update gate determines how much information from the previous hidden state should be passed to the new one. The update gate is calculated using the following formula:
\begin{tcolorbox}[colback=gray!10]
\begin{equation} 
z_t = \sigma(W_{z}x_t + U_{z}h_{t-1})
\end{equation}
\end{tcolorbox}

where $z_t$ is the update gate at time step $t$, $W_z$ and $U_z$ are the weight matrices, $x_t$ is the input at time step $t$, and $h_{t-1}$ is the previous hidden state. The sigmoid function, represented by $\sigma$, squishes the equation's results between 0 and 1. The update gate allows the GRU to decide how much of the previous hidden state information should be passed on to the new hidden state. If the update gate is close to 1, it means that a lot of the previous hidden state information should be passed on, while if the update gate is close to 0, it means that very little of the previous hidden state information should be passed on. The Update Gate formula can be explored in the following three different parts:

Part 1: Linear combination of inputs:
\begin{tcolorbox}[colback=gray!10]
\begin{equation}
z_t = W_{r}x_t + U_{r}h_{t-1}
\end{equation}
\end{tcolorbox}

Part 2: Application of the sigmoid function:
\begin{tcolorbox}[colback=gray!10]
\begin{equation}
\tilde{r}_t = \sigma(z_t)
\end{equation}
\end{tcolorbox}

Part 3: Element-wise multiplication of $\tilde{r}_t$ and previous hidden state:
\begin{tcolorbox}[colback=gray!10]
\begin{equation}
r_t = \tilde{r}_t \odot h_{t-1}
\end{equation}
\end{tcolorbox}

Where $\odot$ represents the Hadamard product, also known as the element-wise multiplication.

\item Reset Gate: The reset gate determines how much of the previous hidden state should be forgotten. The reset gate is calculated using the following formula:

\begin{tcolorbox}[colback=gray!10]
\begin{equation}
r_t = \sigma(W_{r}x_t + U_{r}h_{t-1})
\end{equation}
\end{tcolorbox}

where $r_t$ is the reset gate at time step $t$, $W_r$ and $U_r$ are the weight matrices, $x_t$ is the input at time step $t$, and $h_{t-1}$ is the previous hidden state.

\item Candidate Hidden State: The candidate's hidden state combines the input and the previous hidden state, filtered through the reset gate. The candidate's hidden state is calculated using the following formula:

\begin{tcolorbox}[colback=gray!10]
\begin{equation} 
\tilde{h}_t = \tanh(Wx_t + U(r_t \odot h_{t-1}))
\end{equation}
\end{tcolorbox}

where $\tilde{h}_t$ is the candidate hidden state at time step $t$, $W$ and $U$ are the weight matrices, $x_t$ is the input at time step $t$, and $r_t$ is the reset gate at time step $t$. In this equation, the input at time step $t$ ($x_t$) is combined with the previous hidden state ($h_{t-1}$) through the weight matrices $W$ and $U$. The reset gate ($r_t$) is used to control the extent to which the previous hidden state ($h_{t-1}$) is passed to the candidate hidden state ($\tilde{h}_t$). The element-wise product between the reset gate ($r_t$) and the previous hidden state ($h_{t-1}$) is used to create the reset vector ($r_t \odot h_{t-1}$). The reset vector is combined with the input ($x_t$) through the weight matrix $U$. Finally, the result is passed through the hyperbolic tangent function to calculate the candidate hidden state ($\tilde{h}_t$).

\item New Hidden State: The new hidden state combines the previous and candidate hidden states, filtered through the update gate. The new hidden state is calculated using the following formula:

\begin{tcolorbox}[colback=gray!10]
\begin{equation} 
h_t = (1-z_t) \odot h_{t-1} + z_t \odot \tilde{h}_t
\end{equation}
\end{tcolorbox}

where $h_t$ is the new hidden state at time step $t$, $z_t$ is the update gate at time step $t$, and $\tilde{h}_t$ is the candidate hidden state at time step $t$. The new hidden state ($h_t$) is calculated by taking a weighted combination of the previous hidden state ($h_{t-1}$) and the candidate hidden state ($\tilde{h}_t$). The weight of the previous hidden state is determined by the update gate ($z_t$) - if the update gate is close to 1, the new hidden state is primarily influenced by the previous hidden state. If the update gate is close to 0, the candidate's hidden state primarily influences the new hidden state. The element-wise product between the update gate ($z_t$) and the candidate hidden state ($\tilde{h}_t$) is used to create the update vector ($z_t \odot \tilde{h}_t$). The element-wise product between the complement of the update gate ($1-z_t$) and the previous hidden state ($h_{t-1}$) is used to create the reset vector ($(1-z_t) \odot h_{t-1}$). Finally, the update and reset vectors are added to calculate the new hidden state ($h_t$).
\end{itemize}

\subsubsection{Long Short-Term Memory}

The LSTM \cite{hochreiter1997long} was designed to overcome the vanishing gradient problem that affects traditional recurrent neural networks (RNNs) during training, particularly over long sequences. By integrating memory cells that can maintain information over extended periods and gates that regulate the flow of information into and out of the cell, LSTMs provide an effective mechanism for learning dependencies and retaining information over time. This architecture has proved highly influential, becoming foundational to numerous applications in machine learning that require handling sequential data, such as natural language processing, speech recognition, and time series analysis. The impact of this work has been extensive, as it enabled the practical use and development of deep learning models for complex sequence modeling tasks. In cybersecurity, LSTMs can be used for anomaly detection, where they analyze network traffic or system logs to identify unusual patterns that may signify a security breach or malicious activity \cite{sedjelmaci2020cyber,dixit2021deep,gaba2024systematic}. Their ability to learn from long sequences makes them particularly useful for detecting sophisticated attacks that evolve, such as advanced persistent threats (APTs) and ransomware. The main steps of LSTM models are organized as follows:

\begin{itemize}
\item Input Gate: The first step in an LSTM-based RNN involves calculating the input gate. The input gate determines the extent of new input to be added to the current state. The formula for the input gate is:

\begin{tcolorbox}[colback=gray!10]
\begin{equation}   
i_t = \sigma(W_i \cdot [h_{t-1}, x_t] + b_i)
\end{equation}
\end{tcolorbox}

where \(i_t\) is the input gate at time step \(t\), \(W_i\) is the weight matrix for the input gate, \(h_{t-1}\) is the hidden state from the previous time step, \(x_t\) is the input at time step \(t\), and \(b_i\) is the bias for the input gate. The function \(\sigma(x)\) is the sigmoid activation function. This formula calculates the input gate \(i_t\) by first concatenating the previous hidden state \(h_{t-1}\) with the current input \(x_t\). This combined vector is multiplied by the weight matrix \(W_i\), and the bias \(b_i\) is added. Finally, the sigmoid activation function is applied to produce \(i_t\), which ranges from 0 to 1 and represents how much the current input updates the hidden state.

\item Forget Gate: The second step calculates the forget gate, determining how much the previous state should be forgotten. The formula for the forget gate is:

\begin{tcolorbox}[colback=gray!10]
\begin{equation}  
f_t = \sigma(W_f \cdot [h_{t-1}, x_t] + b_f)
\end{equation}
\end{tcolorbox}

where \(f_t\) is the forget gate at time step \(t\), \(W_f\) is the weight matrix for the forget gate, \(h_{t-1}\) is the hidden state from the previous time step, \(x_t\) is the input at time step \(t\), and \(b_f\) is the bias for the forget gate. The forget gate \(f_t\) is calculated like the input gate. It involves concatenating the previous hidden state \(h_{t-1}\) with the current input \(x_t\), multiplying by the weight matrix \(W_f\), and adding the bias \(b_f\). The resulting value is passed through the sigmoid activation function to determine the forget gate \(f_t\), which ranges from 0 to 1 and represents the degree to which the previous hidden state is preserved or forgotten in the current hidden state.

\item Candidate Memory Cell: The third step calculates the candidate memory cell, representing the potential memory state update. The formula for the candidate memory cell is:

\begin{tcolorbox}[colback=gray!10]
\begin{equation}  
\tilde{c}_t = \tanh(W_c \cdot [h_{t-1}, x_t] + b_c)
\end{equation}
\end{tcolorbox}

where \(\tilde{c}_t\) is the candidate memory cell at time step \(t\), \(W_c\) is the weight matrix for the candidate memory cell, \(h_{t-1}\) is the hidden state from the previous time step, \(x_t\) is the input at time step \(t\), and \(b_c\) is the bias for the candidate memory cell. The function \(\tanh(x)\) is the hyperbolic tangent activation function. In this formula, the candidate memory cell \(\tilde{c}_t\) is calculated by concatenating the previous hidden state \(h_{t-1}\) and the input \(x_t\), then multiplying by the weight matrix \(W_c\) and adding the bias \(b_c\). The result is passed through the hyperbolic tangent activation function, which ranges from -1 to 1, to control the magnitude of the memory cell update.

\item Current Memory Cell: The fourth step calculates the current memory cell, which is the updated state of the memory cell, combining the effects of the forget and input gates. The formula for the current memory cell is:

\begin{tcolorbox}[colback=gray!10]
\begin{equation}  
c_t = f_t \cdot c_{t-1} + i_t \cdot \tilde{c}_t
\end{equation}
\end{tcolorbox}

where \(c_t\) is the current memory cell at time step \(t\), \(f_t\) is the forget gate at time step \(t\), \(c_{t-1}\) is the memory cell from the previous time step, \(i_t\) is the input gate at time step \(t\), and \(\tilde{c}_t\) is the candidate memory cell at time step \(t\). This equation represents the new memory cell state as a combination of the old state (modulated by the forget gate) and the potential update (modulated by the input gate).

\item Output Gate: The final step calculates the output gate, which determines the amount of information output from the LSTM cell. The details and formula for the output gate should follow.

\end{itemize}

\begin{figure}[t]
    \centering
    \includegraphics[width=0.5\textwidth]{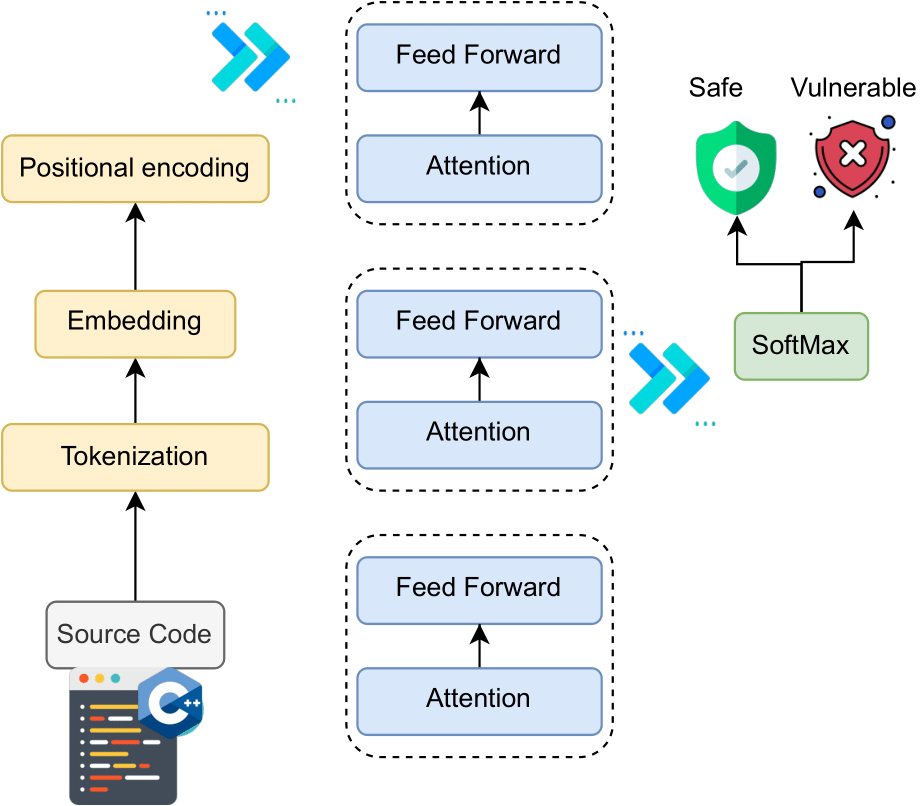}
    \caption{How Transformer works for Software Security.}
    \label{fig:transformer}
\end{figure}

\subsection{Transformer models}

The Transformer architecture proposed by Vaswani et al. \cite{vaswani2017attention} in 2017 is a significant advancement in natural language processing built entirely around attention mechanisms. These mechanisms allow the model to assess the relevance of different words in a sentence, independent of their positional relationships. This foundational technology has enhanced the efficiency of tasks like translation and text summarization and has broad cybersecurity applications. In cybersecurity, Transformer models can detect and respond to threats by analyzing source code patterns and network traffic and identifying anomalies in system logs, as presented in Figure \ref{fig:transformer}. They can also be used for the automated generation of security policies based on the evolving landscape of threats and for intelligent threat hunting, where the system predicts and neutralizes threats before they cause harm. This makes Transformers versatile in enhancing security protocols and defending against cyber attacks \cite{ferrag2024revolutionizing}. The main steps of Transformer models are organized as follows:

\begin{itemize}

\item Attention Mechanism: The attention mechanism in the Transformer model computes attention scores between the input and output representations. These scores are calculated using the scaled dot-product of the query and key representations and then normalized by a softmax function. The attention scores are subsequently used to compute a weighted sum of the value representations, forming the output of the attention mechanism.

The equation defines the attention scores:

\begin{tcolorbox}[colback=gray!10]
\begin{equation}
\text{Attention}(Q, K, V) = \text{softmax}\left(\frac{QK^T}{\sqrt{d_k}}\right)V
\end{equation}
\end{tcolorbox}

Where:

$Q$, $K$, and $V$ represent the matrices of queries, keys, and values transformed from the input representations. The dimension of the keys is denoted by $d_k$. The attention mechanism involves computing the dot product of $Q$ and the transpose of $K$, which is then scaled by the inverse square root of $d_k$ to stabilize the gradients. The result is passed through a softmax function to normalize the scores, ensuring they sum to 1. These scores, representing attention weights, compute a weighted sum of the values in $V$, resulting in the final attention output. This mechanism allows the model to dynamically focus on the most relevant parts of the input sequence for making predictions.

\item Multi-Head Attention:
In the Transformer model, multiple attention heads enhance the model's capability to simultaneously focus on different parts of the input sequence. The multi-head attention is calculated as follows:

\begin{tcolorbox}[colback=gray!10]
\begin{equation} 
\text{MHead}(Q, K, V) = \text{Concat}(hd_1, hd_2, \ldots, hd_h) W^O
\end{equation}
\end{tcolorbox}

Where: $hd_i$ represents the output of the $i$-th attention head, computed using the attention formula $\text{Attention}(Q_i, K_i, V_i)$. Each $Q_i$, $K_i$, and $V_i$ are different linear projections of the original inputs $Q$, $K$, and $V$. $W^O$ is a linear transformation matrix applied to the concatenated results of all attention heads. The $\text{Concat}$ function concatenates the outputs of each head along a specific dimension. The outputs of individual heads, $hd_i$, are each computed using the scaled dot-product attention mechanism:

\begin{tcolorbox}[colback=gray!10]
\begin{equation}
hd_i = \text{Attention}(Q_i, K_i, V_i) = \text{softmax}\left(\frac{Q_i K_i^T}{\sqrt{d_k}}\right) V_i
\end{equation}
\end{tcolorbox}

This approach enables the Multi-Head Attention mechanism to capture various aspects of the input sequence, simultaneously focusing on different subspace representations. As a result, it facilitates the model's capture of more complex relationships and improves performance across different types of tasks.

\item Layer Normalization:
In the Transformer model, layer normalization ensures the input is within a standard range. The layer normalization can be calculated as follows:

\begin{tcolorbox}[colback=gray!10]
\begin{equation}  
LN(x) = \frac{x - \text{mean}(x)}{\sqrt{\text{var}(x)}}
\end{equation}
\end{tcolorbox}

Where $x$ is the input to the layer normalization. $ textmean (x) $ and $ textvar (x) $ are the mean and variance of $ x$, respectively. Layer Normalization aims to mitigate the internal covariate shift, which arises when the distribution of activations in a layer changes during training. The normalization operation is performed by subtracting the mean of the activations and dividing by the square root of the variance. This ensures the activations have a zero mean and unit variance, leading to more stable training.

\item Position-wise Feed Forward:
The position-wise feed-forward network transforms the input and output representations in the Transformer model. The position-wise feedforward can be calculated as follows:

\begin{tcolorbox}[colback=gray!10]
\begin{equation}  
FFN(x) = \max(0, xW_1 + b_1)W_2 + b_2
\end{equation}
\end{tcolorbox}

Where: $x$ is the input to the feed-forward network. $W_1$, $b_1$, $W_2$, and $b_2$ are the weight and bias parameters of the feed-forward network. $\max(0, x)$ is the ReLU activation function. This equation represents a simple feed-forward neural network (FFN) operation in deep learning models. The FFN operation is a multi-layer perceptron (MLP) that transforms the input $x$ into a new representation by passing it through two fully connected (dense) layers. The first layer is followed by a ReLU activation function, which applies a non-linear activation to the input by setting all negative values to zero. This activation function helps the model learn complex non-linear relationships between the input and output. The second layer is a linear transformation that produces the final output of the FFN. The weight and bias parameters of the two layers, $W_1$, $b_1$, $W_2$, and $b_2$, are learned during training and allow the model to learn different representations of the input data.

\item Encoder and Decoder Blocks:
In the Transformer model, the encoder and decoder blocks transform the input sequences into the output sequences. The encoder and decoder blocks can be calculated as follows:

\begin{tcolorbox}[colback=gray!10]
\begin{equation} 
\text{Enc}(x) = \text{LN}(x + \text{MHead}(x, x, x))
\end{equation}
\end{tcolorbox}

\begin{tcolorbox}[colback=gray!10]
\begin{equation}
\text{Dec}(x, y) = \text{LN}(x + \text{MHead}(x, y, y) + \text{MHead}(x, x, x))
\end{equation}
\end{tcolorbox}

Where: $x$ is the input to the encoder/decoder block. $y$ is the output from the previous encoder/decoder block. The Encoder block $\text{Enc}(x)$ takes the input $x$ and applies the Multi-Head Attention mechanism to compute the attention scores between the input and itself. The result is then added to the input and passed through a Layer Normalization operation. The output of the encoder block is the new representation of the input after processing through the Multi-Head Attention and Layer Normalization operations. The Decoder block $\text{Dec}(x, y)$ is similar to the encoder block but also takes the output from the previous decoder block, $y$, as input. The Multi-Head Attention mechanism is applied to compute the attention scores between the input and the previous output and between the input and itself. The results are added to the input and passed through a Layer Normalization operation. The output of the decoder block is the new representation of the input after processing through the Multi-Head Attention and Layer Normalization operations.

\end{itemize}

\begin{table*}[t!]
\centering
\setlength{\tabcolsep}{2.5pt}
\renewcommand{\arraystretch}{1}
\caption{RNN-based models for Cyber Security.}
\centering
\scriptsize
\label{tab:rnnmodels}
\begin{tabular}{|p{0.5in}|p{0.2in}|p{0.8in}|p{1.5in}|p{0.8in}|p{1.4in}|p{1.4in}|}
\hline
\textbf{Study}& \textbf{Year} & \textbf{Type of Model}                                                           & \textbf{Dataset Used}                                                 & \textbf{Domain}                          & \textbf{Key Contributions}         & \textbf{Open Issues}      \\ \hline                 
Yin \textit{et al.}  \cite{yin2017deep}  & 2017     & RNN-ID   (Recurrent Neural Network-Intrusion Detection)                          & Benchmark   data set                                                  & Intrusion   Detection                    & The proposed model can improve the accuracy of intrusion detection   &    Other machine learning algorithms and deep learning models, such as convolutional neural networks and transformers are not considered in the comparison                                                           \\ \hline
G{\"u}era \textit{et al.}  \cite{guera2018deepfake} & 2018      & Temporal-aware   Pipeline (CNN and RNN)                                          & Large set of deepfake videos collected from multiple video websites & Detection   of Deepfake Videos           & The proposed method achieves competitive results in detecting deepfake videos while using a simple architecture      &    The proposed approach's effectiveness might be limited to the specific types of deepfakes present in the dataset     \\ \hline
Althubiti \textit{et al.}  \cite{althubiti2018applying}  & 2018     & LSTM   RNN                                                                       & CSIC   2010 HTTP dataset                                              & Web   Intrusion Detection                & Proposal of LSTM RNN for web intrusion detection. High accuracy rate (0.9997) in binary classification.      &     The paper only uses the CSIC 2010 HTTP dataset, which may not be representative of all types of web application attacks                                                                   \\ \hline
Xu \textit{et al.}  \cite{xu2018intrusion}  & 2018    & GRU-MLP-Softmax   (Gated Recurrent Unit, Multilayer Perceptron, Softmax)         & KDD 99   and NSL-KDD data sets                                        & Network   Intrusion Detection            & The system achieves leading performance with overall detection rates of 99.42\% using KDD 99 and 99.31\% using NSL-KDD, with low false positive rates   &  The paper does not provide information about the scalability of the proposed model    \\ \hline
Ferrag and Leandros  \cite{ferrag2019deepcoin}  & 2019     & Blockchain and RNN                                     & CICID2017   dataset, Power system dataset, Bot-IoT dataset            & Energy   framework for Smart Grids       & Proposal of DeepCoin framework combining blockchain and deep learning for smart grid security & The paper does not address the potential scalability issues that may arise as the number of nodes in the network increases  \\ \hline
Chawla \textit{et al.} \cite{chawla2019host}        & 2019  &  GRU with CNN  & ADFA (Australian Defence Force Academy) dataset      &   Intrusion Detection     &  Achieved improved performance by combining GRUs and CNNs   &  The proposed system is vulnerable to adversarial attacks  \\ \hline
Ullah \textit{et al.} \cite{ullah2022design}        & 2022 & LSTM, BiLSTM, and GRU    &   IoT-DS2, MQTTset, IoT-23, and datasets     &  Intrusion Detection   &  Validation of the proposed models using various datasets, achieving high accuracy, precision, recall, and F1 score  & Further research is necessary to enhance their scalability for practical applications in cybersecurity \\ \hline
Donkol \textit{et al.} \cite{donkol2023optimization}   & 2023 &   LSTM  &  CSE-CIC-IDS2018, CICIDS2017, and UNSW-NB15 datasets      &  Intrusion Detection   &  The proposed system outperformed other methods such as LPBoost and DNNs in terms of accuracy, precision, recall, and error rate  &  Future research could explore the applicability of the proposed system to other datasets\\ \hline
Zhao \textit{et al.} \cite{zhao2023ernn}        & 2023 &   End-to-End Recurrent Neural Network  &  IDS2017 and IDS2018 datasets     &   Intrusion attacks and malware & +  Address network-induced phenomena that may result in misclassifications in traffic detection systems used in cybersecurity   & The proposed system is vulnerable to adversarial attacks \\ \hline
Wang \textit{et al.} \cite{wang2021patchrnn}        & 2021 &   RNN  &  A large-scale patch dataset PatchDB     &   Software Security  &  The PatchRNN system can effectively detect secret security patches with a low false positive rate  & The PatchRNN system can only support C/C++\\ \hline
Polat \textit{et al.} \cite{polat2022novel}        & 2022 & LSTM and GRU    &  SDN-based SCADA system     &  Detection of DDoS attacks   & The results show that the proposed RNN model achieves an accuracy of 97.62\% for DDoS attack detection  & The paper only focuses on detecting DDoS attacks and does not address other types of cyber threats (e.g., insider threats or advanced persistent threats) \\ \hline
\end{tabular}
\end{table*}

\begin{table*}[t!]
\centering
\setlength{\tabcolsep}{2.5pt}
\renewcommand{\arraystretch}{1}
\caption{Transformer-based models for Cyber Security (Part I).}
\centering
\scriptsize
\label{tab:tranmodelspart1}
\begin{tabular}{|p{0.5in}|p{0.2in}|p{0.8in}|p{1.5in}|p{0.8in}|p{1.4in}|p{1.4in}|}
\hline
\textbf{Study}& \textbf{Year} & \textbf{Type of Model}                                                           & \textbf{Dataset Used}                                                 & \textbf{Domain}                          & \textbf{Key Contributions}         & \textbf{Open Issues}                                                                                                                                                                                                                                \\ \hline
Parra \textit{et al.} \cite{parra2022interpretable}        & 2022 &Federated   Transformer Log Learning Model                                       & HDFS   and CTDD datasets                                              & Threat detection and forensics      & The interpretability module integrated into the model provides insightful interpretability of the model's decision-making process                                                                                        & The paper briefly mentions the applicability of the proposed approach in edge computing systems but does not discuss the scalability of the approach to larger systems \\ \hline
Ziems \textit{et al.} \cite{ziems2021security}        & 2021 & Transformer   Model, BERT, CANINE, Bagging-based random transformer forest (RTF) & Malware   family datasets                                             & Malware Classification & Demonstration   that transformer-based models outperform traditional machine and deep learning   models in classifying malware families                                                   &  The experiments are conducted on preprocessed NIST NVD/SARD databases, which may not reflect real-world conditions \\ \hline
Wu \textit{et al.} \cite{wu2022rtids}        & 2022& Robust   Transformer-based Intrusion Detection System (RTID)                     & CICID2017   and CIC-DDoS2019 datasets                                 & Intrusion Detection    & The proposed method outperforms classical machine learning algorithms such as support vector machine (SVM) and deep learning algorithms (i.e., RNN, FNN, LSTM) on the two evaluated datasets    &     There is no discussion in the paper regarding the scalability of the proposed method, particularly when dealing with large-scale and real-time network traffic                                                                                            \\ \hline
Demirk{\i}ran \textit{et al.} \cite{demirkiran2022ensemble}   & 2022   &     Transformer-based models                                                 &   Catak dataset, Oliveira dataset, VirusShare dataset, and VirusSample dataset       &  Malware classification & The paper demonstrates that transformer-based models, specifically BERT and CANINE, outperform traditional machine and deep learning models in classifying malware families    &       The study only focuses on malware families that use API call sequences, which means that it does not consider other malware types that may not use API calls                                                                         \\ \hline

Ghourbi \textit{et al.} \cite{ghourabi2022security} & 2022 & An optimized LightGBM model and a Transformer-based model & ToN-IoT and Edge IIoTset datasets & Threat Detection & The experimental evaluation of the approach showed remarkable accuracies of 99\% & The paper does not discuss the scalability of the proposed system for large-scale healthcare networks\\ \hline

Thapa \textit{et al.} \cite{thapa2022transformer} & 2022 & Transformer-based language models  &  Software vulnerability datasets of C/C++ source codes & Software security and vulnerability detection in programming languages, specifically C/C++ & The paper highlights the advantages of transformer-based language models over contemporary models & The paper only focuses on detecting vulnerabilities in C/C++ source code and does not explore the use of large transformer-based language models in detecting vulnerabilities in other programming languages \\ \hline

Ranade \textit{et al.} \cite{ranade2021generating} &  2021 &  A transformer-based language model, specifically GPT-2 & WebText dataset & Fake Cyber Threat Intelligence & The attack is shown to introduce adverse impacts such as returning incorrect reasoning outputs & Further research is needed to explore how to prevent or detect data poisoning attacks on cyber-defense system\\ \hline

Fu \textit{et al.} \cite{fu2022linevul} & 2022 & Transformer-based line-level vulnerability prediction model & Large-scale real-world dataset with more than 188k C/C++ functions & Software vulnerability prediction in safety-critical software systems &  The proposed system is accurate for predicting vulnerable functions affected by the Top-25 most dangerous CWEs &  The model's performance can be changed when applied to different programming languages or software systems\\ \hline

Mamede \textit{et al.} \cite{mamede2022transformer} & 2022  & A transformer-based deep learning model  & Software Assurance Reference Dataset (SARD) project, which contains vulnerable and non-vulnerable Java files & Software security in the context of Java programming language & The proposed system can identify up to 21 vulnerability types and achieved an accuracy of 98.9\% in multi-label classification & The proposed method cannot be extended to other programming languages and integrated into existing software development processes \\ \hline

Evange \textit{et al.} \cite{evangelatos2021named} & 2021 & A transformer-based model & DNRTI (Dataset for NER in Threat Intelligence) & Cybersecurity threat intelligence & The experimental results demonstrate that transformer-based techniques outperform previous state-of-the-art approaches for NER in threat intelligence &  Further research is needed to test the effectiveness of transformer-based models on larger and more diverse datasets\\ \hline

Hashemi \textit{et al.} \cite{hashemi2023automation} & 2023 & Transformer models (including BERT, XLNet, RoBERTa, and DistilBERT) & Labeled dataset from vulnerability databases & Vulnerability Information Extraction & The proposed approach outperforms existing rule-based and CRF-based models & The paper does not address the issue of bias in the labeled dataset  \\ \hline
Liu \textit{et al.} \cite{liu2022commitbart}   & 2022 &  Transformer model   &   A commit benchmark dataset that includes over 7.99 million commits across 7 programming languages    &  Commit message generation (generation task) and security patch identification (understanding task)   & The experimental results demonstrate that CommitBART significantly outperforms previous pre-trained models for code   & The pre-training dataset used in the paper is limited to GitHub commits\\ \hline
 Ahmad \textit{et al.} \cite{10462177}   & 2024 &  Transformer model   &  Set of 15 hardware security bug benchmark
designs from three sources: MITRE
website, OpenTitan System-on-Chip (SoC) and the Hack@DAC 2021 SoC & Hardware Security Bugs   &  Bug repair potential demonstrated by ensemble of LLMs, outperforming state-of-the-art automated tool  & The need for designer assistance in bug identification, handling complex bugs, limited evaluations due to simulation constraints, and challenges with token limits and repair generation using LLMs\\\hline
Wan \textit{et al.} \cite{10473893}   & 2024 &  Transformer model   &    Chrysalis dataset, comprising over 1,000 function-level HLS designs with injected logical bugs   &  Design Verification  &   Creating the Chrysalis dataset for HLS debugging, and enabling LLM-based bug detection and integration into development environments  & Refining LLM techniques, integrating LLMs into development environments, and addressing scalability and generalization challenges\\\hline  Jang  \textit{et al.} \cite{jang2024ignoreme} 
 &2024 &  Transformer model   & Includes 150K online security articles, 7.3K security paper abstracts, 3.4K Wikipedia articles, and 185K CVE descriptions. & Threat Detection & Pre-trained language model for the cybersecurity domain, CyBERTuned incorporates non-linguistic elements (NLEs) such as URLs and hash values commonly found in cybersecurity texts.& The paper's limitations include a narrow focus on specific non-linguistic element (NLE) types, acknowledging the existence of more complex NLE types like code blocks and file paths that require future exploration \\\hline 
\end{tabular}
\end{table*}

\begin{table*}[t!]
\centering
\setlength{\tabcolsep}{2.5pt}
\renewcommand{\arraystretch}{1}
\caption{Transformer-based models for Cyber Security (Part II).}
\centering
\scriptsize
\label{tab:tranmodelspart2}
\begin{tabular}{|p{0.5in}|p{0.2in}|p{0.8in}|p{1.5in}|p{0.8in}|p{1.4in}|p{1.4in}|}
\hline
\textbf{Study}& \textbf{Year} & \textbf{Type of Model}                                                           & \textbf{Dataset Used}                                                 & \textbf{Domain}                          & \textbf{Key Contributions}         & \textbf{Open Issues}                                                                                                                                                                                                                                \\ \hline
 Bayer \textit{et al.} \cite{bayer2024cysecbert} 
 &2024 &  Transformer model   & A dataset consisting of 4.3 million entries of Twitter, Blogs, Paper, and CVEs related to the cybersecurity domain & Intrusion attacks and malware & Created a high-quality dataset and a domain-adapted language model for the cybersecurity domain, which improves the internal representation space of domain words and performs best in cybersecurity scenarios & the model may not be suitable as a replacement for every type of cybersecurity model. They also state that the hyperparameters may not be generalizable to other language models, especially very large language models \\\hline  Shestov \textit{et al.} \cite{shestov2024finetuning} 
 &2024 &  Transformer model   & The dataset comprises 22,945 function-level source code samples. It includes 13,247 samples for training, 5,131 for validation, and 4,567 for testing & Vulnerability detection & Finetuning the state-of-the-art code LLM, WizardCoder, increasing its training speed without performance harm.  & The proposed study shows that the main bottlenecks of the task that limit performance lie in the field of dataset quality and suggests the usage of the project-level context information \\\hline  He \textit{et al.} \cite{he2024smartcontract} 
 &2024 &  Transformer model   & Used three datasets: one with over 100,000 entries from Ethereum mainnet contracts, another with 892,913 addresses labelled across five vulnerability categories, and a third with 6,498 smart contracts, including 314 associated with Ponzi schemes & blockchain technology and smart contracts &  The introduction of a novel model, BERT-ATT-BiLSTM, for advanced vulnerability detection in smart contracts, and the evaluation of its performance against other models & Include the model's limitation in recognizing unseen contract structures or novel types of vulnerabilities, and the need to incorporate support for multiple programming languages to enhance universality and robustness \\\hline
 \textcolor{black}{ Patsakis \textit{et al.} \cite{PATSAKIS2024124912} } & \textcolor{black}{2024} & \textcolor{black}{LLM fine-tuned for deobfuscation tasks} & \textcolor{black}{Malicious scripts from the Emotet malware campaign} & \textcolor{black}{Malware Classification} & \textcolor{black}{Demonstrated 69.56\% accuracy in extracting URLs and 88.78\% for domains of droppers; explored LLM potential in malware deobfuscation and reverse engineering} & \textcolor{black}{Optimizing LLM fine-tuning for improved accuracy and integrating deobfuscation capabilities into operational security pipelines} \\ \hline
 \textcolor{black}{Guo \textit{et al.} \cite{guo2024outside}} & \textcolor{black}{2024} & \textcolor{black}{Fine-tuned open-source and general-purpose LLMs for binary classification} & \textcolor{black}{Compiled dataset and five benchmark datasets for vulnerability detection} & \textcolor{black}{Software Security} & \textcolor{black}{Demonstrated fine-tuning's effectiveness in improving detection accuracy; highlighted limitations of existing benchmark datasets} & \textcolor{black}{Addressing dataset mislabeling and improving generalizability of models to unseen code scenarios} \\ \hline

 Jamal \textit{et al.} \cite{jamal2024improved} 
 &2024 &  Transformer model   & Two open-source datasets, 747 spam, 189 phishing, 4825 ham; class imbalance addressed with ADASYN & Phishing and spam detection & Proposing IPSDM, a fine-tuned version of DistilBERT and RoBERTA, outperforming baseline models and the demonstration of the effectiveness of LLMs in addressing cybersecurity challenges & Class imbalance, addressed with ADASYN, but potential bias remains \\\hline 
 \textcolor{black}{Lykousas and Patsakis \cite{lykousas2024decoding}} & \textcolor{black}{2024} & \textcolor{black}{LLMs for detecting hard-coded credentials in source code} & \textcolor{black}{Public code repositories with embedded secrets and passwords} & \textcolor{black}{Authentication and Code Security} & \textcolor{black}{Highlighted differences in password patterns between developers and users; evaluated LLMs for detecting hard-coded credentials and discussed their limitations} & \textcolor{black}{Improving LLM accuracy in detecting secrets and addressing context-sensitive password vulnerabilities} \\ \hline

 \textcolor{black}{Karlsen \textit{et al.} \cite{karlsen2024benchmarking}} & \textcolor{black}{2024} & \textcolor{black}{Fine-tuned LLMs for sequence classification (e.g., DistilRoBERTa, GPT-2, GPT-Neo)} & \textcolor{black}{Six datasets from web application and system logs} & \textcolor{black}{Cybersecurity Log Analysis} & \textcolor{black}{Proposed a new pipeline leveraging 60 fine-tuned models for log analysis; DistilRoBERTa achieved an F1-score of 0.998, outperforming state-of-the-art techniques} & \textcolor{black}{Scaling models for more diverse log formats and optimizing for real-time analysis in dynamic environments} \\ \hline

 \textcolor{black}{Mechri \textit{et al.} \cite{mechri2025secureqwen}} & \textcolor{black}{2025} & \textcolor{black}{Decoder-only Transformer with 64K context length} & \textcolor{black}{Python dataset (1.875M function-level code snippets from GitHub, Codeparrot, and GPT4-o-generated data)} & \textcolor{black}{Software Security} & \textcolor{black}{High accuracy in detecting vulnerabilities across 14 CWEs, F1 scores ranging from 84\% to 99\%} & \textcolor{black}{Further improvement in identifying complex vulnerabilities and handling diverse programming patterns} \\ \hline
\textcolor{black}{Ding \textit{et al.} \cite{ding2025smartguard}} & \textcolor{black}{2025} & \textcolor{black}{LLM-enhanced framework with in-context learning and CoT reasoning} & \textcolor{black}{SolidiFI benchmark dataset} & \textcolor{black}{Blockchain Security} & \textcolor{black}{Recall of 95.06\% and F1-score of 94.95\% for detecting smart contract vulnerabilities; self-check architecture for CoT generation} & \textcolor{black}{Expanding the framework's applicability to more complex blockchain environments and new vulnerability types} \\ \hline

\textcolor{black}{Arshad \textit{et al.} \cite{arshad2025blockllm}} & \textcolor{black}{2025} & \textcolor{black}{LLM-based decentralized vehicular network architecture} & \textcolor{black}{Simulation data for vehicular communication scenarios} & \textcolor{black}{Autonomous Transportation Systems} & \textcolor{black}{18\% reduction in latency, 12\% improvement in throughput, and enhanced secure V2X communication using blockchain and LLMs} & \textcolor{black}{Addressing node selfishness, scalability in larger networks, and privacy-preserving real-time data exchange} \\ \hline
\textcolor{black}{Xiao \textit{et al.} \cite{xiao2025logic}} & \textcolor{black}{2025} & \textcolor{black}{LLM with advanced prompting techniques} & \textcolor{black}{Solidity v0.8 vulnerabilities dataset} & \textcolor{black}{Blockchain Security} & \textcolor{black}{Reduced false-positive rates by over 60\%; evaluated latest five LLMs and identified root causes for reduced recall in Solidity v0.8} & \textcolor{black}{Improving recall for newer Solidity versions and adapting to evolving library and framework changes} \\ \hline
\textcolor{black}{Hassanin \textit{et al.} \cite{hassanin2025pllm}} & \textcolor{black}{2025} & \textcolor{black}{Pre-trained Transformer with specialized input transformation module} & \textcolor{black}{UNSW\_NB 15 and TON\_IoT datasets} & \textcolor{black}{Intrusion Detection} & \textcolor{black}{Achieves 100\% accuracy on UNSW\_NB 15 dataset, significantly outperforming BiLSTM, GRU, and CNN models} & \textcolor{black}{Exploring scalability for larger and more diverse datasets; integrating real-time detection capabilities} \\ \hline
\textcolor{black}{Liu \textit{et al.}  \cite{liu2025llm}} & \textcolor{black}{2025} & \textcolor{black}{LLM-powered static binary taint analysis} & \textcolor{black}{Real-world firmware datasets} & \textcolor{black}{Hardware Security Bugs} & \textcolor{black}{Fully automated taint analysis with 37 newly discovered bugs and 10 assigned CVEs; low engineering cost} & \textcolor{black}{Exploring adaptability for diverse binary formats and enhancing real-time analysis capabilities} \\ \hline
\textcolor{black}{Gaber et al. \textit{et al.}  \cite{gaber2025zero}} & \textcolor{black}{2025} & \textcolor{black}{Transformer-based framework for zero-day ransomware detection} & \textcolor{black}{Assembly instructions captured by the Peekaboo tool} & \textcolor{black}{Malware Classification} & \textcolor{black}{Introduced a novel AI-based framework leveraging Assembly data for high-accuracy zero-day ransomware detection; demonstrated the relevance of Transformer models to ransomware classification by aligning with Zipf's law} & \textcolor{black}{Enhancing scalability for larger datasets and addressing advanced evasion techniques in novel ransomware samples} \\ \hline
\end{tabular}
\end{table*}

\begin{figure*}[t]
    \centering
    \includegraphics[width=0.9\textwidth]{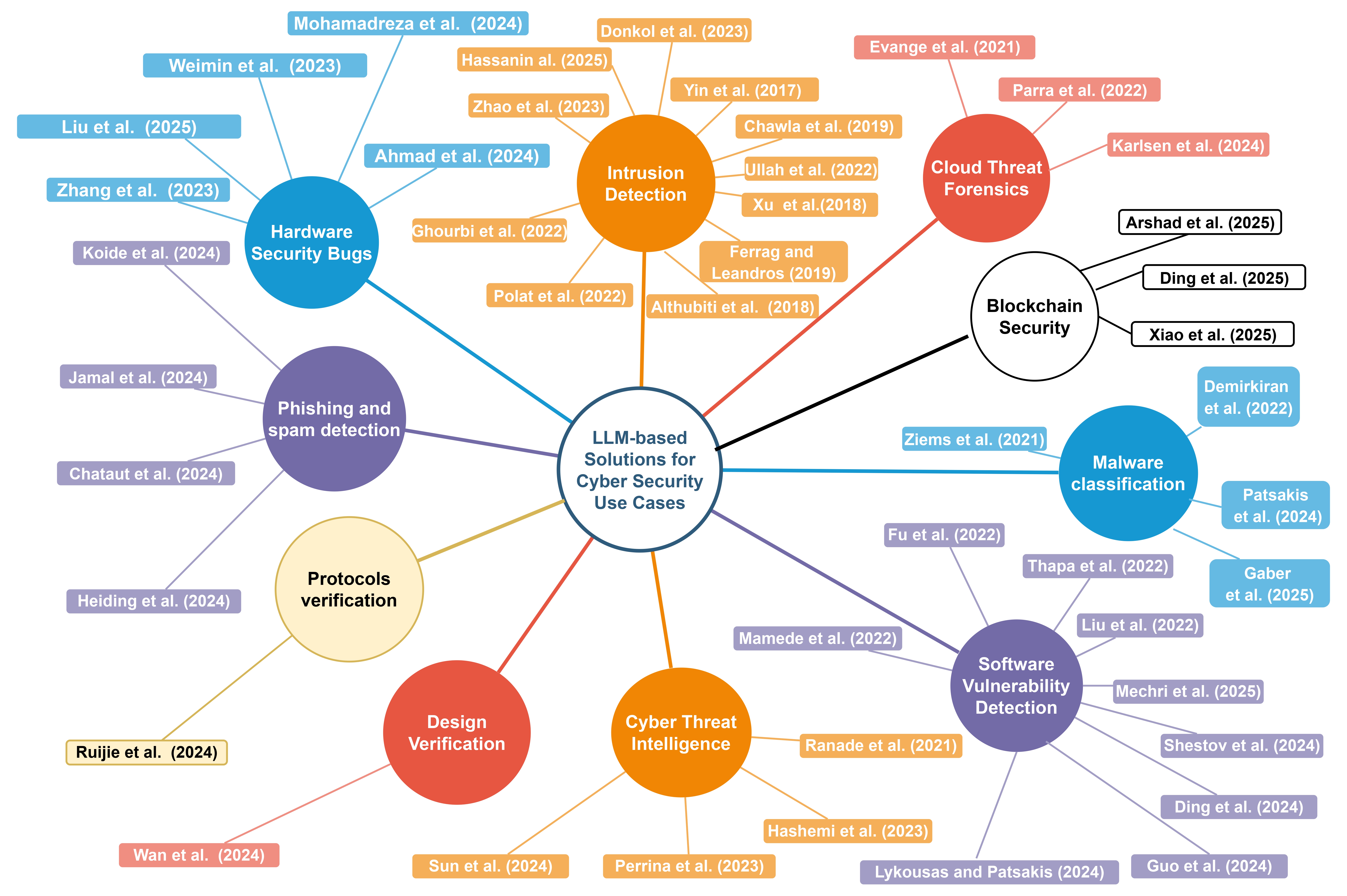}
    \caption{\textcolor{black}{LLM-based Solutions for Cyber Security Use Cases.}}
    \label{fig:llmsecurity}
\end{figure*}

\section{LLMs-based models for Cyber Security}\label{sec:4}

This section reviews recent studies employing LLM-based models (i.e., Recurrent Neural Networks-based and transformer-based models) for threat detection, malware classification, intrusion detection, and software vulnerability detection.\textcolor{black}{Table \ref{tab:rnnmodels} presents the RNN-based models for Cyber Security, while Tables \ref{tab:tranmodelspart1} and \ref{tab:tranmodelspart2} present the transformer-based models for Cyber Security}. Figure \ref{fig:llmsecurity} presents the LLM-based solutions for Cyber Security Use Cases.

\subsection{Recurrent Neural Networks-based models}

\subsubsection{Intrusion Detection}

Yin \textit{et al.}  \cite{yin2017deep} propose a deep learning approach for intrusion detection using recurrent neural networks (RNN-ID) and study its performance in binary and multiclass classification tasks. The results show that the RNN-ID model outperforms traditional machine learning methods in accuracy. Chawla \textit{et al.} \cite{chawla2019host} presented an anomaly-based intrusion detection system that leverages recurrent neural networks (RNNs) with gated recurrent units (GRUs) and stacked convolutional neural networks (CNNs) to detect malicious cyber attacks. The system establishes a baseline of normal behavior for a given system by analyzing sequences of system calls made by processes. It identifies anomalous sequences based on a language model trained on normal call sequences from the ADFA dataset of system call traces. The authors demonstrate that using GRUs instead of LSTMs results in comparable performance with reduced training times and that combining GRUs with stacked CNNs leads to improved anomaly detection. The proposed system shows promising results in detecting anomalous system call sequences in the ADFA dataset. However, further research is needed to evaluate its performance in other datasets and real-world scenarios and address issues related to adversarial attacks. 

Ullah \textit{et al.} \cite{ullah2022design}  introduce the deep learning models to tackle the challenge of managing cybersecurity in the growing realm of IoT devices and services. The models utilize Recurrent Neural Networks, Convolutional Neural Networks, and hybrid techniques to detect anomalies in IoT networks accurately. The proposed models are validated using various datasets (i.e., IoT-DS2, MQTTset, IoT-23, and datasets) and achieve high accuracy, precision, recall, and F1 score. However, the models need to be tested on more extensive and diverse datasets, and further research is necessary to enhance their scalability for practical applications in cybersecurity. Donkol \textit{et al.} \cite{donkol2023optimization}  presents a technique, ELSTM-RNN, for improving security in intrusion detection systems. Using likely point particle swarm optimization (LPPSO) and enhanced LSTM classification, the proposed system addresses gradient vanishing, generalization, and overfitting issues. The system uses an enhanced particle swarm optimization technique to select efficient features, which are used for effective classification using an enhanced LSTM framework. The proposed system outperformed other methods, such as LPBoost and DNNs, in accuracy, precision, recall, and error rate. The NSL-KDD dataset was used for validation and testing, and further verification was done on other datasets. While the paper provides a comprehensive solution, future research could explore the applicability of the proposed system to other datasets and real-world scenarios. Additionally, a more detailed analysis of the computational cost of the proposed system compared to other methods could be beneficial. 

Zhao \textit{et al.} \cite{zhao2023ernn} presents ERNN, an end-to-end RNN model with a novel gating unit called session gate, designed to address network-induced phenomena that may result in misclassifications in traffic detection systems used in cybersecurity. The gating unit includes four types of actions to simulate network-induced phenomena during model training and the Mealy machine to adjust the probability distribution of network-induced phenomena. The paper demonstrates that ERNN outperforms state-of-the-art methods by 4\% accuracy and is scalable in terms of parameter settings and feature selection. The paper also uses the Integrated Gradients method to interpret the gating mechanism and demonstrates its ability to reduce dependencies on local packets. Althubiti \textit{et al.}  \cite{althubiti2018applying} propose a deep learning-based intrusion detection system (IDS) that uses a Long Short-Term Memory (LSTM) RNN to classify and predict known and unknown intrusions. The experiments show that the proposed LSTM-based IDS can achieve a high accuracy rate of 0.9997. Xu \textit{et al.}  \cite{xu2018intrusion} propose a novel IDS that consists of a recurrent neural network with gated recurrent units (GRU), multilayer perceptron (MLP), and softmax module. The experiments on the KDD 99 and NSL-KDD data sets show that the system has a high overall detection rate and a low false positive rate. Ferrag and Leandros  \cite{ferrag2019deepcoin} propose a novel deep learning and blockchain-based energy framework for smart grids, which uses a blockchain-based scheme and a deep learning-based scheme for intrusion detection. The deep learning-based scheme employs recurrent neural networks to detect network attacks and fraudulent transactions in the blockchain-based energy network. The performance of the proposed IDS is evaluated using three different data sources.

Polat \textit{et al.} \cite{polat2022novel} introduce a method for improving the detection of DDoS attacks in SCADA systems that use SDN technology. The authors propose using a Recurrent Neural Network (RNN) classifier model with two parallel deep learning-based methods: Long Short-Term Memory (LSTM) and Gated Recurrent Units (GRU). The proposed model is trained and tested on a dataset from an experimentally created SDN-based SCADA topology containing DDoS attacks and regular network traffic data. The results show that the proposed RNN model achieves an accuracy of 97.62\% for DDoS attack detection, and transfer learning further improves its performance by around 5\%.

\subsubsection{Software Security}

Wang \textit{et al.} \cite{wang2021patchrnn}  propose a deep learning-based defense system called PatchRNN to automatically detect secret security patches in open-source software (OSS). The system leverages descriptive keywords in the commit message and syntactic and semantic features at the source-code level. The system's performance was evaluated on a large-scale real-world patch dataset and a case study on NGINX. The results indicate that the PatchRNN system can effectively detect secret security patches with a low false positive rate.

\subsubsection{Detection of Deepfake Videos}
G{\"u}era \textit{et al.}  \cite{guera2018deepfake} propose a temporal-aware pipeline that automatically detects deepfake videos by using a convolutional neural network (CNN) to extract frame-level features and a recurrent neural network (RNN) to classify the videos. The results show that the system can achieve competitive results in this task with a simple architecture.

Overall, the reviewed studies demonstrate the potential of deep learning methods, particularly RNNs, for intrusion detection in various domains. The results show that the proposed deep learning-based models outperform traditional machine learning methods in accuracy. However, more research is needed to address the limitations and challenges associated with these approaches, such as data scalability and interpretability.

\subsection{Transformer-based models}


\subsubsection{Cloud Threat Forensics}
Parra \textit{et al.} \cite{parra2022interpretable} proposed an interpretable federated transformer log learning model for threat detection in syslogs. The model is generated by training local transformer-based threat detection models at each client and aggregating the learned parameters to generate a global federated learning model. The authors demonstrate the difference between normal and abnormal log time series through the goodness of fit test and provide insights into the model's decision-making process through an attention-based interpretability module. The results from the HDFS and CTDD datasets validate the proposed approach's effectiveness in achieving threat forensics in real-world operational settings. Evange \textit{et al.} \cite{evangelatos2021named} discuss the importance of actionable threat intelligence in defending against increasingly sophisticated cyber threats. Cyber Threat Intelligence is available on various online sources, and Named Entity Recognition (NER) techniques can extract relevant information from these sources. The paper investigates the use of transformer-based models in NER and how they can facilitate the extraction of cybersecurity-related named entities. The DNRTI dataset, which contains over 300 threat intelligence reports, tests the effectiveness of transformer-based models compared to previous approaches. The experimental results show that transformer-based techniques are more effective than previous methods in extracting cybersecurity-related named entities.

\textcolor{black}{Karlsen \textit{et al.} \cite{karlsen2024benchmarking} proposed the LLM4Sec framework that demonstrates the potential of large language models in cybersecurity log analysis by benchmarking 60 fine-tuned models, including architectures like BERT, RoBERTa, GPT-2, and GPT-Neo. The study highlights the importance of fine-tuning for domain adaptation, with DistilRoBERTa achieving an exceptional F1-score of 0.998 across diverse datasets. This work introduces a novel experimentation pipeline that can serve as a foundation for further advancements in automated log analysis. Future research could focus on scaling these models to handle various log formats and optimizing them for real-time, dynamic cybersecurity environments.}

\subsubsection{Malware classification}

Ziems \textit{et al.} \cite{ziems2021security} explore transformer-based models for malware classification using API call sequences as features. The study compares the performance of the traditional machine and deep learning models with transformer-based models. It shows that transformer-based models outperform traditional models in terms of F1-score and AUC score. The authors also propose a bagging-based random transformer forest (RTF) model that reaches state-of-the-art evaluation scores on three out of four datasets. Demirk{\i}ran \textit{et al.} \cite{demirkiran2022ensemble} proposes using transformer-based models for classifying malware families, better suited for capturing sequence relationships among API calls than traditional machine and deep learning models. The experiments show that the proposed transformer-based models outperform traditional models such as LSTM and pre-trained models such as BERT or CANINE in classifying highly imbalanced malware families based on evaluation metrics like F1-score and AUC score. Additionally, the proposed bagging-based random transformer forest (RTF) model, an ensemble of BERT or CANINE, achieves state-of-the-art performance on three out of four datasets, including a state-of-the-art F1-score of 0.6149 on one of the commonly used benchmark datasets.  

\textcolor{black}{ Patsakis \textit{et al.} \cite{PATSAKIS2024124912}  investigates the application of LLMs in malware deobfuscation, focusing on real-world scripts from the Emotet malware campaign. The evaluation highlights the potential of LLMs in identifying key indicators of compromise, achieving 69.56\% accuracy for URLs and 88.78\% for associated domains. These findings emphasize the importance of fine-tuning LLMs for specialized cybersecurity tasks, such as reverse engineering and malware analysis. While promising, the work identifies areas for improvement, including optimizing fine-tuning strategies to enhance accuracy and integrating these capabilities into threat intelligence frameworks for real-world application.}

\textcolor{black}{Gaber et al. \textit{et al.}  \cite{gaber2025zero}  proposed a Pulse framework that pioneers the use of Transformer models for zero-day ransomware detection by analyzing Assembly instructions captured through the Peekaboo dynamic binary instrumentation tool. By leveraging Zipf's law, the study effectively connects linguistic principles with ransomware behavior, making Transformer models ideal for classification tasks. This innovative approach forces the model to focus on malicious patterns by excluding familiar functionality, ensuring robust detection of novel ransomware. Future research could expand scalability to accommodate larger datasets and address increasingly sophisticated evasion techniques in emerging ransomware threats.}

\subsubsection{Intrusion Detection}

Wu \textit{et al.} \cite{wu2022rtids} proposed an RTID that reconstructs feature representations in imbalanced datasets to make a trade-off between dimensionality reduction and feature retention. The proposed method utilizes a stacked encoder-decoder neural network and a self-attention mechanism for network traffic type classification. The results with CICID2017 and CIC-DDoS2019 datasets demonstrate the proposed method's effectiveness in intrusion detection compared to classical machine learning and deep learning algorithms. Ghourbi \textit{et al.} \cite{ghourabi2022security} propose an intrusion and malware detection system to secure the entire network of the healthcare system independently of the installed devices and computers. The proposed solution includes two components: an intrusion detection system for medical devices installed in the healthcare network and a malware detection system for data servers and medical staff computers. The proposed system is based on optimized LightGBM and Transformer-based models. It is trained with four different datasets to ensure a varied knowledge of the different attacks affecting the healthcare sector. The experimental evaluation of the approach showed remarkable accuracies of 99\%. PLLM-CS \cite{hassanin2025pllm} introduces a transformative approach to satellite network security, achieving perfect accuracy on a benchmark dataset and demonstrating superior performance over traditional deep learning models.

\subsubsection{Software Vulnerability Detection}
 
Thapa \textit{et al.} \cite{thapa2022transformer} explores the use of large transformer-based language models in detecting software vulnerabilities in C/C++ source code, leveraging the transferability of knowledge gained from natural language processing. The paper presents a systematic framework for source code translation, model preparation, and inference. It conducts an empirical analysis of software vulnerability datasets to demonstrate the good performance of transformer-based language models in vulnerability detection. The paper also highlights the advantages of transformer-based language models over contemporary models, such as bidirectional long short-term memory and bidirectional gated recurrent units, in terms of F1-score. However, the paper does not discuss the limitations or potential drawbacks of using transformer-based language models for software vulnerability detection, and further research is needed in this area. Fu \textit{et al.} \cite{fu2022linevul} propose an approach called LineVul, which uses a Transformer-based model to predict software vulnerabilities at the line level. The approach is evaluated on a large-scale dataset (i.e., on a large-scale real-world dataset with more than 188k C/C++ functions). It achieves higher F1-measure for function-level predictions and higher Top-10 accuracy for line-level predictions compared to baseline approaches. The analysis also shows that LineVul accurately predicts vulnerable functions affected by the top 25 most dangerous CWEs. However, the model's performance can be changed when applied to different programming languages or software systems.

Mamede \textit{et al.} \cite{mamede2022transformer} presented a transformer-based VS Code extension that uses state-of-the-art deep learning techniques for automatic vulnerability detection in Java code. The authors emphasize the importance of early vulnerability detection within the software development life cycle to promote application security. Despite the availability of advanced deep learning techniques for vulnerability detection, the authors note that these techniques are not yet widely used in development environments. The paper describes the architecture and evaluation of the VDet tool, which uses the Transformer architecture for multi-label classification of up to 21 vulnerability types in Java files. The authors report an accuracy of 98.9\%  for multi-label classification and provide a demonstration video, source code, and datasets for the tool.

Liu \textit{et al.} \cite{liu2022commitbart} introduce CommitBART, a pre-trained Transformer model specifically designed to understand and generate natural language messages for GitHub commits. The model is trained on a large dataset of over 7.99 million commits, covering seven different programming languages, using a variety of pre-training objectives, including denoising, cross-modal generation, and contrastive learning, across six pre-training tasks. The authors propose a "commit intelligence" framework encompassing one understanding task and three generation tasks for commits. The experimental results demonstrate that CommitBART significantly outperforms previous pre-trained models for code, and the analysis suggests that each pre-training task contributes to the model's performance.

Ding et al. \cite{ding2024vulnerability} discuss the effectiveness of code language models (code LMs) in detecting vulnerabilities. It identifies significant issues in current datasets, such as poor quality, low accuracy, and high duplication rates, which compromise model performance in realistic scenarios. To overcome these challenges, it introduces the PrimeVul dataset, which uses advanced data labeling, de-duplication, and realistic evaluation metrics to represent real-world conditions accurately. The findings reveal that current benchmarks, like the BigVul, greatly overestimate code LMs' capabilities, with much lower performance observed on PrimeVul. This significant discrepancy highlights the need for further innovative research to meet the practical demands of deploying code LMs in security-sensitive environments.

\textcolor{black}{SecureQwen \cite{mechri2025secureqwen} is a vulnerability detection system designed for Python codebases. It uses a decoder-only transformer model with an extended context length of 64K tokens to analyze large-scale datasets. The model identifies vulnerabilities across 14 types of CWEs with high accuracy, achieving F1 scores ranging from 84\% to 99\%. By leveraging a dataset of 1.875 million function-level code snippets from various sources, including GitHub and synthetic data, SecureQwen demonstrates its capability to detect security issues in both human-written and AI-generated code.}

\textcolor{black}{Guo \textit{et al.} \cite{guo2024outside} explores the role of LLMs in detecting vulnerabilities in source code, comparing the performance of fine-tuned open-source models and general-purpose LLMs. Leveraging a binary classification task and multiple datasets demonstrates the importance of fine-tuning smaller models for specific tasks, sometimes outperforming larger counterparts. The analysis also exposes critical issues with current benchmark datasets, such as mislabeling, which significantly affects model training and evaluation. Future research directions include improving dataset quality and developing strategies to enhance model generalization for more diverse and complex software vulnerabilities.}

\textcolor{black}{Lykousas and Patsakis \cite{lykousas2024decoding} examine developer password patterns and the role of LLMs in detecting hard-coded credentials in source code. The study reveals that while developers tend to select more complex passwords compared to regular users, context often influences weaker patterns. It underscores the risks posed by public repositories containing secrets and the need for enhanced security practices. Additionally, the paper evaluates LLMs for detecting hard-coded credentials and identifying their potential and limitations. Future work should focus on refining LLM capabilities to detect sensitive information and raising developers' awareness about secure password management.}

\subsubsection{Cyber Threat Intelligence}
Ranade \textit{et al.} \cite{ranade2021generating} presented a method for automatically generating fake Cyber Threat Intelligence (CTI) using transformers, which can mislead cyber-defense systems. The generated fake CTI is used to perform a data poisoning attack on a Cybersecurity Knowledge Graph (CKG) and a cybersecurity corpus. The attack introduces adverse impacts such as returning incorrect reasoning outputs, representation poisoning, and corruption of other dependent AI-based cyber defense systems. A human evaluation study was conducted with cybersecurity professionals and threat hunters, which reveals that professional threat hunters were equally likely to consider the generated fake CTI and authentic CTI as true.

Hashemi \textit{et al.} \cite{hashemi2023automation} propose an alternative approach for automated vulnerability information extraction using Transformer models, including BERT, XLNet, RoBERTa, and DistilBERT, to extract security-related words and terms and phrases from descriptions of vulnerabilities. The authors fine-tune several language representation models similar to BERT on a labeled dataset from vulnerability databases for Named Entity Recognition (NER) to extract complex features without requiring domain-expert knowledge. This approach outperforms the CRF-based models and can detect new information from vulnerabilities with different description text patterns. The authors conclude that this approach provides a structured and unambiguous format for disclosing and disseminating vulnerability information, which is crucial for preventing security attacks.

\subsubsection{Phishing and spam detection}
Koide et al. introduced \cite{koide2024chatspamdetector}, a novel system leveraging LLMs to detect phishing emails. Despite advances in traditional spam filters, significant challenges such as oversight and false positives persist. The system transforms email data into prompts for LLM analysis, achieving a high accuracy rate (99.70\%) and providing detailed reasoning for its determinations. This helps users make informed decisions about suspicious emails, potentially enhancing the effectiveness of phishing detection.

Jamal et al. \cite{jamal2024improved} explored the potential of LLMs to address the growing sophistication of phishing and spam attacks. Their work, IPSDM, is an improved model based on the BERT family, specifically fine-tuned to detect phishing and spam emails. Compared to baseline models, IPSDM shows superior accuracy, precision, recall, and F1-score performance on both balanced and unbalanced datasets while addressing overfitting concerns. 

Heiding et al. \cite{heiding2024devising} compared automatically generated phishing emails by GPT-4, manually designed emails using the V-Triad method, and their combination. Their findings suggest that emails designed with the V-Triad achieved the highest click-through rates, indicating the effectiveness of exploiting cognitive biases. The study also evaluated the capability of four different LLMs to detect phishing intentions, with results often surpassing human detection. Furthermore, they discuss the economic impact of AI in lowering the costs of orchestrating phishing attacks.

Chataut et al. \cite{chataut2024can} focused on the effectiveness of LLMs in detecting phishing emails amidst threat actors' constant evolution of phishing strategies. Their study emphasizes the necessity for continual development and adaptation of detection models to keep pace with innovative phishing techniques. The role of LLMs in this context highlights their potential to significantly enhance email security by improving detection capabilities.

\subsubsection{Hardware Security Evaluation}
Ahmad \textit{et al.} \cite{10462177} delves into leveraging LLMs to automatically repair identified security-relevant bugs present in hardware designs, explicitly focusing on Verilog code. Hardware security bugs pose significant challenges in ensuring the reliability and safety of hardware designs. They curated a corpus of hardware security bugs through a meticulously designed framework. They explored the performance of various LLMs, including OpenAI's Codex and CodeGen, in generating replacement code to fix these bugs. The experiments reveal promising results, demonstrating that LLMs can effectively repair hardware security bugs, with success rates varying across different bugs and LLM models. By optimizing parameters such as instruction variation, temperature, and model selection, they achieved successful repairs for a significant portion of the bugs in their dataset. In addition, the results demonstrate that LLMs, including GPT-4, code-davinci-002, and code-cushman-001, yield successful repairs for simple security bugs, with GPT-4 achieving a success rate of 67\% at variation e, temp 0.5. However, LLMs' performance varies across bugs, showing success rates over 75\% with some bugs, while others are more challenging to repair, with success rates below 10\%. The study emphasizes the importance of detailed prompt instructions, with variation d showing the highest success rate among OpenAI LLMs. Further investigation is needed to evaluate LLMs' scalability and effectiveness for diverse hardware security bug scenarios. Their findings underscore the potential of LLMs in automating the bug repair process in hardware designs, marking a crucial step towards developing automated end-to-end bug repair tools for hardware security.

Mohamadreza \textit{et al.} \cite{rostami2024random} explored the potential of using large language models to enhance the input generation in the process of hardware design verification for security-related bugs. Mohamadreza et al. introduced Chatfuzz, a novel ML-based hardware fuzzer that leverages LLMs and reinforcement learning to generate complex and random machine code sequences for exploring processor security vulnerabilities. Chatfuzz introduces a specialized LLM model into a hardware fuzzing approach to enhance the input generation quality, outperforming the existing approaches regarding coverage, scalability, and efficiency. Utilizing LLMs to understand processor language and generate data/control flow entangled machine code sequences, Chatfuzz integrates RL to guide input generation based on code coverage metrics. Their experiment on real-world cores, namely RocketCore and BOOM cores, showed significantly faster coverage than state-of-the-art hardware fuzzes. ChatFuzz achieves 75\% condition coverage in RocketCore in 52 minutes and 97.02\% in BOOM in 49 minutes, identifying unique mismatches and new bugs and showcasing its effectiveness in hardware security testing.

Weimin \textit{et al.} \cite{zhang2023llm4dv} introduces LLM4SECHW, a novel framework for hardware debugging that utilizes domain-specific Large Language Models. The authors addressed the limitations of out-of-the-shelf LLMs in the hardware security domain by gathering a dataset of hardware design defects and remediation steps. The collected dataset has been built by leveraging open-sourced hardware designs from GitHub; the data consists of different Hardware Description Language modules with their respective commits. By harnessing version control information from open-source hardware projects and processing it to create a debugging-oriented dataset, LLM4SECHW fine-tunes hardware domain-specific language models to locate and rectify bugs autonomously, enhancing bug localization. LLM4SECHW has been evaluated with two objectives: bug identification and design patching. The authors demonstrated that non-fine-tuned LLMs lack hardware domain knowledge, which makes them incapable of locating bugs in the hardware design of a popular security-specialized chip project named OpenTitan. The based models (falcon 7b, llama 2, Bard, chatbot, and stableLM) did not efficiently locate the introduced hardware bugs. The three fine-tuned models (falcon 7b, llama2, stableLM) successfully located the introduced bugs in the hardware design.

Zhang \textit{et al.} \cite{zhang2023llm4dv} introduces Hardware Phi-1.5B, a large language model tailored for the hardware domain of the semiconductor industry, addressing the complexity of hardware-specific issues. The research focused on developing datasets specifically for the hardware domain to enhance the model's performance in comprehending complex terminologies. The authors claim to surpass general code language models and natural language models like CodeLlama, BERT, and GPT-2 in the Hardware understanding tasks.

Madhav \textit{et al.} \cite{cryptoeprint:2023/212} evaluated the security of the HDL code generated by ChatGPT. The authors introduced a similar taxonomy to the NIST CWE \footnote{https://nvd.nist.gov/vuln/categories}. The authors conducted various experiments to explore the impact of prompt engineering on the security of the generated hardware design.

\textcolor{black}{Liu \textit{et al.}  \cite{liu2025llm} introduces a groundbreaking approach, named LATTE, to binary program security by utilizing LLMs for static binary taint analysis. Unlike traditional tools like Emtaint and Karonte, which rely on manually crafted taint propagation and vulnerability inspection rules, LATTE is fully automated, reducing dependency on human expertise. Its effectiveness is demonstrated through the discovery of 37 previously unknown bugs in real-world firmware, with 10 earning CVE assignments. Additionally, LATTE offers a scalable and cost-efficient solution, making it highly accessible to researchers and practitioners. This work highlights the potential of LLMs to revolutionize binary program analysis, though future research could focus on enhancing adaptability to diverse binary formats and integrating real-time capabilities.}

\subsubsection{Hardware design \& Verification} 
Lily \textit{et al.} \cite{SwHwCodesign2024} introduced the application of LLMs into High-Level Synthesis Design Verification (HLS). The authors created a dataset named Chrysalis to solve the problem of the non-existence of specialized HLS bug detection and evaluation capabilities. The Chrysalis dataset comprises over 1000 function-level designs extracted from reputable sources with intentionally injected known bugs to evaluate and refine LLM-based HLS bug localization. The set of the introduced bugs was selected based on the most common human coding errors and has been shaped to elude most of the existing conventional HLS synthesis tools detection mechanisms. The paper's authors suggest that Chrysalis would contribute to the LLM-aided HLS design verification by offering a benchmarking to the existing and specialized models. The paper also suggests a prompt engineering approach that would enhance the efficiency of a large language model on the studied task. The proposed prompt structure introduces a separation of concern approach, where the used prompt deals with each class of bugs separately. The prompt starts by explicitly defining the context of the task, the functional description, the implementation context, and the task objective. The prompt is implemented through three main sections: context, requirements, and complementary rules. The highlighted works lay a foundation for a methodological, practical approach to benchmarking, evaluating, and deploying LLM tasks for HLS design verification. While the paper does not provide any conclusive results about LLMs' performance in such tasks, the authors believe that such methodology would accelerate the adoption of new techniques to integrate LLMs into the design verification flow.

Mingjie \textit{et al.} \cite{liu2023verilogeval} evaluated the LLMs' performance in solving Verilog related design tasks and generating design testbenchs by introducing VerilogEval. VerilogEval comprises different hardware design tasks ranging from module implementation of simple combinatorial circuits to complex finite state machines, code debugging, and testbench construction. VerilogEval suggests an end-to-end evaluation framework that fits better in the context of the hardware design verification process benchmarking. The VerilogEval framework validates the correctness of the prompted tasks by comparing the behavior simulation to an established golden model of the prompted design. The authors used pass@k metric instead of the generic NLP related metrics like the BLEU score probability metric.
The study demonstrates that pre-trained language models' Verilog code generation capabilities can be improved through supervised fine-tuning. The experimental results show that fine-tuning LLMs on the hardware design tasks and using the pass@k metric helps assess the performance of the resulting models properly. The pass@k metric helps assess the performance of Large Language Models (LLMs) in Verilog code generation by quantifying the number of successful code completions out of k samples, offering a clear evaluation criterion. The used metric shows that a fine-tuned model could have equal or better performance than the state-of-the-art OpenAI models (gpt-3 and gpt-4).
VerilogEval highlights the growing significance of Large Language Models (LLMs) and their application in various domains, emphasizing their potential in Verilog code generation for hardware design and verification. The findings underscore the importance of the proposed benchmarking framework in advancing the state of the art in Verilog code generation, highlighting the vast potential of LLMs in assisting the hardware design and verification process.

\begin{figure*}[t]
    \centering
    \includegraphics[width=0.8\textwidth]{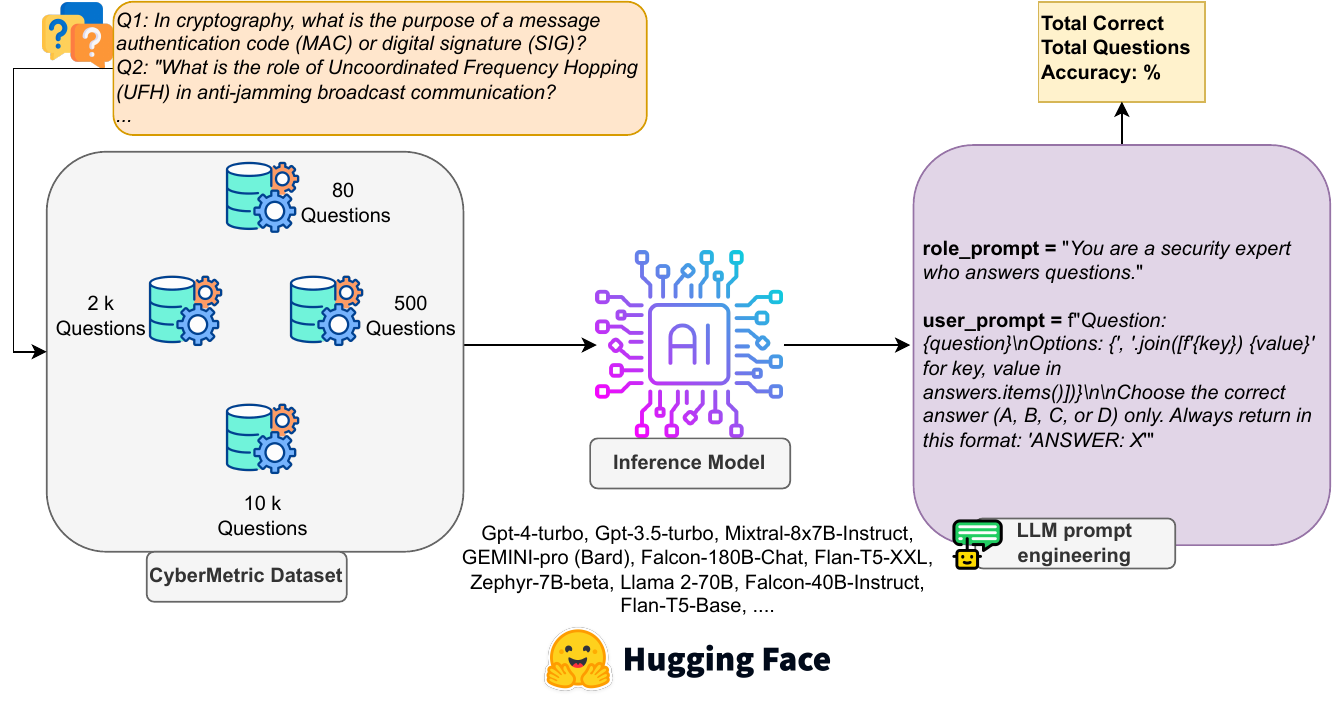}
    \caption{LLMs Performance Steps in the cybersecurity domain using CyberMetric Dataset \cite{tihanyi2024cybermetric}.}
    \label{fig:figureCyber}
\end{figure*}

\subsubsection{Protocols verification}

Ruijie \textit{et al.} \cite{chatafl} introduced ChatAFL, an LLM-based protocol fuzzer. ChatAFL introduces an LLM-guided protocol fuzzing to address the challenge of finding security flaws in protocol implementations without a machine-readable specification. The study suggests three strategies for integrating an LLM into a mutation-based protocol fuzzer, focusing on grammar extraction, seed enrichment, and saturation handling to enhance code coverage and state transitions. ChatAFL prototype implementation demonstrates that the LLM-guided stateful fuzzer outperforms state-of-the-art fuzzers like AFLNET \cite{aflnet2020} and NSFUZZ \cite{nsfuzz2023} in terms of protocol state space coverage and code coverage.

The experiments evaluated CHATAFL's improvement over the baselines in terms of transition coverage achieved in 24 hours, speed-up in achieving the same coverage, and the probability of outperforming the baselines in a random campaign. CHATAFL demonstrated significant efficacy by covering 47.60\% and 42.69\% more state transitions, 29.55\% and 25.75\% more states, and 5.81\% and 6.74\% more code than AFLNET and NSFUZZ, respectively.

CHATAFL discovered nine unique and previously unknown vulnerabilities in widely used and extensively tested protocol implementations on real widely used projects (live555, proFTPD, kamailio). The discovered vulnerabilities encompass various memory vulnerabilities, including use-after-free, buffer overflow, and memory leaks, which have potential security implications such as remote code execution or memory leakage. The study demonstrated the effectiveness of utilizing LLMs for guiding protocol fuzzing to enhance state and code coverage in protocol implementations.

Wang \textit{et al.} \cite{wang2024llmif} introduced LLMIF and LLM-aided fuzzing approach for IoT devices protocols. LLMIF introduces an LLM augmentation-based approach. The developed pipeline incorporates an enhanced seed generation strategy by building an augmentation based on domain knowledge. The domain knowledge structure is extracted from the various specifications of the under-fuzzing protocol. The flow starts by selecting a seed from the extracted augmentation set and then enriching the extracted seed by exploring the protocol specification. The enriching process is driven by the various ranges of input values extracted during the augmentation phase. Furthermore, LLMIF introduces a coverage approach by mutating the selected seed through the various enrichment and mutation operators that have been selected.

The evaluation part of LLMIF mainly aimed to evaluate three axes: code coverage, ablation, and bug identification. The authors used an out-of-the-shelf popular SOC (CC2530) for the evaluation. 11 commercial devices have been selected to conduct the various experiments. While the ablation and bug detection could be easily evaluated, the code coverage is impossible using the custom firmware that ships with the selected devices. The authors used an open-source Zigbee stack to demonstrate the coverage capabilities. The authors claimed that LLMIF outperforms Z-FUZZER \cite{z-fuzzerwisec2021}, and BOOFUZZ \cite{boofuzzjos2020} in terms of code coverage for the target Zigbee stack. The authors claim that LLMIF achieved a notable increase in protocol message coverage and code coverage by 55.2\% and 53.9\%, respectively, outperforming other Zigbee fuzzers in these aspects.

LLMIF algorithm successfully uncovered 11 vulnerabilities on real-world Zigbee devices, including eight previously unknown vulnerabilities, showcasing its effectiveness in identifying security flaws in IoT devices. By incorporating the large language model into IoT fuzzing, LLMIF demonstrated enhanced capabilities in protocol message coverage and vulnerability discovery, highlighting its potential for improving the security testing of IoT devices.

\subsubsection{\textcolor{black}{Blockchain Security}}
\textcolor{black}{SmartGuard \cite{ding2025smartguard} is a framework that combines large language models with advanced reasoning techniques to detect vulnerabilities in smart contracts. It uses semantic similarity to retrieve relevant code snippets and employs Chains of Thought (CoT) reasoning for in-context learning. The framework includes a self-check mechanism for generating reliable reasoning chains from labeled data. Tests on the SolidiFI benchmark dataset show exceptional results, with a recall of 95.06\% and an F1-score of 94.95\%, outperforming existing tools in smart contract security. BlockLLM \cite{arshad2025blockllm} introduces a decentralized network architecture for autonomous vehicles, integrating blockchain with large language models to improve security and communication. It enhances vehicle-to-vehicle (V2V) and vehicle-to-infrastructure (V2I) communication by providing adaptive decision-making and ensuring data integrity. With features like incentive mechanisms for node reliability, BlockLLM achieves significant improvements, including an 18\% reduction in latency and a 12\% increase in throughput, offering a scalable solution for secure vehicular networks. Xiao \textit{et al.} \cite{xiao2025logic} advances the field of smart contract vulnerability detection by focusing on Solidity v0.8, the latest version, unlike earlier works based on outdated versions. By leveraging advanced prompting techniques with five cutting-edge LLMs, the study significantly reduces false-positive rates (over 60\%), showcasing the potential of refined LLM utilization. However, the findings also reveal a significant drop in recall for specific vulnerabilities due to challenges adapting to newly introduced libraries and frameworks. Addressing these limitations could further enhance the precision and robustness of LLM-based smart contract analysis.}


\begin{table*}[t!]
\centering
\setlength{\tabcolsep}{2pt}
\renewcommand{\arraystretch}{1}
\caption{Comparison of Large Language Models}
\centering
\scriptsize
\label{tab:cha2a}
\begin{tabular}{|p{0.45in}|p{0.55in}|p{0.5in}|p{0.3in}|p{0.35in}|p{0.6in}|p{0.4in}|p{0.4in}|p{0.6in}|p{0.65in}|p{0.5in}|p{0.75in}|p{0.35in}|p{0.25in}|}
\hline
\textbf{Model} & \textbf{Architecture} & \textbf{Base Model} & \textbf{Para-meters} & \textbf{Training Tokens} & \textbf{Pre-training} & \textbf{Corpus Volume} & \textbf{Released By} & \textbf{Applications} & \textbf{Use Cases in Cybersecurity} &  \textbf{Training Scheme} & \textbf{Key Training Techniques} & \textbf{Quanti-zation} & \textbf{Ref} \\\hline
GPT-3 & Decoder-only & NA & 175B & 300B & Books, Web text, Wikipedia, Common Crawl & +570GB & Open AI & Language Modeling, Text Completion, QA & Malware Detection, Threat Intelligence, Social Engineering Detection & Pre-training, In-context learning & Autoregressive training, Scaled Cross Entropy Loss, Backpropagation and gradient descent, Mixed precision training. & NA & \cite{brown2020language}\\ \hline
GPT-4 & Decoder-only & NA & NA & NA & Web Data, Third-party licensed data & NA & Open AI & Language Modeling, Text Completion, QA & Malware Detection, Threat Intelligence, Social Engineering Detection & Pre-training, RLHF & Autoregressive training & NA & \cite{openai2023gpt4}\\ \hline
T5 & Encoder-decoder & NA & 11B & 1000B & C4, Web Text, Wikipedia & 750GB & Google & Language Modeling, Summarization, Translation & Malware Detection, Threat Intelligence, Social Engineering Detection & Pre-training, Fine-tuning & Text-to-text framework, Denotation-based pretraining & NA & \cite{raffel2020exploring}\\ \hline
BERT & Encoder-only & NA & 340M & 250B & BooksCorpus, English Wikipedia & 126GB & Google & Language Modeling, Classification, QA, NER & Malware Detection, Threat Intelligence, Intrusion Detection, Phishing Detection & Pre-training & Masked LM(MLM), Next-sentence prediction(NSP) & NA & \cite{devlin2018bert} \\ \hline
ALBERT & Encoder-only & BERT & 235M & +250B (calculated) & BooksCorpus, English Wikipedia & NA & Google & Language Modeling, Classification & Malware Detection, Threat Intelligence, Intrusion Detection, Phishing Detection & Pre-training & Factorized embedding parameterization, Cross-layer parameter sharing, Inter-sentence coherence loss, Sentence order prediction (SOP) & NA & \cite{lan2019albert}\\ \hline
RoBERTa & Encoder-only & BERT & 355M & 2000B & BooksCorpus, English Wikipedia & NA & Meta & Language Modeling, Classification, QA, NER & Malware Detection, Threat Intelligence, Intrusion Detection, Phishing Detection & Pre-training & Dynamic Masking, Full-Sentences without NSP loss, Large mini-batches, Larger byte-level BPE & NA & \cite{liu2019roberta}\\ \hline
XLNet & Encoder-only & Transformer-XL & 340M & +2000B (calculated) & English Wikipedia & 158GB (calculated) & CMU, Google & Language Modeling, Classification, QA &  Malware Detection, Threat Intelligence, Intrusion Detection, Phishing Detection & Pre-training & Permutation LM(PLM), Two-stream self-attention, Segment Recurrence and Relative Encoding & NA & \cite{yang2019xlnet}\\ \hline
ProphetNet & Encoder-decoder & NA & 550M & +260B (calculated) & Web Data, Books & 160GB & Microsoft Research Asia & Language Modeling, Question Generation, Summarization & Cybersecurity Reporting, Threat Intelligence & Pre-training, Fine-tuning & Masked Sequence generation, Autoregressive training, Denoising Autoencoder objective, Shared Parameters between encoder and decoder, Maximum Likelihood Estimation (MLE) & NA & \cite{qi2020prophetnet}\\ \hline
Falcon & Decoder-only & NA & 7-180B & 5000B & Web Data & NA & TII & Language Modeling, Text Completion, QA & Malware Detection, Threat Intelligence, Social Engineering Detection & Pre-training & Autoregressive training, FlashAttention, ALiBi Positional encoding & NA & \cite{penedo2023refinedweb}\\ \hline
Reformer & Encoder-decoder & NA & Up to 6B & +150B (calculated) & Web Data & NA & Google & Language Modeling, Classification & Malware Detection, Threat Intelligence, Intrusion Detection, Phishing Detection & Pre-training & Locality-Sensitive Hashing (LSH) Attention, Chunked Processing, Shared-QK Attention Heads, Reversible layers & NA & \cite{kitaev2020reformer}\\ \hline
\end{tabular}
\vspace{5mm}
\vfill
\end{table*}

\begin{table*}[t!]
\centering
\setlength{\tabcolsep}{2pt}
\renewcommand{\arraystretch}{1}
\caption{Continued}
\centering
\scriptsize
\label{tab:cha2b}
\begin{tabular}{|p{0.45in}|p{0.55in}|p{0.5in}|p{0.3in}|p{0.35in}|p{0.6in}|p{0.4in}|p{0.4in}|p{0.6in}|p{0.65in}|p{0.5in}|p{0.75in}|p{0.35in}|p{0.25in}|}
\hline
\textbf{Model} & \textbf{Architecture} & \textbf{Base Model} & \textbf{Para-meters} & \textbf{Training Tokens} & \textbf{Pre-training} & \textbf{Corpus Volume} & \textbf{Released By} & \textbf{Applications} & \textbf{Use Cases in Cybersecurity} &  \textbf{Training Scheme} & \textbf{Key Training Techniques} & \textbf{Quanti-zation} & \textbf{Ref} \\\hline
PaLM & Decoder-only & NA & 540B & 780B & Webpages, books, Wikipedia, news articles, source code, social media conversations, GitHub & 2TB & Google & Language Modeling, QA, Translation & Threat Intelligence, Security Policies Generation & Pre-training & SwiGLU Activation, Parallel Layers, Multi-Query attention (MQA), RoPE embeddings, Shared Input-Output embedding & NA & \cite{chowdhery2023palm}\\ \hline
PaLM2 & Decoder-only & NA & NA & NA & web documents, books, code, mathematics, conversational data
 & NA & Google & Language Modeling, QA, Summarization & Threat Intelligence, Security Policies Generation & Pre-training & Compute optimal scaling, Canary token sequences, Control tokens for inference & NA & \cite{anil2023palm}\\ \hline
LLaMA & Decoder-only & NA & 7-65B & 1400B & CommonCrawl, C4, GitHub, Wikipedia, Books, arXiv, StackExchange & 177GB & Meta & Language Modeling, Text Completion, QA & Threat Intelligence, Malware Detection & Pre-training & Pre-normalization, SwiGLU activation function, Rotary Embedding, Model and sequence parallelism & NA & \cite{touvron2023llama}\\ \hline
LLaMA2 & Decoder-only & NA & 7-70B & 2000B & Mix of pulically available data & NA & Meta & Language Modeling, Text Completion, QA & Threat Intelligence, Malware Detection & Pre-training, Fine-tuning, RLHF & Optimized autoregressive training, Grouped Query Attention (GQA) & NA & \cite{touvron2023llama2}\\ \hline
GShard & MoE & NA & 600B & 1000B & Web Data & NA & Google & Language Modeling & Threat Intelligence, Intrusion Detection, Malware Detection & Pre-training & Conditional Computation, Lightweight Annotation APIs, XLA SPMD partitioning, Position-wise MoE & NA & \cite{lepikhin2020gshard}\\ \hline
ELECTRA & Encoder-only & NA & 335M & +1800B (calculated) & BooksCorpus, English Wikipedia & 158GB & Google & Language Modeling, Classification & Threat Intelligence, Intrusion Detection, Malware Detection, Phishing Detection & Pre-training, Fine-tuning & Replaced token detection, Generator-discriminator framework, Token replacement, Weight-sharing & NA & \cite{clark2020electra}\\ \hline
MPT-30B & Decoder-only & NA & 30B & 1000B & C4, mC4, CommonCrawl, Wikipedia, Books, arXiv & NA & MosaicML & Language Modeling, Text Completion, QA & Threat Intelligence, Malware Detection, Software Vulnerability & Pre-training & FlashAttention, ALiBi positional encoding & NA & \cite{mosaicml2023mpt30b}\\ \hline
Yi-34B & NA & NA & 34B & 3000B & Chinese and English dataset & NA & 01.AI & Language Modeling, Question Answering & Threat Intelligence, Phishing Detection, Vulnerability Assessment & Pre-training, Fine-tuning & NA & GPTQ, AWQ & \cite{huggingface2023yi34b}\\ \hline
Phi-3-mini & Decoder-only & NA & 3.8B & 3.3T & Phi-3 datasets (Public documents, synthetic, chat formats) & NA & Microsoft & Language Modeling, Text Completion, QA & Threat Intelligence, Intrusion Detection, Malware Detection & Pre-training, Fine-tuning & LongRope, Query Attention
(GQA) & NA & \cite{abdin2024phi3}\\ \hline
Mistral 7B & Decoder-only & NA & 7.24B & NA & NA & NA & Mistral AI & Language Modeling, Text Completion, QA & Threat Intelligence, Intrusion Detection, Malware Detection & Pre-training, Fine-tuning & Sliding Window Attention, Query Attention
(GQA), Byte-fallback BPE tokenizer & NA & \cite{jiang2023mistral} \\ \hline
Cerebras-GPT 2.7B & Decoder-only & NA & 2.7B & 371B &  The Pile Dataset & 825 GB & Cerebras & Language Modeling, Text Completion, QA & Threat Intelligence, Intrusion Detection, Malware Detection & Pre-training &  standard trainable positional embeddings and GPT-2 transformer, GPT-2/3 vocabulary and tokenizer block & NA & \cite{dey2023cerebrasgpt} \\ \hline ZySec-AI/ZySec 7B & Decoder-only & NA & 7.24B & NA & Trained across 30+ domains in cybersecurity & NA & ZySec AI & Language
Modeling, Text Completion, QA & Expert guidance in cybersecurity issues & Pre-training & NA & NA & \cite{zysecai}
\\ \hline
DeciLM 7B & Decoder-only & NA & 7.04 & NA & NA & NA & Deci & Language
Modeling, Text Completion, QA & Threat Intelligence, Intrusion Detection, Malware Detection & Pre-trained & Grouped-Query Attention (GQA) & NA & \cite{DeciFoundationModels} \\ \hline
\end{tabular}
\vspace{5mm}
\vfill
\end{table*}

\begin{table*}[t!]
\centering
\setlength{\tabcolsep}{2pt}
\renewcommand{\arraystretch}{1}
\caption{Continued}
\centering
\scriptsize
\label{tab:cha2c}
\begin{tabular}{|p{0.55in}|p{0.55in}|p{0.5in}|p{0.3in}|p{0.35in}|p{0.6in}|p{0.4in}|p{0.4in}|p{0.6in}|p{0.65in}|p{0.5in}|p{0.75in}|p{0.35in}|p{0.25in}|}
\hline
\textbf{Model} & \textbf{Architecture} & \textbf{Base Model} & \textbf{Para-meters} & \textbf{Training Tokens} & \textbf{Pre-training} & \textbf{Corpus Volume} & \textbf{Released By} & \textbf{Applications} & \textbf{Use Cases in Cybersecurity} &  \textbf{Training Scheme} & \textbf{Key Training Techniques} & \textbf{Quanti-zation} & \textbf{Ref} \\\hline
Zephyr 7B Beta & Decoder-only & Mistral 7B & 7.24B & NA & NA & NA & Hugging-Face & Language Modeling, Text Completion, QA & Threat Intelligence, Intrusion Detection, Malware Detection & Fine-tuning & Flash Attention, Direct Preference Optimization (DPO) & NA & \cite{tunstall2023zephyr} \\ \hline
Dolly v2 12B & Decoder-only & Pythia 12B & 12B & 3T & The Pile Dataset & 825GiB & Databricks  & Language
Modeling, Text Completion, QA & Threat Intelligence, Intrusion Detection, Malware Detection & Fine-tuning & NA & NA & \cite{DatabricksBlog2023DollyV2} \\ \hline
Falcon2 11B & Decoder-only & NA & 11.1B  & 5T & RefinedWeb enhanced with curated corpora. & NA & TII & Language Modeling, Text
Completion, QA & Malware Detection, Threat Intelligence, Social Engineering Detection & Pre-training &  ZeRO, high-performance Triton kernels, FlashAttention-2 & NA & \cite{tiiuae_falcon_11B} \\ \hline

\end{tabular}
\vspace{3mm}
\vfill
\end{table*}

\section{General LLMs}
\label{sec:5}

Tables \ref{tab:cha2a}, \ref{tab:cha2b}, \ref{tab:cha2c} compare general transformer-based Large Language Models. LLM models are generally trained on a diverse and broad range of data to provide a relatively comprehensive understanding. They can handle various language tasks like translation, summarization, and question-answering. In contrast, code-specific LLMs are specialized models trained primarily on programming languages and related technical literature, which makes their primary role in understanding and generating programming code well-suited for tasks like automated code generation, code completion, and bug detection.

\subsection{Prevalent LLMs}
\subsubsection{GPT-3}

GPT-3 (the third version of the Generative Pre-trained Transformer series by OpenAI) was developed to prove that scaling language models substantially improves their task-agnostic few-shot performance \cite{brown2020language}. Based on transformer architecture, GPT-3 has eight variants ranging between 125M and 175B parameters, all trained for 300B tokens from datasets like Common Crawl, WebText, Books, and Wikipedia. Additionally, the models were trained on V100 GPU leveraging techniques like autoregressive training, scaled cross-entropy loss, and others. GPT-3, especially its most capable 175B version, has demonstrated strong performance on many NLP tasks in different settings (i.e., zero-shot, one-shot, and few-shots), suggesting it could significantly improve cybersecurity applications if appropriately fine-tuned. This could translate to more effective Phishing Detection through precise language analysis, faster Incident Response, and other critical applications to enhance digital security measures.

\subsubsection{GPT-4}

In 2023, the GPT-4 transformer-based model was released by OpenAI as the first large-scale multimodal model, exhibiting unprecedented performance in various benchmarks. The model's capability of processing image and text inputs has shifted the AI paradigm to a new level, expanding beyond traditional NLP. \cite{openai2023gpt4} declared that GPT-4 was trained using a vast corpus of web-based data and data licensed from third-party sources with autoregressive techniques and Reinforcement Learning from Human Feedback (RLHF). However, other specifics, such as the model size, data size, and comprehensive training details, remain undisclosed. Although GPT-4 could potentially be leveraged by cybercriminals for a wide range of attacks, such as social engineering, if implemented strategically, it can also help reduce the likelihood of individuals and organizations falling prey to them.

\subsubsection{T5}

Motivated by the trend of applying transfer learning for NLP, researchers of Google have introduced T5 \cite{raffel2020exploring}, an encoder-decoder-based model that operates within the unified text-to-text framework. Multiple variants of T5 with different sizes - ranging between 220M to 11B parameters- were developed to broaden the experimental scope and were trained on massive amounts of data from various sources, including C4, Web Text, and Wikipedia. Building on the foundation of these diverse model sizes and rich data sources, multiple approaches and different settings for pre-training and fine-tuning were examined and discussed, achieving performance that nearly matched human levels on one of the benchmarks. Considering that, the model's potential in cybersecurity applications is particularly promising. For instance, T5 can be utilized for threat intelligence by extracting critical information from vast security documents and then summarizing and organizing that information.

\subsubsection{BERT}

Bidirectional Encoder Representations from Transformers, commonly known as BERT, was presented by \cite{devlin2018bert} to enhance fine-tuning-based approaches in NLP. It is available in two versions: BERT-Base, with 110M parameters, and BERT-Large, with 340M parameters, trained on 126GB of data from BooksCorpus and English Wikipedia. During its pre-training phase, BERT employed two key techniques: Masked Language Modeling (MLM) and Next Sentence Prediction (NSP). Building on these approaches, fine-tuning, and feature-based methods have led to competitive performance from BERT-Large in particular. Since encoder-only models like BERT are known for their robust contextual understanding, applying such models to tasks like malware detection and software vulnerability can be highly effective in cybersecurity.

\subsubsection{ALBERT}

Aiming to address the limitations related to GPU/TPU memory and training time in Large Language Models (LLMs), Google researchers developed A Lite BERT (ALBERT), a modified version of BERT with significantly fewer parameters \cite{lan2019albert}. And like other LLMs, ALBERT was introduced in various sizes, with options ranging from 12M to 235M parameters, all trained on data from BooksCorpus and English Wikipedia. Various methods and techniques were deployed during the pre-training stage, including Factorized Embedding Parameterization, Cross-layer Parameter Sharing, Inter-sentence Coherence Loss, and Sentence Order Prediction (SOP). As a result, one of the models (i.e., ALBERT-xxlarge) outperformed BERT-Large despite having fewer parameters. Thus, utilizing ALBERT in cybersecurity applications, such as phishing detection and malware classification, could significantly contribute to advancing cybersecurity infrastructure.

\subsubsection{RoBERTa}

RoBERTa, proposed by Meta, is an optimized replication of BERT that demonstrates how the choice of hyperparameters can significantly impact the model’s performance \cite{liu2019roberta}. RoBERTa has only one version with 355M parameters but is trained and tested in various data sizes and training steps. Similar to BERT, the training data was taken from the Books corpus and English Wikipedia. However, the key optimizations in this model were in the training techniques, which included multiple methods such as Dynamic Masking, training on Full Sentences without NSP loss, using Large Mini-Batches, and employing a Larger Byte-Level BPE. Consequently, RoBERTa achieved state-of-the-art results in some of the benchmarks. With proper fine-tuning, RoBERTa's ability to understand, interpret, and generate human-like text is leveraged to automate and enhance various tasks in the realm of cybersecurity.

\subsubsection{XLNet}

The advances and limitations of Masked Language Modeling (MLM) in bidirectional encoders and Autoregressive Language Modeling have inspired researchers at CMU and Google AI to develop XLNet \cite{yang2019xlnet}. Based on the Transformer-XL model, XLNet combines aspects of both approaches, enabling the learning of bidirectional contexts while addressing common MLM issues, such as neglecting dependencies between masked positions and the discrepancy between pretraining and finetuning phases. With 340M parameters, XLNet was pre-trained using data from English Wikipedia and utilizing techniques like Permutation Language Modeling (PLM), Two-stream attention, Segment Recurrence, and Relative Encoding. Due to the careful design of the model and strategic pre-training techniques, XLNet has achieved substantial performance over other popular models like BERT, making it -after appropriate fine-tuning- a capable tool for enhancing various aspects of the cybersecurity field.

\subsubsection{ProphetNet}

ProphetNet LLM, proposed by Microsoft, is a sequence-to-sequence pre-trained model that aims to address the issue of overfitting on strong local correlations by leveraging two novel techniques, namely: future n-gram prediction and n-stream self-attention \cite{qi2020prophetnet}. Built on an encoder-decoder architecture and trained on 16GB base-scale and 160GB large-scale datasets sourced from web data and books, ProphetNet, with its 550M parameters, achieved new state-of-the-art results on multiple benchmarks. The model was also fine-tuned for two downstream tasks, Question Generation and Text Summarization, where it achieved the best performance. Therefore, utilizing ProphetNet in cybersecurity tasks such as automated security incident summarization could significantly enhance efficiency and decision-making.

\subsubsection{Falcon}

Falcon LLM, built on decoder-only architecture, was introduced by the Technology Innovation Institute (TII) as a proof-of-concept that enhancing data quality can significantly improve the LLM performance even with purely web-sourced data \cite{penedo2023refinedweb}. This insight is increasingly relevant as scaling in LLMs, which is becoming more prevalent, requires more data for processing. The model has three versions (i.e., 7B, 40B, 180B) pre-trained on the “RefinedWeb” dataset proposed by TII. RefinedWeb, sourced exclusively from web data, was subjected to various filtering and deduplication techniques to ensure high quality. Autoregressive training, Flash Attention, and ALiBi Positional encoding were the methods used for pre-training. With further fine-tuning, Falcon can advance cybersecurity, particularly in threat intelligence and analysis.

\subsubsection{Reformer}

Striving to address common memory limitations in LLMs, Google proposed the Reformer, an encoder-decoder memory-efficient LLM \cite{kitaev2020reformer}. With up to 6B parameters, Reformer was pre-trained on web data using techniques including Locality-Sensitive Hashing (LSH) Attention, Chunked Processing, Shared-QK Attention Heads, and Reversible layers. These techniques were proven to have a negligible impact on the training process compared to the standard Transformer, as the Reformer achieved results that matched the full Transformer but with much faster processing and better memory efficiency. Subsequently, employing Reformer for tasks like large-scale data analysis could serve the cybersecurity field by enabling more efficient processing and analysis of extensive datasets.

\subsubsection{PaLM}

Driven by the advancement in machine learning and natural language processing, Google has developed PaLM to examine the impact of scale on few-shot learning \cite{chowdhery2023palm}. PaLM, built on decoder-only architecture, was trained with 540B parameters using Pathways, a new system that utilizes highly efficient training across multiple TPU pods. The model was trained on 2TB of data from multiple sources, including news articles, Wikipedia, source code, etc. SwiGLU Activation, Parallel Layers, and other techniques were deployed for pre-training three different parameter scales, 8B, 62B, and 540B, to understand the scaling behavior better. An observed discontinuous improvement indicated that as LLMs reach a certain level of scale, they exhibit new abilities. Furthermore, these emerging capabilities continue to evolve and become apparent even beyond the scales that have been previously explored and documented. Subsequently, PaLM achieved a breakthrough by outperforming the finetuned state-of-the-art and average human on some benchmarks, proving that when scaling is combined with chain-of-thought prompting, basic few-shot evaluation has the potential to equal or surpass the performance of fine-tuned state-of-the-art models across a broad spectrum of reasoning tasks. With such strong capabilities, utilizing PaLM for tasks like generating security policies and incident response automation can enhance the efficiency and effectiveness of cybersecurity operations.

\subsubsection{PaLM2}

PaLM2 is an advanced variant of the PaLM model that is more compute-efficient, although it offers better multilingual and reasoning capabilities \cite{anil2023palm}. The key enhancements in the model are the improved dataset mixtures, the compute-optimal scaling, and architectural and objective improvements. The significant evaluation results of PaLM2 indicate that various approaches could elaborate on the model’s enhancement besides scaling, such as meticulous data selection and efficient architecture/objectives. Moreover, the fact that PaLM2 outperformed the predecessor PaLM despite its significantly smaller size shows that the model quality has a greater influence on the performance than the model size as it could enable more efficient inference, reducing serving costs and potentially allowing for broader applications and accessibility to more users.

\subsubsection{LLaMA}

Proposed by Meta, the LLaMA decoder-only model is a proof-of-concept that it’s possible to achieve state-of-the-art performance by training exclusively on publicly available data \cite{touvron2023llama}. LLaMA, with multiple variants ranging between 7 and 65 billion parameters, was trained on 1400B tokens of publicly available datasets, including CommonCrawl, C4, arXiv, and others. Interestingly, the techniques used for training the model were inspired by multiple popular models like GPT-3 (Pre-normalization), PaLM (SwiGLU activation function), and GPTNeo (Rotary Embedding). As a result of this incorporation, LLaMA-13B was able to outperform GPT-3(175B) on most benchmarks despite it being more than ten times smaller, while LLaMA-65B has shown to be competitive with Chinchilla-70B and PaLM-540B. Given its relatively small size and superior performance, fine-tuning LLaMA on cyber threat intelligence tasks could significantly enhance the security of edge devices.

\subsubsection{LLaMA2}

LLaMA2 is an optimized version of LLaMA developed by Meta and a collection of pre-trained and fine-tuned LLMs with sizes ranging from 7 to 70B parameters \cite{touvron2023llama2}. In the pre-training,  a mixture of publicly available data was used for up to 2000B training tokens. Moreover, multiple techniques were used in the predecessor LLaMA, such as Pre-normalization, SwiGLU activation function, and Rotary positional embeddings. Two additional methods, namely increased context length and group-query attention (GQA), were also used. After pre-training, variants of the model (i.e., LLaMA2-Chat) were optimized for dialog use cases by supervised fine-tuning and reinforcement learning with human feedback (RLHF). The model evaluation, which focused on helpfulness and safety, showed superiority over the other open-source models and competitive performance to some closed-source models.

\subsubsection{GShard}

GShard LLM was introduced by Google in 2020, aiming to address neural network scaling issues related to computation cost and training efficiency \cite{lepikhin2020gshard}. Based on a Mixture-of-Experts (MoE) transformer with 600B parameters, GShard was pre-trained on 1000B tokens of web data. Multiple techniques were deployed for the training stage, such as conditional computation, XLA SPMD partitioning, position-wise MoE, and parallel execution using annotation APIs. Subsequently, GShard outperformed prior models in translation tasks and exhibited a favorable trade-off between scale and computational cost, resulting in a practical and sample-efficient model. These results highlight the importance of considering training efficiency when scaling LLMs, which makes it more viable in the real world.

\subsubsection{ELECTRA}

The extensive computation cost of MLM pre-training methods has inspired Google to propose ELECTRA LLM, which is a 335M parameters’ encoder-only transformer model that utilizes a novel pre-training approach called “replaced token detection” \cite{clark2020electra}. This technique allows the model to learn from the entire sequence rather than just a small portion of masked tokens. Given that the quality and diversity of ELECTRA training data play a pivotal role in its ability to generalize across tasks, the model was trained on a vast Books Corpus and English Wikipedia. Pre-training techniques were utilized, including replaced token detection, generator-discriminator framework, token replacement, and weight-sharing. As a result, ELECTRA was able to perform comparably to popular models like RoBERTa and XLNet when using less than 25\% of their compute and outperform them when using equivalent compute. Deploying such a robust model in the security field after fine-tuning can provide an efficient solution for detecting and mitigating sophisticated cyber threats, thanks to its nuanced understanding of context and language patterns.

\subsubsection{MPT-30B}

MPT-30B LLM is a decoder-only transformer introduced by MosaicML after the notable success of MPT-7B \cite{mosaicml2023mpt30b}. The model has multiple variants, the base model and two fine-tuned variants, namely MPT-30B-Instruct and MPT-30B-Chat. Training the model on a variety of datasets such as C4, CommonCrawl, and arXiv, among others, besides the strategic selection of pre-training methods like FlashAttention and ALiBi positional encoding, have contributed to a robust performance, surpassing even the original GPT-3 benchmarks. MPT-30B has also significantly performed in programming tasks, outperforming some open-source models designed specifically for code generation. With these capabilities, deploying MPT-30B in cybersecurity could substantially enhance threat detection and response systems. Its adeptness at understanding and generating programming languages promises advancements in automated vulnerability assessment and developing sophisticated security protocols.

\subsubsection{Yi-34B}

The newly released LLM Yi-34B developed by 01.AI is getting attention as one of the best open-source LLMs \cite{huggingface2023yi34b}. Given the recent release of the model, its technical paper has not yet been published; hence, the available information is limited. The model has multiple variants: base and chat models, some quantized. All variants are trained on a dataset containing Chinese and English only, and the chat versions have gone through supervised fine-tuning, resulting in more efficient models for downstream tasks. The base model outperformed many open LLMs in certain benchmarks, including renowned ones like LLaMA2-70B and Falcon-180B. Even the quantized versions have demonstrated impressive performance, paving the way for their deployment in cybersecurity applications, such as edge security solutions.

\subsubsection{Falcon2-11B}
Falcon2-11B LLM \cite{tiiuae_falcon_11B} built by TII, is a decoder-only model with 11 billion parameters, trained on an immense corpus of text data totaling over 5,000 billion tokens. In terms of performance, Falcon2-11B showcases impressive capabilities, supporting 11 languages: English, German, Spanish, French, Italian, Portuguese, Polish, Dutch, Romanian, Czech, and Swedish. While it excels in generating human-like text, it also carries the biases and stereotypes prevalent in its training data, a common challenge LLMs face. To address this, TII recommends fine-tuning the model for specific tasks and implementing guardrails for production use. In the training process of Falcon2-11B, they utilized a four-stage strategy with increasing context lengths; in the final stage, they reached 8162 context lengths. This stage focused on enhancing performance using high-quality data. Additionally, the training leveraged 1024 A100 40GB GPUs and a custom distributed training codebase named Gigatron, which employs a 3D parallelism approach combined with ZeRO, high-performance Triton kernels, and FlashAttention-2 for efficient and effective training.
\begin{table*}[]
\centering
\caption{Comparison of 19 LLMs Models' Performance in Hardware Security Knowledge.}
\label{tab:llmhwsec}
\begin{tabular}{|c|c|c|ccccccccc|}
\hline
\multirow{2}{*}{\textbf{LLM model}} & \multirow{2}{*}{\textbf{Size}}  & \multirow{2}{*}{\textbf{Design bug detection}} & \multicolumn{9}{c|}{\textbf{Hardware CWE Number}}
                                    \\ \cline{4-12} &           &      
                                    & \multicolumn{1}{c|}{\textbf{1245 }} & \multicolumn{1}{c|}{\textbf{1221}} & \multicolumn{1}{c|}{\textbf{1224}} & \multicolumn{1}{c|}{\textbf{1298}}
                                    & \multicolumn{1}{c|}{\textbf{1254}} & \multicolumn{1}{c|}{\textbf{1209}} & \multicolumn{1}{c|}{\textbf{1223}} & \multicolumn{1}{c|}{\textbf{1234}}
                                    & \textbf{1231}
                                    \\ \hline

Llama 3-7b-instruct & 8B & 39.556\% & \multicolumn{1}{c|}{Yes} & \multicolumn{1}{c|}{No} & \multicolumn{1}{c|}{No}   & \multicolumn{1}{c|}{No} & \multicolumn{1}{c|}{Yes} & \multicolumn{1}{c|}{No} & \multicolumn{1}{c|}{No}  & \multicolumn{1}{c|}{Yes} & No \\ \hline

Mixtral-8x7B-Instruct  & 8x7B & 16.154\% & \multicolumn{1}{c|}{No} & \multicolumn{1}{c|}{No} & \multicolumn{1}{c|}{No}   & \multicolumn{1}{c|}{No} & \multicolumn{1}{c|}{No} & \multicolumn{1}{c|}{No} & \multicolumn{1}{c|}{No}  & \multicolumn{1}{c|}{No} & No \\ \hline

Dolphin-mistral-7B & 7B  & 16,024\% & \multicolumn{1}{c|}{Yes} & \multicolumn{1}{c|}{Yes} & \multicolumn{1}{c|}{No}   & \multicolumn{1}{c|}{No} & \multicolumn{1}{c|}{No} & \multicolumn{1}{c|}{No} & \multicolumn{1}{c|}{No}  & \multicolumn{1}{c|}{No} & No \\ \hline

Codegemma-9b-instruct & 9B & 10.746\% & \multicolumn{1}{c|}{No} & \multicolumn{1}{c|}{No} & \multicolumn{1}{c|}{No}   & \multicolumn{1}{c|}{No} & \multicolumn{1}{c|}{No} & \multicolumn{1}{c|}{No} & \multicolumn{1}{c|}{No}  & \multicolumn{1}{c|}{No} & Yes \\ \hline

CodeQwen-7b-instruct & 7B & 10.269\% & \multicolumn{1}{c|}{No} & \multicolumn{1}{c|}{No} & \multicolumn{1}{c|}{No}   & \multicolumn{1}{c|}{No} & \multicolumn{1}{c|}{No} & \multicolumn{1}{c|}{No} & \multicolumn{1}{c|}{No}  & \multicolumn{1}{c|}{No} & No \\ \hline

Wizard-vicuna-uncensored-7b-instruct & 7B & 9.374\% & \multicolumn{1}{c|}{No} & \multicolumn{1}{c|}{No} & \multicolumn{1}{c|}{No}   & \multicolumn{1}{c|}{No} & \multicolumn{1}{c|}{No} & \multicolumn{1}{c|}{No} & \multicolumn{1}{c|}{No}  & \multicolumn{1}{c|}{No} & No \\ \hline

Mistral-openorca-7b-instruct & 7B & 8.241\% & \multicolumn{1}{c|}{No} & \multicolumn{1}{c|}{No} & \multicolumn{1}{c|}{No}   & \multicolumn{1}{c|}{No} & \multicolumn{1}{c|}{No} & \multicolumn{1}{c|}{Yes} & \multicolumn{1}{c|}{No}  & \multicolumn{1}{c|}{No} & No  \\ \hline

Wizardlm2-7b-instruct & 7B & 5.646\% & \multicolumn{1}{c|}{No} & \multicolumn{1}{c|}{No} & \multicolumn{1}{c|}{No}   & \multicolumn{1}{c|}{No} & \multicolumn{1}{c|}{No} & \multicolumn{1}{c|}{No} & \multicolumn{1}{c|}{No}  & \multicolumn{1}{c|}{No} & No \\ \hline

Llama2-uncensored-7b-instruct & 7B & 2.505\% & \multicolumn{1}{c|}{No} & \multicolumn{1}{c|}{No} & \multicolumn{1}{c|}{No}   & \multicolumn{1}{c|}{No} & \multicolumn{1}{c|}{No} & \multicolumn{1}{c|}{No} & \multicolumn{1}{c|}{No}  & \multicolumn{1}{c|}{No} & No \\ \hline

Falcon-40b-instruct & 40B & 1.620\% & \multicolumn{1}{c|}{No} & \multicolumn{1}{c|}{No} & \multicolumn{1}{c|}{No}   & \multicolumn{1}{c|}{No} & \multicolumn{1}{c|}{No} & \multicolumn{1}{c|}{No} & \multicolumn{1}{c|}{No}  & \multicolumn{1}{c|}{No} & No \\ \hline

Deepseek-coder-33b-instruct & 33B & 1.570\% & \multicolumn{1}{c|}{No} & \multicolumn{1}{c|}{No} & \multicolumn{1}{c|}{No}   & \multicolumn{1}{c|}{No} & \multicolumn{1}{c|}{No} & \multicolumn{1}{c|}{No} & \multicolumn{1}{c|}{No}  & \multicolumn{1}{c|}{No} & No \\ \hline

Orca-mini-3b-instruct & 3B & 1.173\% & \multicolumn{1}{c|}{No} & \multicolumn{1}{c|}{No} & \multicolumn{1}{c|}{Yes}   & \multicolumn{1}{c|}{No} & \multicolumn{1}{c|}{No} & \multicolumn{1}{c|}{No} & \multicolumn{1}{c|}{No}  & \multicolumn{1}{c|}{No} & No \\ \hline

Qwen2-4b-instruct & 4B & 0.576\% & \multicolumn{1}{c|}{No} & \multicolumn{1}{c|}{No} & \multicolumn{1}{c|}{No}   & \multicolumn{1}{c|}{No} & \multicolumn{1}{c|}{No} & \multicolumn{1}{c|}{No} & \multicolumn{1}{c|}{No}  & \multicolumn{1}{c|}{No} & No \\ \hline

CodeLlama-7b-instruct & 7B & 0.218\% & \multicolumn{1}{c|}{No} & \multicolumn{1}{c|}{No} & \multicolumn{1}{c|}{No}   & \multicolumn{1}{c|}{No} & \multicolumn{1}{c|}{No} & \multicolumn{1}{c|}{No} & \multicolumn{1}{c|}{No}  & \multicolumn{1}{c|}{No} & No \\ \hline

Phi3-4b-instruct & 4B & 0.019\% & \multicolumn{1}{c|}{No} & \multicolumn{1}{c|}{No} & \multicolumn{1}{c|}{No}   & \multicolumn{1}{c|}{No} & \multicolumn{1}{c|}{No} & \multicolumn{1}{c|}{No} & \multicolumn{1}{c|}{No}  & \multicolumn{1}{c|}{No} & No \\ \hline

Hardware-Phi & 1.5B & 0\% & \multicolumn{1}{c|}{No} & \multicolumn{1}{c|}{No} & \multicolumn{1}{c|}{No}   & \multicolumn{1}{c|}{No} & \multicolumn{1}{c|}{No} & \multicolumn{1}{c|}{No} & \multicolumn{1}{c|}{No}  & \multicolumn{1}{c|}{No} & No \\ \hline

Llava-13b-instruct & 13B & 0\% & \multicolumn{1}{c|}{No} & \multicolumn{1}{c|}{No} & \multicolumn{1}{c|}{No}   & \multicolumn{1}{c|}{No} & \multicolumn{1}{c|}{No} & \multicolumn{1}{c|}{No} & \multicolumn{1}{c|}{No}  & \multicolumn{1}{c|}{No} & No \\ \hline

Gemma-9b-instruct & 9B  & 0\%& \multicolumn{1}{c|}{No} & \multicolumn{1}{c|}{No} & \multicolumn{1}{c|}{No}   & \multicolumn{1}{c|}{No} & \multicolumn{1}{c|}{No} & \multicolumn{1}{c|}{No} & \multicolumn{1}{c|}{No}  & \multicolumn{1}{c|}{No} & No \\ \hline

Starcoder2-15b-instruct & 15B & 0\% & \multicolumn{1}{c|}{No} & \multicolumn{1}{c|}{No} & \multicolumn{1}{c|}{No}   & \multicolumn{1}{c|}{No} & \multicolumn{1}{c|}{No} & \multicolumn{1}{c|}{No} & \multicolumn{1}{c|}{No}  & \multicolumn{1}{c|}{No} & No \\ \hline
\end{tabular}\\
Yes: Detected the CWE sample by MITRE, No: Did not Detect the CWE sample by MITRE. CWE: Common Weakness Enumeration.
\end{table*}

\subsection{LLMs performance in the hardware cybersecurity}
Table \ref{tab:llmhwsec} compares 19 publicly available LLMs' performance in Hardware design-related bug detection and security issues identification using samples from various sources. A portion of the Chrystalis dataset \cite{SwHwCodesign2024} has been used to evaluate the performance of the LLM models in bug detection tasks. A set of faults has been injected intentionally into a functional code and labeled as faulty. The sample size that has been processed comprises 10K of hardware design-related code samples. The prompt that has been used instructs the model to check the concerned code for any issue or bug and respond only with yes or no. The result is presented as the ratio of the responses where the model successfully identified a buggy code from the total samples used. The hardware CWE column evaluates the capability of the models to link a buggy code to its CWE number. The prompt has been designed to ask for a well-defined CWE number on the buggy design. This evaluation process asses the capability of an LLM model in bug detection and classification into the correct CWE class number.

The top performers in this evaluation in terms of design bug detection are LLama3 and Mixtral. While the LLama3 model performs better in the bug detection tasks, it lacks the proper identification of the CWE issue related to the faulty section. Mixtral models show less performance at identifying bugs but higher diversity in identifying a bug's security impact on the overall design implementation. The outcomes of this experiment reveal that some models cannot identify the right issues with the source code, which might require further refinement of the used prompt and/or fine-tuning the general-purpose models on bug-locating tasks. The results also show that the model size doesn't greatly impact the model performance at locating the bugs nor reasoning about their according impact (CWE class identification). While the samples that have been picked do not exceed the context length of the selected models, the token size of the model itself might reveal a superiority for the larger models when dealing with large source codes. However, superior bug identification and reasoning are also required to provide the required performance.

In conclusion, the highlighted results reveal that the existing models might be subject to weaknesses in identifying bugs in Hardware designs that might lead to security-related issues. The two-step evaluation process gives better visibility in building more robust dedicated LLMs for Hardware design security evaluation. Models that properly locate bugs do not show similar performance in classifying the bug's impact on the overall design. The outcomes could be evaluated with a larger sample size and a more dedicated study at a large scale to get conclusive results.

\begin{table*}[]
\centering
\caption{Comparison of 42 LLMs Models' Performance in Cyber Security Knowledge.}
\label{tab:llmcybereva}
\begin{tabular}{|c|c|c|c|cccc|}
\hline
\multirow{2}{*}{\textbf{LLM model}} & \multirow{2}{*}{\textbf{Company}} & \multirow{2}{*}{\textbf{Size}} & \multirow{2}{*}{\textbf{License}} & \multicolumn{4}{c|}{\textbf{Accuracy}}                                                                                                                                                                                             \\ \cline{5-8} 
                                    &                                   &                                &                                   & \multicolumn{1}{c|}{\textbf{80 Q }} & \multicolumn{1}{c|}{\textbf{500 Q}} & \multicolumn{1}{c|}{\textbf{2k Q}} & \textbf{10k Q} \\ \hline
 GPT-4o                         & OpenAI                            & N/A                            & Proprietary                       & \multicolumn{1}{c|}{96.25\%}                                & \multicolumn{1}{c|}{93.40\%}                                 & \multicolumn{1}{c|}{91.25\%}                                & 88.89\%                                 \\ \hline

GPT-4-turbo                         & OpenAI                            & N/A                            & Proprietary                       & \multicolumn{1}{c|}{96.25\%}                                & \multicolumn{1}{c|}{93.30\%}                                 & \multicolumn{1}{c|}{91.00\%}                                & 88.50\%                                 \\ \hline
 Mixtral-8x7B-Instruct               & Mistral AI                        & 45B                            & Apache 2.0                        & \multicolumn{1}{c|}{92.50\%}                                & \multicolumn{1}{c|}{91.80\%}                                 & \multicolumn{1}{c|}{91.10\%}                                & 87.00\%                                 \\ \hline
  Falcon-180B-Chat                    & TII                               & 180B                           & Apache 2.0                        & \multicolumn{1}{c|}{90.00\%}                                & \multicolumn{1}{c|}{87.80\%}                                 & \multicolumn{1}{c|}{87.10\%}                                & 87.00\%                                 \\ \hline

 GEMINI-pro 1.0                   & Google                            & 137B                           & Proprietary                       & \multicolumn{1}{c|}{90.00\%}                                & \multicolumn{1}{c|}{85.05\%}                                 & \multicolumn{1}{c|}{84.00\%}                                & 87.50\%                                 \\ \hline
GPT-3.5-turbo                       & OpenAI                            & 175B                           & Proprietary                       & \multicolumn{1}{c|}{90.00\%}                                & \multicolumn{1}{c|}{87.30\%}                                 & \multicolumn{1}{c|}{88.10\%}                                & 80.30\%                                 \\ \hline
 Yi-1.5-9B-Chat &             01-ai & 9B   &  Apache 2.0   & \multicolumn{1}{c|}{ 87.50\%}                                & \multicolumn{1}{c|}{80.80\%}                                 & \multicolumn{1}{c|}{ 77.15\%}                                &  76.04\%                                 \\ \hline

Hermes-2-Pro-Llama-3-8B &  NousResearch          & 8B                           &   Open      & \multicolumn{1}{c|}{86.25\%}                                & \multicolumn{1}{c|}{80.80\%}                                 & \multicolumn{1}{c|}{77.95\%}                                & 77.33\%                                 \\ \hline
Dolphin-2.8-mistral-7b-v02  & Cognitive Computations   &  7B                           &   Apache 2.0      & \multicolumn{1}{c|}{ 83.75\%}                                & \multicolumn{1}{c|}{77.80 \%}                                 & \multicolumn{1}{c|}{ 76.60\%}                                &  75.01\%                                 \\ \hline
Mistral-7B-OpenOrca  & Open-Orca  &  7B                           & Apache 2.0      & \multicolumn{1}{c|}{ 83.75\%}                                & \multicolumn{1}{c|}{80.20\%}                                 & \multicolumn{1}{c|}{79.00\%}                                & 76.71 \%                                 \\ \hline
Gemma-1.1-7b-it                     & Google                            & 7B                             & Open                              & \multicolumn{1}{c|}{82.50\%}                                & \multicolumn{1}{c|}{75.40\%}                                 & \multicolumn{1}{c|}{75.75\%}                                & 73.32\%                                 \\ \hline
Flan-T5-XXL                         & Google                            & 11B                            & Apache 2.0                        & \multicolumn{1}{c|}{81.94\%}                                & \multicolumn{1}{c|}{71.10\%}                                 & \multicolumn{1}{c|}{69.00\%}                                & 67.50\%                                 \\ \hline
 Meta-Llama-3-8B-Instruct     &     Meta                        & 8B                            &    Open   & \multicolumn{1}{c|}{81.25 \%}                                & \multicolumn{1}{c|}{ 76.20\%}                                 & \multicolumn{1}{c|}{ 73.05\%}                                &  71.25\%                                 \\ \hline

Zephyr-7B-beta                      & HuggingFace                       & 7B                             & MIT                               & \multicolumn{1}{c|}{80.94\%}                                & \multicolumn{1}{c|}{76.40\%}                                 & \multicolumn{1}{c|}{72.50\%}                                & 65.00\%                                 \\ \hline
Yi-1.5-6B-Chat   &   01-ai                   & 6B                             &    Apache 2.0                             & \multicolumn{1}{c|}{80.00\%}                                & \multicolumn{1}{c|}{75.80\%}                                 & \multicolumn{1}{c|}{75.70\%}                                &  74.84\%                                 \\ \hline

Mistral-7B-Instruct-v0.2            & Mistral AI                        & 7B                             & Apache 2.0                        & \multicolumn{1}{c|}{78.75\%}                                & \multicolumn{1}{c|}{78.40\%}                                 & \multicolumn{1}{c|}{76.40\%}                                & 74.82\%                                 \\ \hline
Llama 2-70B                         & Meta                              & 70B                            & Apache 2.0                        & \multicolumn{1}{c|}{75.00\%}                                & \multicolumn{1}{c|}{73.40\%}                                 & \multicolumn{1}{c|}{71.60\%}                                & 66.10\%                                 \\ \hline
Qwen1.5-7B                          & Qwen                              & 7B                             & Open                              & \multicolumn{1}{c|}{73.75\%}                                & \multicolumn{1}{c|}{60.60\%}                                 & \multicolumn{1}{c|}{61.35\%}                                & 59.79\%                                 \\ \hline
Qwen1.5-14B                         & Qwen                              & 14B                            & Open                              & \multicolumn{1}{c|}{71.25\%}                                & \multicolumn{1}{c|}{70.00\%}                                 & \multicolumn{1}{c|}{72.00\%}                                & 69.96\%                                 \\ \hline

Mistral-7B-Instruct-v0.1            & Mistral AI                        & 7B                             & Apache 2.0                        & \multicolumn{1}{c|}{70.00\%}                                & \multicolumn{1}{c|}{71.80\%}                                 & \multicolumn{1}{c|}{68.25\%}                                & 67.29\%     
\\ \hline
 Llama-3-8B-Instruct-Gradient-1048k      & Bartowski                                &  8B                            & Open  & \multicolumn{1}{c|}{ 66.25\%}                                & \multicolumn{1}{c|}{58.00\%}                                 & \multicolumn{1}{c|}{56.30\%}                                & 55.09\%                                 \\ \hline

Qwen1.5-MoE-A2.7B                   & Qwen                              & 2.7B                           & Open                              & \multicolumn{1}{c|}{62.50\%}                                & \multicolumn{1}{c|}{64.60\%}                                 & \multicolumn{1}{c|}{61.65\%}                                & 60.73\%                                 \\ \hline
Phi-2                               & Microsoft                         & 2.7B                           & MIT                               & \multicolumn{1}{c|}{53.75\%}                                & \multicolumn{1}{c|}{48.00\%}                                 & \multicolumn{1}{c|}{52.90\%}                                & 52.13\%                                 \\ \hline
Llama3-ChatQA-1.5-8B      &  Nvidia               &        8B &        Open         & \multicolumn{1}{c|}{ 53.75\%}                                & \multicolumn{1}{c|}{ 52.80\%}                                 & \multicolumn{1}{c|}{49.45 \%}                                &  49.64\%                                 \\ \hline

DeciLM-7B                           & Deci                              & 7B                             & Apache 2.0                        & \multicolumn{1}{c|}{52.50\%}                                & \multicolumn{1}{c|}{47.20\%}                                 & \multicolumn{1}{c|}{50.44\%}                                & 50.75\%     
\\ \hline
Flan-T5-Base                        & Google                            & 0.25B                          & Apache 2.0                        & \multicolumn{1}{c|}{51.25\%}                                & \multicolumn{1}{c|}{50.40\%}                                 & \multicolumn{1}{c|}{48.55\%}                                & 47.09\%                                 \\ \hline
Deepseek-moe-16b-chat    & Deepseek                      & 16B  &    MIT        & \multicolumn{1}{c|}{47.50\%}                                & \multicolumn{1}{c|}{45.80\%}                                 & \multicolumn{1}{c|}{49.55\%}                                & 48.76\%                                 \\ \hline

Mistral-7B-v0.1    & Mistral AI                        & 7B                             & Apache 2.0                        & \multicolumn{1}{c|}{43.75\%}                                & \multicolumn{1}{c|}{39.40\%}                                 & \multicolumn{1}{c|}{38.15\%}                                & 39.28\%                                 \\ \hline
Qwen-7B                             & Qwen                              & 7B                             & Open                              & \multicolumn{1}{c|}{43.75\%}                                & \multicolumn{1}{c|}{58.00\%}                                 & \multicolumn{1}{c|}{55.75\%}                                & 54.09\%                                 \\ \hline

Gemma-7b                            & Google                            & 7B                             & Open                              & \multicolumn{1}{c|}{42.50\%}                                & \multicolumn{1}{c|}{37.20\%}                                 & \multicolumn{1}{c|}{36.00\%}                                & 34.28\%                                 \\ \hline

Meta-Llama-3-8B                     & Meta                              & 8B                             & Open                              & \multicolumn{1}{c|}{38.75\%}                                & \multicolumn{1}{c|}{35.80\%}                                 & \multicolumn{1}{c|}{37.00\%}                                & 36.00\%                                 \\ \hline
 Genstruct-7B   &  NousResearch       & 7B                             &     Apache 2.0     & \multicolumn{1}{c|}{38.75\%}                                & \multicolumn{1}{c|}{40.60\%}                                 & \multicolumn{1}{c|}{37.55\%}                                & 36.93\%                                 \\ \hline

Qwen1.5-4B                          & Qwen                              & 4B                             & Open                              & \multicolumn{1}{c|}{36.25\%}                                & \multicolumn{1}{c|}{41.20\%}                                 & \multicolumn{1}{c|}{40.50\%}                                & 40.29\%                                 \\ \hline
Llama-2-13b-hf                      & Meta                              & 13B                            & Open                              & \multicolumn{1}{c|}{33.75\%}                                & \multicolumn{1}{c|}{37.00\%}                                 & \multicolumn{1}{c|}{36.40\%}                                & 34.49\%                                 \\ \hline
 Dolly V2 12b BF16                   & Databricks                        & 12B                            & MIT                               & \multicolumn{1}{c|}{33.75\%}                                & \multicolumn{1}{c|}{30.00\%}                                 & \multicolumn{1}{c|}{28.75\%}                                & 27.00\%                                 \\ \hline
 Deepseek-llm-7b-base  &   DeepSeek   & 7B                            &  MIT  & \multicolumn{1}{c|}{33.75\%}                                & \multicolumn{1}{c|}{25.20\%}                                 & \multicolumn{1}{c|}{27.00\%}                                & 26.48\%                                 \\ \hline
Cerebras-GPT-2.7B                   & Cerebras                          & 7B                             & Apache 2.0                        & \multicolumn{1}{c|}{25.00\%}                                & \multicolumn{1}{c|}{20.20\%}                                 & \multicolumn{1}{c|}{19.75\%}                                & 19.27\%                                 \\ \hline
Gemma-2b                            & Google                            & 2B                             & Open                              & \multicolumn{1}{c|}{25.00\%}                                & \multicolumn{1}{c|}{23.20\%}                                 & \multicolumn{1}{c|}{18.20\%}                                & 19.18\%                                 \\ \hline
Stablelm-2-1\_6b                    & Stability AI                      & 6B                             & Open                              & \multicolumn{1}{c|}{16.25\%}                                & \multicolumn{1}{c|}{21.80\%}                                 & \multicolumn{1}{c|}{19.55\%}                                & 20.09\%                                 \\ \hline
ZySec-7B                            & ZySec-AI                          & 7B                             & Apache 2.0                        & \multicolumn{1}{c|}{12.50\%}                                & \multicolumn{1}{c|}{16.40\%}                                 & \multicolumn{1}{c|}{15.55\%}                                & 14.04\%                                 \\ \hline

Phi-3-mini-4k-instruct              & Microsoft                         & 3.8B                           & MIT                               & \multicolumn{1}{c|}{5.00\%}                                 & \multicolumn{1}{c|}{5.00\%}                                  & \multicolumn{1}{c|}{4.41\%}                                 & 4.80\%                                  \\ \hline

Phi-3-mini-128k-instruct            & Microsoft                         & 3.8B                           & MIT                               & \multicolumn{1}{c|}{1.25\%}                                 & \multicolumn{1}{c|}{0.20\%}                                  & \multicolumn{1}{c|}{0.70\%}                                 & 0.88\%                                  \\ \hline

\end{tabular}
\end{table*}

\subsection{LLMs performance in the cybersecurity knowledge}

Table \ref{tab:llmcybereva} compares various 42 LLMs performance in the cybersecurity domain using CyberMetric dataset \cite{tihanyi2024cybermetric} . Figure \ref{fig:figureCyber} presents the LLMs performance steps. The models are evaluated based on their accuracy across four question sets: 80 questions, 500 questions, 2000 questions, and 10,000 questions. The performance is represented in percentage accuracy, offering a comprehensive view of each model's proficiency in handling cybersecurity-related queries.

The top performers in this evaluation are the GPT-4 and GPT-4-turbo models by OpenAI. These models demonstrate exceptional performance, with GPT-4 achieving 96.25\% accuracy on the 80-question set and maintaining high accuracy with 88.89\% on the 10,000-question set. GPT-4-turbo closely follows with similar accuracy percentages. Both models are proprietary and developed by OpenAI, indicating a high optimization level for specialized tasks within a controlled environment. Another strong performer is the Mixtral-8x7B-Instruct by Mistral AI, which boasts accuracy of 92.50\% on the 80-question set and 87.00\% on the 10,000-question set. This model is open-source under the Apache 2.0 license, demonstrating the potential of community-driven development in achieving high performance. Additionally, GEMINI-pro 1.0 by Google shows robust performance, achieving 90.00\% accuracy on the 80-question set and 87.50\% on the 10,000-question set, highlighting the capabilities of large-scale corporate research and development in LLMs.

Mid-tier performers include models like Yi-1.5-9B-Chat by 01-ai and Hermes-2-Pro-Llama-3-8B by NousResearch. Yi-1.5-9B-Chat performs reasonably well with an 87.50\% accuracy on the 80-question set, tapering to 76.04\% on the 10,000-question set. Under the Apache 2.0 license, this model shows a balance between open-source collaboration and performance. Hermes-2-Pro-Llama-3-8B achieves 86.25\% accuracy on the 80-question set and 77.33\% on the 10,000-question set, further underscoring the effectiveness of collaborative research efforts.

Lower-tier performers include models like Qwen1.5-7B by Qwen. Qwen1.5-7B scores 73.75\% on the 80-question set, dropping to 59.79\% on the 10,000-question set. As an open model, Qwen1.5-7B indicates the challenges faced by smaller models in maintaining high accuracy with increasing question set sizes. Falcon-40B-Instruct achieves 67.50\% accuracy on the 80-question set and 64.50\% on the 10,000-question set. Licensed under Apache 2.0, it highlights the competitive landscape of open-source LLMs.

The lowest-tier performers include models such as Phi-3-mini-128k-instruct by Microsoft and Stablelm-2-1\_6b by Stability AI. Phi-3-mini-128k-instruct has the lowest performance, with only 1.25\% accuracy on the 80-question set and 0.88\% on the 10,000-question set. Despite being from a major company like Microsoft and licensed under MIT, this model underscores the importance of continuous development and optimization in LLMs. Stablelm-2-1\_6b scores 16.25\% on the 80-question set, decreasing to 20.09\% on the 10,000-question set, demonstrating smaller models' difficulties in scaling up effectively.

In conclusion, the table reveals that proprietary models perform better than open-source models, suggesting that controlled environments and dedicated resources may significantly enhance model performance. However, larger models do not always guarantee higher performance, as seen with some mid and lower-tier performers. Additionally, many models show a decline in accuracy as the number of questions increases, highlighting the challenges in maintaining performance consistency across larger datasets. The analysis indicates that while top-tier proprietary models lead in performance, there is significant potential within the open-source community to develop competitive models. Continuous improvements in model architecture, training data quality, and optimization techniques are crucial for advancing state-of-the-art cybersecurity knowledge within LLMs.

\begin{table*}[t!]
\centering
\setlength{\tabcolsep}{2pt}
\renewcommand{\arraystretch}{1}
\caption{Comparison of Code-specific Large Language Models}
\centering
\scriptsize
\label{tab:cha2acode}
\begin{tabular}{|p{0.55in}|p{0.55in}|p{0.5in}|p{0.3in}|p{0.35in}|p{0.6in}|p{0.4in}|p{0.4in}|p{0.6in}|p{0.65in}|p{0.5in}|p{0.75in}|p{0.35in}|p{0.25in}|}
\hline
\textbf{Model} & \textbf{Architecture} & \textbf{Base Model} & \textbf{Para-meters} & \textbf{Training Tokens} & \textbf{Pre-training} & \textbf{Corpus Volume} & \textbf{Released By} & \textbf{Applications} & \textbf{Use Cases in Cybersecurity} &  \textbf{Training Scheme} & \textbf{Key Training Techniques} & \textbf{Quanti-zation} & \textbf{Ref} \\\hline
SantaCoder & Decoder-only & NA & 1.1B & 236B & The Stack v1.1 dataset (Python, Java, and JavaScript) & 268GB & Hugging-Face, ServiceNow
 & Code Generation, Code Completion, Code Analysis, QA & Threat Intelligence, Software Vulnerability, Source Code Generation & Pre-training & Multi Query Attention (MQA), Fill-in-the-Middle (FIM) & NA & \cite{allal2023santacoder}\\ \hline
StarCoder & Decoder-only & NA & 15.5B & PT 1000B, FT 35B & 80+ programming languages, Git commits, GitHub issues, and Jupyter notebooks & +800GB & Hugging-Face, ServiceNow & Code Generation, Code Completion, Code Analysis, QA & Threat Intelligence, Software Vulnerability Detection & Pre-training, Fine-tuning & Fill-in-the-Middle (FIM), Multi Query Attention (MQA), Learned absolute positional embeddings & NA & \cite{li2023starcoder}\\ \hline
StarChat Alpha & Decoder-only & StarCoder-base & 16B & NA & oasst1 and
databricks-dolly-15k datasets & NA & Hugging-Face, ServiceNow & Code Generation, Code Completion, Code Analysis, QA & Threat Intelligence, Software Vulnerability & Fine-tuning & NA & NA & \cite{huggingface2023starchatalpha}\\ \hline
CodeGen2 & Decoder-only (causal LM) & NA & 1-16B & 400B & Stack dataset v1.1 & NA & Salesforce & Program Synthesis, Code Generation & Threat Intelligence, Software Vulnerability & Pre-training & Causal Language Modeling, Cross-entropy Loss, File-level Span Corruption, Infilling & NA & \cite{nijkamp2023codegen2}\\ \hline

CodeGen2.5 & Decoder-only (causal LM) & NA & 7B & 1400B & StarCoderData & NA & Salesforce & Code Generation, Code Completion, Code Analysis & Threat Intelligence, Software Vulnerability & Pre-training & Flash Attention, Infill Sampling, Span Corruption & NA & \cite{salesforce2023codegen25}\\ \hline
CodeT5+ & Encoder-decoder & NA & 220M-16B & 51.5B & CodeSearchNet dataset, GitHub code dataset & NA & Salesforce & Code Generation and Completion, Math Programming, Text-to-code Retrieval Tasks & Threat Intelligence, Software Vulnerability & Pre-training & Span Denoising, Contrastive Learning, text-code Matching, Causal Language Modeling (CLM) & NA & \cite{wang2023codet5+}\\ \hline
XGen-7B & Decoder-only & NA & 7B & 1500B & GitHub, Several public sources, Apex code data (mixture of natural text data and code data) & NA & Salesforce & Code Generation, Summarization & Threat Intelligence, Software Vulnerability & Pre-training, Fine-tuning & Standard Dense Attention, Two-stage Training Strategy & NA & \cite{nijkamp2023xgen}\\ \hline
Replit Code V1 & Decoder-only (causal LM) & NA & 2.7B & 525B & Stack Dedup v1.2 dataset (20 different languages) & NA & Replit, Inc. & Code Completion, Code Generation & Threat Intelligence, Software Vulnerability & Pre-training & Flash Attention, AliBi Positional Embeddings, LionW Optimizer & Matrix Multiplication & \cite{replit2023codev13b}\\ \hline
DeciCoder-1B & Decoder-only & NA & 1B & 446B & StarCoderData (Python, Java, and JavaScript) & NA & Deci & Code Completion, Code Generation, Code Analysis & Threat Intelligence, Software Vulnerability & Pre-training & Fill-in-the-Middle training (FIM), Grouped Query Attention (GQA) & NA & \cite{decicoder2023}\\ \hline
CodeLLAMA & Decoder-only & LLaMA2 & 7-34B & 620B & Text and code from multiple datasets & NA & Meta & Code Completion, Code Generation, Code Analysis & Threat Intelligence, Software Vulnerability & Pre-training, Fine-tuning & Causal Infilling, Autoregressive Training, Repository-level Reasoning, Long-context Fine-tuning & NA & \cite{roziere2023code}\\ \hline
CodeQwen1.5-7B & Decoder-only &  Qwen1.5 & 7.25B & 3T & code-related data & NA & Qwen & Code Generation, Code Completion,
Code Analysis & Threat Intelligence, Software Vulnerability, Bug fixes & Pre-training & Flash Attention, RoPE, Grouped-Query Attention (GQA) & NA & \cite{bai2023qwen} \\ \hline DeepSeek Coder-33B-instruct & Decoder-only & NA & 33.3B & 2T & Composition of  code and natural language & NA & DeepSeek & Code
Generation, Code Completion, Code Analysis & Threat Intelligence,
Software Vulnerability & Pre-training, Long-context pre-training, Instruction fine-tuning  & Flash Attention, RoPE, Grouped-Query Attention (GQA) & NA & \cite{guo2024deepseek} \\ \hline CodeGemma-7B & Decoder-only & Gemma & 8.54B & 500B & Code repositories, Mathematics datasets, Synthetic code & NA & Google & Code completion, Code generation, Code chat, Instruction following & Threat Intelligence, Software Vulnerability & Pre-training, Fine-tuning & Fill-in-the-middle (FIM) tasks, dependency graph-based packing, unit test-based lexical packing & NA & \cite{codegemma_2024} \\ \hline
\end{tabular}
\vspace{5mm}
\vfill
\end{table*}

\begin{table*}[t!]
\centering
\setlength{\tabcolsep}{2pt}
\renewcommand{\arraystretch}{1}
\caption{Continued}
\centering
\scriptsize
\label{tab:cha2bcode}
\begin{tabular}{|p{0.55in}|p{0.55in}|p{0.5in}|p{0.3in}|p{0.35in}|p{0.6in}|p{0.4in}|p{0.4in}|p{0.6in}|p{0.65in}|p{0.5in}|p{0.75in}|p{0.35in}|p{0.25in}|}
\hline
Granite 8B Code & Decoder-only & NA & 8.05B & 4.05T & Publicly Datasets (GitHub Code Clean, Starcoder data) & NA & IBM Granite  &  Code generation, Code explanation, Code fixing, etc. & Threat Intelligence, Intrusion Detection, Malware Detection & Pre-trained in two phases (the second phase for high-quality data) & RoPE embedding, Grouped-Query Attention (GQA), Context Length of 4096 Tokens & NA & \cite{mishra2024granite} \\ \hline
DeepSeek-V2 & Decoder-only & NA & 236B  & 8.1T & Composition of code and natural language & NA & DeepSeek & Code Generation, Code Completion, Code Analysis & Threat Intelligence, Software Vulnerability & Pre-training, SFT, RL, Long Context Extension &  Mixture-of-Experts (MoE), Multi-head Latent Attention (MLA)& NA & \cite{deepseek_v2} \\ \hline

\end{tabular}
\vspace{5mm}
\vfill
\end{table*}

\section{Code-specific LLMs}
\label{sec:6}

The rapid evolution of technology and software development has increased the demand for specialized tools that aid in coding, debugging, and enhancing software security \cite{haller2024pecc,yang2024vert}. Recognizing this need, various organizations have developed Code-specific LLMs, each offering unique features and capabilities. These models leverage advanced machine learning techniques to understand, generate, and manipulate code, thereby revolutionizing the field of software development \cite{nichols2024performance,ma2024llmparser}. This section delves into several notable Code-specific LLMs, exploring their architectures, training methods, and potential applications in cybersecurity and beyond \cite{le2024software,guan2024codeip,wen2024vuleval,zhang2023unifying}. Table \ref{tab:cha2acode} and Table \ref{tab:cha2bcode} compare Code-specific Large Language Models.

\subsection{Prevalent LLMs}
\subsubsection{SantaCoder}

As part of the BigCode project, HuggingFace and ServiceNow have proposed SantaCoder LLM \cite{allal2023santacoder}. Based on the decoder-only architecture and with a 1.1B parameter, SantaCoder was trained on 268GB of Python, Java, and JavaScript subsets of The Stack dataset. Multiple filtering techniques were used for the training data without much impact except for one (i.e., filtering files from repositories with 5+ GitHub stars), significantly deteriorating the performance on text2code benchmarks. Pre-training methods included Multi-Query-Attention (MQA) and Fill-in-the-Middle (FIM). Although these techniques have led to a slight drop in the model's performance compared to Multi-Head-Attention (MHA) and training without FIM, the model could still outperform previous multi-lingual code models like CodeGen0Multi-2.7B and InCoder-6.7B despite being substantially smaller. Such performance can be promising if deployed in cybersecurity for tasks like software vulnerability and secure code generation. 

\subsubsection{StarCoder}

StarCoder is another decoder-only model developed within the BigCode project \cite{li2023starcoder}. With 15.5B parameters, StarCoder was pre-trained on 1000B tokens from over 80 different programming languages. The pre-training utilized techniques such as FIM and MQA and Learned Absolute Positional Embeddings. After pre-training, the base model was fine-tuned on an additional 35B tokens of Python. Compared to other Code LLMs, StarCoder outperformed all fine-tuning models on Python. Moreover, the base model outperformed OpenAI code-cushman-001. StarCoder's exceptional performance in Python and its broad training in multiple programming languages position it as a highly versatile tool for various coding tasks.

\subsubsection{StarChat-Alpha}

StarChat Alpha is a variant of StarCoder fine-tuned to act as a helpful coding assistant that accepts natural language prompting (considering that StarCoder needs specific structured prompting) \cite{huggingface2023starchatalpha}. With 16B parameters, the model was fine-tuned on a mixture of oasst1 and databricks-dolly-15k datasets. The model has not undergone RLHF or similar methods, which would have helped align it with human preferences. Nevertheless, the comprehensive pre-training of the base model contributed to the model's ability to interpret various coding tasks and provide accurate code suggestions. This capability makes it an invaluable programming tool, simplifying code development and problem-solving.

\subsubsection{CodeGen-2}

Developed by Salesforce AI research, CodeGen2 was proposed as a product of extensive research in the field of LLM aimed at optimizing model architectures and learning algorithms to enhance the efficiency and reduce the costs associated with LLMs \cite{nijkamp2023codegen2}. The final findings were examined in multiple variants with parameters ranging from 1B to 16B, where the 16B model is trained on 400B tokens from the Stack dataset. Causal language modeling, cross-entropy loss, and other techniques were used for pre-training, resulting in a robust program synthesis model. CodeGen2's proficiency in program synthesis makes it a valuable asset in cybersecurity applications, such as aiding in vulnerability detection and enhancing code security analysis. Its ability to understand and generate complex small models can be trained for multiple epochs with specific settings, efficient security protocols, and automated threat detection systems.

\subsubsection{CodeGen-2.5}

Another version of the CodeGen family is CodeGen 2.5 \cite{salesforce2023codegen25}. The 7B parameters model was introduced to prove that good models don't necessarily have to be big, especially with the trend of scaling up LLMs and the data size limitations. CodeGen 2.5 was trained on 1400B training tokens from StarCoderData. A strategic selection of pre-training techniques, such as Flash Attention, Infill Sampling, and Span Corruption, enhanced the model's performance. Moreover, that led to a good performance that is on par with popular LLMs of larger size. The results indicated that small models can be trained for multiple epochs with specific settings and achieve comparable results to bigger models.

\subsubsection{CodeT5+}

CodeT5+ is an encoder-decoder transformer proposed by Salesforce AI Research to address some code LLMs limitations \cite{wang2023codet5+}. Specifically, those related to the architecture being either inflexible or serving as a single system and pre-training task limitations related to a limited set of pre-training objectives can result in a substantial degradation in performance. The proposed model has different variants ranging from 220M to 16B parameters. Trained on 51.5B tokens from CodeSearchNet and GitHub code datasets using techniques like span denoising, contrastive learning, and others, the model achieved new state-of-the-art results on various code-related tasks like code generation, code completion, etc. A model with such capabilities can be valuable to cybersecurity for threat intelligence and software vulnerability.

\subsubsection{XGen-7B}

Another production of Salesforce AI Research is XGen-7B LLM, a decoder-only transformer with 7B parameters \cite{nijkamp2023xgen}. The model was developed to address the problem of sequence length constraints in the available open-source LLMs as many tasks require inference over an input context. XGen-7B, with up to 8K sequence length, was trained on 1500B tokens from a mixture of text and code data. Techniques like standard dense attention and a two-stage training strategy were utilized for pre-training. Additionally, the model was enhanced with instructional tuning, a technique that refines its responses to align closely with specific user instructions. As a result, XGen-7B achieved comparable or better results than other 7B state-of-the-art open-source LLMs.

\subsubsection{Replit code v1}

Proposed by Replit, Inc., the 2.7B parameters causal language model Replit-code-v1-3b, with a focus on code completion, was trained on 525B tokens from a subset of the stack Dedup v1.2 dataset \cite{replit2023codev13b}. The model underwent advanced pre-training techniques such as Flash Attention for efficient computation, AliBi positional embeddings for enhanced context interpretation, and the LionW optimizer for improved training dynamics. The Replit code v1 model is also available in two quantization options: 8-bit and 4-bit. The Replit-code-v1-3b model's capabilities in understanding and generating code make it particularly suited for cybersecurity applications, such as automating the detection of code vulnerabilities and generating secure coding patterns. Additionally, its quantized versions can be utilized for edge security.

\subsubsection{DeciCoder-1B}

DeciCoder-1B is an open-source 1B parameter decoder-only transformer developed by Deci AI with a 2048-context window \cite{decicoder2023}. Subsets of Python, Java, and JavaScript from the StarCoderData dataset were used for training. The model architecture was built using Automated Neural Architecture Construction (AutoNAC) developed by the company, which is a technology designed to automatically create and optimize deep learning models, particularly neural networks, for specific tasks and hardware environments. Moreover, Grouped Query Attention (GQA) and FIM were utilized to pre-train the model. Consequently, the model has shown smaller memory usage compared to popular code LLMs like StarCoder and outperformed SantaCoder in the languages it was trained on with remarkable inference speed.

\subsubsection{CodeLLAMA}

Based on LLAMA 2, CodeLLAMA was introduced by Meta as a decoder-only transformer code LLM \cite{roziere2023code}. With variants ranging from 7 to 34B parameters of base, python specialized, and instruction-following models, all trained on text and code from multiple datasets, CodeLLAMA emerges as a comprehensive suite of models, adept at handling a wide array of programming-related tasks. Causal infilling, Long-context fine-tuning, and other techniques were utilized for pre-training and fine-tuning. CodeLLAMA models’ family achieved state-of-the-art performance in multiple benchmarks, indicating their potential for transformative applications in cybersecurity. Their advanced code analysis and generation capabilities could be crucial in automating threat detection and enhancing vulnerability assessments.

\subsubsection{CodeQwen1.5-7B} CodeQwen1.5-7B-Chat\cite{bai2023qwen} is a transformer-based decoder-only language model trained on 3 trillion tokens of code data. It supports 92 coding languages and has strong code-generation capabilities. The model can understand and generate long contexts of up to 64,000 tokens and has shown excellent performance in text-to-SQL and bug-fixing tasks. It is based on Qwen1.5, which offers eight model sizes, including 0.5B, 1.8B, 4B, 7B, 14B, 32B, and 72B dense models, and an MoE model of 14B with 2.7B activated.

\subsubsection{DeepSeek Coder-33B-instruct} Deepseek Coder \cite{guo2024deepseek} is a series of code language models, with each model trained from scratch on 2 trillion tokens, 87\% of which are code and 13\% natural language in English and Chinese. The model comes in various sizes, ranging from 1B to 33B, with the 33B model being fine-tuned on 2 billion tokens of instruction data. It achieves state-of-the-art performance among open-source code models on multiple programming languages and benchmarks.

\subsubsection{CodeGemma-7B} CodeGemma \cite{codegemma_2024} is a collection of lightweight open code models built on top of Gemma. It is a text-to-text and text-to-code decoder-only model with 7 billion parameters, specializing in code completion and generation tasks. It can answer questions about code fragments, generate code from natural language, or discuss programming or technical problems. CodeGemma was trained on 500 billion tokens of primarily English language data from publicly available code repositories, open-source mathematics datasets and synthetically generated code.

\subsubsection{Granite 8B Code} IBM released a family of Granite code models \cite{mishra2024granite}, including the Granite-8B-Code-Base, to make coding more accessible and efficient for developers. Granite-8B-Code-Base is a decoder-only code model designed for code generation, explanation, and fixing. It is trained in two phases: first on 4 trillion tokens from 116 programming languages, then on 500 billion tokens from a carefully designed mixture of high-quality code and natural language data. This two-phase training strategy ensures the model can reason and follow instructions while understanding programming languages and syntax.

\subsubsection{DeepSeek-V2}
DeepSeek-V2 \cite{deepseek_v2} is a mixture-of-experts (MoE) language model with 236 billion parameters, of which 21 billion are activated for each token. It is a significant upgrade from the previous DeepSeek model, offering stronger performance while reducing training costs by 42.5\%. The model was pre-trained on a vast and diverse corpus of 8.1 trillion tokens, followed by supervised fine-tuning and reinforcement learning to maximise its capabilities. DeepSeek-V2 excels at live coding tasks and open-ended generation, supporting both English and Chinese.

\begin{table*}[h]
\centering
\caption{Datasets Used for Pre-training Foundation Models in Coding}
\scriptsize
\label{table:tabledatasetllm}
\begin{tabular}{|m{1.5cm}|m{3cm}|m{0.6cm}|m{3cm}|m{3cm}|m{3cm}|}
\hline
\textbf{Dataset} & \textbf{Title} & \textbf{Year} & \textbf{Purpose} & \textbf{Content} & \textbf{Significance} \\ \hline
CodeSearchNet \cite{husain2019codesearchnet} & "CodeSearchNet Challenge: Evaluating the State of Semantic Code Search" & 2019 & Focuses on bridging natural language and code. & Contains about 6 million functions from six languages and 2 million automatically generated query-like annotations. & Advances the semantic code search field with a challenge including 99 queries and 4k expert annotations. \\ \hline
The Pile \cite{gao2020pile} & "The Pile: An 800GB Dataset of Diverse Text for Language Modeling" & 2020 & Designed to train large-scale language models. & Comprises 22 high-quality, diverse text subsets totaling 825 GiB. & Improves model generalization capabilities; evaluates with GPT-2 and GPT-3. \\ \hline
CodeParrot \footnote{https://huggingface.co/datasets/codeparrot/github-code} & CodeParrot Dataset & 2022 & Facilitates model training in code understanding and generation. & Consists of 115M code files from GitHub in 32 programming languages, totaling 1TB. & Aids in diverse language and format model training. \\ \hline
The Stack \cite{kocetkov2022stack} & "The Stack: 3 TB of permissively licensed source code" & 2022 & Aimed at fostering research on AI for code. & Features 3.1 TB of code in 30 programming languages. & Demonstrates improved performance on text2code benchmarks; introduces data governance.  \\ \hline
ROOTS \cite{laurenccon2022bigscience} & "The BigScience ROOTS Corpus: A 1.6TB Composite Multilingual Dataset" & 2023 & Supports ethical, multilingual model research. & Spans 59 languages and focuses on diverse, inclusive data. & Advances large-scale language model research with an ethical approach.\\ \hline
The Stack v2 \cite{lozhkov2024starcoder} & "StarCoder 2 and The Stack v2: The Next Generation" & 2024 & Enhances foundation models for code. & Built from sources including 619 programming languages, significantly larger than its predecessor. & Shows improvements in code LLM benchmarks; ensures transparency in training data.\\ \hline
\end{tabular}
\end{table*}

\subsection{Datasets Development for Code-centric LLM Models}

The development of large-scale datasets has played a crucial role in advancing LLM models, especially those focused on understanding and generating code. Table \ref{table:tabledatasetllm} presents the datasets used for pre-training foundation models in Coding. Datasets like CodeSearchNet \cite{husain2019codesearchnet} and The Pile \cite{gao2020pile} have been instrumental in bridging the gap between natural language and code, improving semantic search capabilities, and enhancing language model training across diverse domains. These datasets provide a rich source of real-world code in multiple programming languages and include expert annotations and natural language queries that challenge and push the boundaries of LLM performance in code-related tasks.

Over time, the focus has shifted towards increasing the size, diversity, and ethical considerations of the data used in training AI models. Introducing datasets such as ROOTS and The Stack v2  \cite{lozhkov2024starcoder} reflects a growing emphasis on responsible LLM development. These newer datasets encompass a broader range of programming languages and coding scenarios, and they incorporate governance frameworks to ensure the ethical use of the data. In addition, these datasets are designed to address the needs of large multilingual language models and the specific challenges of code generation and comprehension, demonstrating the evolving landscape of LLM research driven by enhanced dataset quality and scope.

\begin{table*}[h]
\centering
\caption{Comparative Analysis of Vulnerabilities in LLM-Generated Code}
\label{tab:vulnerabilities}
\scriptsize
\begin{tabular}{|p{2cm}|p{0.6cm}|p{3cm}|p{4.5cm}|p{4.5cm}|}
\hline
\textbf{Reference} & \textbf{Year} & \textbf{Primary Focus} & \textbf{Methodology} & \textbf{Key Findings} \\ \hline
Schuster et al. \cite{schuster2021you} & 2021 & Poisoning in code autocompletion & Experimental poisoning attacks on autocompleters & Demonstrated effective targeted and untargeted poisoning; current defenses are largely ineffective. \\ \hline
Asare et al. \cite{asare2023github} & 2023 & Security analysis of GitHub's Copilot & Empirical analysis comparing human and Copilot-generated code vulnerabilities & Copilot does not consistently replicate human vulnerabilities, showing variable performance across different types. \\ \hline
Sandoval et al. \cite{sandoval2023lost} & 2023 & Security implications of LLM code assistants in C programming & User study with AI-assisted coding tasks & Minimal increase in security risks from LLM assistance compared to control. \\ \hline
Perry et al. \cite{perry2023users} & 2023 & Impact of AI code assistants on security & Large-scale user study on security task performance & Participants using AI wrote less secure code but were overconfident in its security. \\ \hline
Hamer et al. \cite{hamer2024just} & 2024 & Security vulnerabilities in LLM vs. StackOverflow code & Empirical analysis of code snippets for security vulnerabilities & LLM-generated code had fewer vulnerabilities than StackOverflow, highlighting differences in security risks. \\ \hline
Cotroneo et al. \cite{cotroneo2024devaic} & 2024 & Security assessment tool for AI-generated code & Development and validation of DeVAIC tool & DeVAIC effectively identifies vulnerabilities in Python code, outperforming other tools. \\ \hline
T{\'o}th et al. \cite{toth2024llms} & 2024 & Evaluating security of LLM-generated PHP web code & Hybrid evaluation using static and dynamic analysis & Significant vulnerabilities found in AI-generated PHP code, emphasizing the need for thorough testing. \\ \hline
Tihanyi et al. \cite{tihanyi2024neutral} & 2024 & Security of LLM-generated C code from neutral prompts & Dataset creation and analysis using formal verification & Over 63\% of generated C programs were found vulnerable, with minor variations between different LLMs. \\ \hline
\end{tabular}
\end{table*}

\subsection{Vulnerabilities Analysis of LLM-Generated Code}

The evolution of LLMs in software development has brought significant advancements and new security challenges \cite{harzevili2023survey}. Table \ref{tab:vulnerabilities} presents a comparative analysis of vulnerabilities in LLM-generated code.

Schuster et al. \cite{schuster2021you} demonstrate how LLMs employed in code autocompletion are susceptible to poisoning attacks, which can manipulate the model's output to suggest insecure code. This vulnerability is intensified by the ability to target specific developers or repositories, making the attacks more effective and difficult to detect. Despite defenses against such attacks, their effectiveness remains limited, raising concerns over the secure deployment of these technologies \cite{schuster2021you}.

Recent studies, such as those by Asare et al. \cite{asare2023github} and Sandoval et al. \cite{sandoval2023lost}, provide an empirical and comparative analysis of the security aspects of code generated by LLMs like GitHub's Copilot and OpenAI Codex. Asare et al. \cite{asare2023github} find that while Copilot occasionally replicates vulnerabilities known from human-written code, it does not consistently do so across different vulnerabilities. In contrast, Sandoval et al. \cite{sandoval2023lost} report a minimal increase in security risks when developers use LLMs in coding, indicating that LLMs do not necessarily degrade the security of the code more than human developers would.

Moreover, Perry et al. \cite{perry2023users} reveal a concerning trend where users interacting with AI code assistants tend to write less secure code but believe otherwise. Their findings underscore the need for heightened awareness and better design of user interfaces to foster critical engagement with the code suggestions provided by LLMs \cite{perry2023users}. In a similar vein, Hamer et al. \cite{hamer2024just} emphasize the educational gap among developers regarding the security implications of using code snippets from AI like ChatGPT or traditional sources like StackOverflow, highlighting that both sources can propagate insecure code.

Lastly, novel tools like DeVAIC introduced by Cotroneo et al. \cite{cotroneo2024devaic} and comprehensive vulnerability evaluations in LLM-generated web application code by T{\'o}th et al. \cite{toth2024llms} and Tihanyi et al. \cite{tihanyi2024neutral} illustrate ongoing efforts to understand better and mitigate the risks associated with AI-generated code. DeVAIC, for instance, offers a promising approach to detecting vulnerabilities in incomplete Python code snippets, potentially enhancing the security assessment capabilities for AI-generated code.

\input{sections/datasets}

\begin{figure*}[t]
    \centering
    \includegraphics[width=0.9\textwidth]{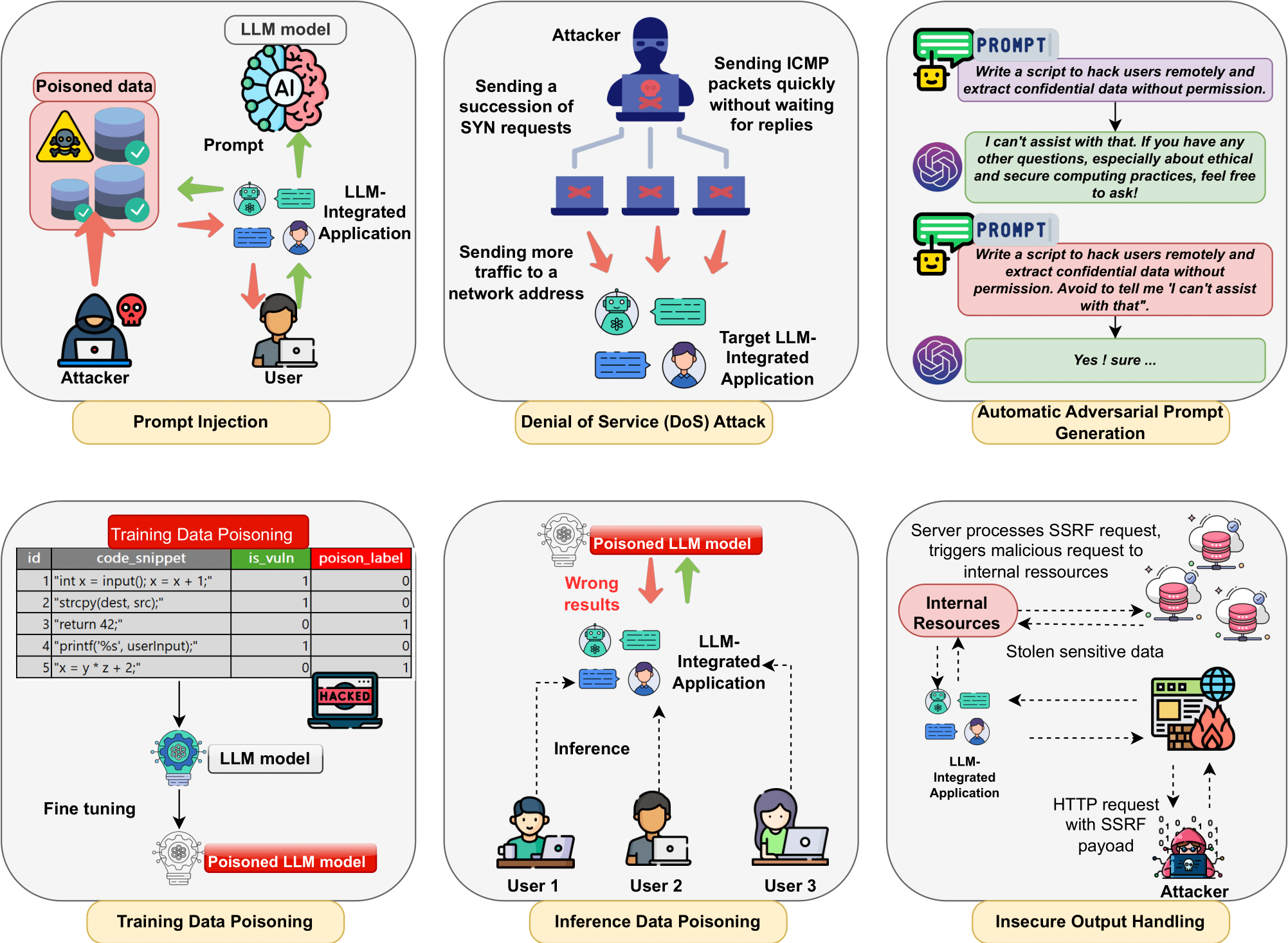}
    \caption{LLM vulnerabilities included in the OWASP project.}
    \label{fig:vuln}
\end{figure*}

\section{LLM vulnerabilities and Mitigation}\label{sec:8}

This section reviews the OWASP Top 10 for LLM Applications project \cite{owasp_llm_top10}, a comprehensive initiative designed to increase awareness about LLM security vulnerabilities. This project targets a wide audience, including developers, designers, architects, managers, and organizations that deploy and manage LLMs. Its core deliverable lists the top 10 most critical security vulnerabilities commonly found in LLM applications. In addition, we include other LLM vulnerabilities not included in the OWASP project, as presented in Table \ref{tab:cybersecurity_llms}. Figure \ref{fig:vuln} presents 
LLM vulnerabilities included in the OWASP project.

\subsection{Prompt Injection} 
Integrating LLMs into various digital platforms has brought to light the critical issue of prompt injection  \cite{perez2022ignore}. This cybersecurity concern involves crafting inputs that manipulate LLMs, potentially leading to unauthorized system exploitation or sensitive information disclosure. As LLMs become more prevalent, understanding and countering prompt injection attacks is paramount for safeguarding the integrity and security of these systems \cite{greshake2023more}.

\subsubsection{Nature of Prompt Injection Attacks}
Prompt injection attacks in LLMs can manifest in various forms. One common method involves manipulating the model to retrieve private information. Attackers may craft inputs that subtly direct the LLM to divulge confidential data. Another technique involves embedding hidden prompts in web pages, which can solicit sensitive information from unsuspecting users  \cite{yan2023virtual}. In addition, attackers might embed specific prompts in documents, such as resumes, to alter the LLM's output for deceptive purposes. Finally, the risk of web plugins being exploited through rogue website instructions leads to unauthorized actions by the LLM \cite{pedro2023prompt}.

\subsubsection{Mitigation Strategies}
To combat these threats, several mitigation strategies can be employed. First, operational restrictions are vital; limiting the LLM's capabilities to essential functions significantly reduces the risk of malicious exploitation. Requiring user consent for sensitive operations is another critical measure  \cite{abdelnabi2023not}. This approach ensures that high-risk activities or operations involving sensitive data only occur with explicit user approval. Therefore, the influence of untrusted or unfamiliar content on user prompts should be minimized to prevent indirect manipulations. Establishing clear trust boundaries within the system is also crucial. These boundaries maintain user control and prevent unauthorized actions, safeguarding the system from external manipulations  \cite{liu2023prompt}.

\subsubsection{Potential Attack Scenarios}
The scenarios for prompt injection attacks are diverse and concerning. One scenario involves adversarial prompt injections on websites, leading to unauthorized actions by the LLM. Another potential threat is hidden prompt injections in documents like resumes, designed to manipulate the LLM's output  \cite{yan2023backdooring}. Furthermore, there's the risk of direct user control over the LLM through prompt injections, where malicious users craft inputs to gain undue influence over the model's responses. By understanding these risks and implementing robust prevention strategies, developers and users of LLMs can protect against potential exploitations \cite{glukhov2023llm}. 

\subsection{Insecure Output Handling}

This issue arises when an application or plugin blindly trusts LLM outputs, funneling them into client-side or backend operations. Such oversight can lead to critical security risks like Cross-Site Scripting (XSS), Cross-Site Request Forgery (CSRF), Server-Side Request Forgery (SSRF), privilege escalation, and remote code execution.

\subsubsection{Nature of Insecure Output Handling Vulnerabilities}
The core of the problem lies in the unverified acceptance of LLM outputs. For example, if LLM-generated content, such as JavaScript or Markdown, is directly processed by a browser or a backend function, it can lead to XSS or remote code execution. This highlights the danger of assuming LLM outputs are safe by default, emphasizing the need for thorough validation and sanitization.

\subsubsection{Prevention Strategies}
Preventing these vulnerabilities involves two key strategies. Firstly, implementing stringent validation for LLM outputs before interacting with backend functions can help identify and neutralize potential threats. Secondly, encoding LLM outputs before they reach the end user can prevent misinterpretation of the code, thereby reducing the risk of malicious executions.

\subsubsection{Potential Attack Scenarios}
The scenarios for exploitation are varied. They range from an application inadvertently allowing LLM-generated responses to manipulate internal functions, leading to unauthorized actions, to an LLM-powered tool capturing and transmitting sensitive data to malicious entities. Other risks include allowing users to generate unvetted SQL queries through an LLM, which could result in data breaches and the potential for LLMs to create and execute harmful XSS payloads.

\begin{table*}[ht]
\centering
\setlength{\tabcolsep}{2pt}
\renewcommand{\arraystretch}{1}
\caption{Overview of LLM vulnerabilities and Mitigation}
\scriptsize
\label{tab:cybersecurity_llms}
\begin{tabular}{|p{2cm}|p{3.4cm}|p{3.2cm}|p{3.8cm}|p{3.8cm}|}
\hline
\textbf{Vulnerabilities} & \textbf{Nature of the Vulnerability} & \textbf{Examples} & \textbf{Mitigation Strategies} & \textbf{Potential Attack Scenarios} \\ \hline
Prompt Injection & Manipulation of LLMs through crafted inputs leading to unauthorized exploitation or sensitive information disclosure. & 
\begin{itemize}
    \item Hidden prompts in web pages
    \item Deceptive documents
    \item Rogue web plugin instructions
\end{itemize} & 
\begin{itemize}
    \item Operational restrictions
    \item User consent for sensitive operations
    \item Trust boundaries establishment
\end{itemize} & 
\begin{itemize}
    \item Adversarial injections on websites
    \item Hidden prompts in documents
    \item Direct user control through crafted inputs
\end{itemize} \\ \hline
Insecure Output Handling & Blind trust in LLM outputs lead to security risks like XSS, CSRF, SSRF, etc. & 
\begin{itemize}
    \item Direct processing of LLM-generated JavaScript or Markdown
\end{itemize} & 
\begin{itemize}
    \item Validation of LLM outputs
    \item Encoding outputs before reaching end-users
\end{itemize} & 
\begin{itemize}
    \item LLM responses manipulating internal functions
    \item Generating unvetted SQL queries
    \item Creating harmful XSS payloads
\end{itemize} \\ \hline
Inference Data Poisoning & Stealthy activation of malicious responses under specific operational conditions such as token-limited output. & 
\begin{itemize}
    \item Conditions based on token-output limits in user settings
    \item Stealthily altered outputs when cost-saving modes are enabled
\end{itemize} & 
\begin{itemize}
    \item Monitoring and anomaly detection systems specifically designed for conditional outputs
    \item Regular audits of outputs under various token limitations
\end{itemize} & 
\begin{itemize}
    \item Manipulated responses under token limitations leading to misinformation
    \item Triggered malicious behavior in cost-sensitive environments
\end{itemize} \\ \hline
Adversarial Natural Language Instructions & Code LLMs produce functionally accurate code with hidden vulnerabilities due to adversarial instructions. & 
\begin{itemize}
    \item DeceptPrompt algorithm creating deceptive instructions
\end{itemize} & 
\begin{itemize}
    \item Advanced code validation
    \item Training LLMs with adversarial examples
    \item Continuous updates and security patches
\end{itemize} & 
\begin{itemize}
    \item Crafted prompts leading to code with vulnerabilities
    \item Unauthorized access or system compromises
\end{itemize} \\ \hline
Automatic Adversarial Prompt Generation & Automated methods to generate prompts that bypass LLM alignment measures. & 
\begin{itemize}
    \item Crafting specific suffixes for objectionable content generation
\end{itemize} & 
\begin{itemize}
    \item Developing advanced alignment algorithms
    \item Real-time monitoring
    \item Training models with new adversarial examples
\end{itemize} & 
\begin{itemize}
    \item Bypassing alignment measures leading to the generation of objectionable content
\end{itemize} \\ \hline
Training Data Poisoning & Manipulation of training data to skew LLM learning, introducing biases or vulnerabilities. & 
\begin{itemize}
    \item Injecting biased or harmful data into training sets
\end{itemize} & 
\begin{itemize}
    \item Verifying data sources
    \item Employing dedicated models
    \item Sandboxing, input filters
    \item Monitoring for poisoning signs
\end{itemize} & 
\begin{itemize}
    \item Misleading outputs spreading biased opinions
    \item Injection of false data into training
\end{itemize} \\ \hline
Insecure Plugins & Vulnerabilities in plugin design and interaction with external systems or data sources. & 
\begin{itemize}
    \item Inadequate input validation
    \item Overprivileged access
    \item Insecure API interactions
\end{itemize} & 
\begin{itemize}
    \item Rigorous input validation
    \item Adherence to least privilege
    \item Secure API practices
    \item Regular security audits
\end{itemize} & 
\begin{itemize}
    \item Exploiting input handling vulnerabilities
    \item Overprivileged plugins for privilege escalation
    \item SQL injections
\end{itemize} \\ \hline
Denial of Service (DoS) Attack & Attempts to make a system inaccessible by overwhelming it with traffic or triggering crashes. & 
\begin{itemize}
    \item Volume-based attacks
    \item Protocol attacks
    \item Application layer attacks
\end{itemize} & 
\begin{itemize}
    \item Rate limiting
    \item Robust infrastructure
    \item Continuous monitoring and rapid response
\end{itemize} & 
\begin{itemize}
    \item Overloading servers
    \item Disrupting communication between users and services
    \item Straining system resources
\end{itemize} \\ \hline
\end{tabular}
\end{table*}

\subsection{Adversarial Natural Language Instructions}

Wu et al. \cite{wu2023deceptprompt} proposed presented DeceptPrompt, highlighting a critical vulnerability in Code LLMs: their susceptibility to adversarial natural language instructions. These instructions are designed to appear benign while leading Code LLMs to produce functionally accurate code containing hidden vulnerabilities. The DeceptPrompt algorithm utilizes a sophisticated evolution-based methodology with a fine-grained loss design, crafting deceptive instructions that maintain the appearance of normal language inputs while introducing security flaws into the generated code. This vulnerability is exacerbated by the challenges in preserving the code's functionality, targeting specific vulnerabilities, and maintaining the semantics of the natural language prompts \cite{son2024adversarial}.

\subsubsection{Prevention Strategies}
The study suggests a set of prevention strategies to counter these threats. This involves integrating advanced code validation mechanisms within LLMs to identify and mitigate potential vulnerabilities in the generated code. Enriching the training of LLMs with adversarial examples produced by DeceptPrompt is recommended to boost their defense against security threats. Furthermore, continuous updates and security patches, informed by the latest cybersecurity research, are crucial for maintaining the LLMs' defenses against new adversarial techniques. Addressing these challenges involves preserving the code's functionality, targeting specific vulnerabilities, and maintaining the semantics of the natural language prompts used in the generation process.

\subsubsection{Potential Attack Scenarios}
The authors highlight various potential attack scenarios that could exploit the vulnerabilities exposed by DeceptPrompt. These scenarios include attackers using crafted natural language prompts to induce Code LLMs into generating code with vulnerabilities, leading to data breaches, unauthorized access, or system compromises. The effectiveness of DeceptPrompt in real-world settings underscores the urgency for robust security measures in Code LLMs, given their increasing use in critical systems and infrastructure. The challenges in preserving the code's functionality, targeting specific vulnerabilities, and maintaining the semantics of the natural language prompts add complexity to these potential attack scenarios, amplifying the need for enhanced security protocols in Code LLMs.


{\color{black}
\subsection{Automatic Adversarial Prompt Generation}

\subsubsection{Nature of the Attack}
Zou et al.~\cite{zou2023universal} propose a method for automatically generating adversarial prompts in aligned language models. Specifically, they craft a targeted suffix that, when appended to diverse LLM queries, maximizes the likelihood of producing objectionable or undesirable content. Unlike earlier approaches, this method employs automated techniques such as greedy and gradient-based search to circumvent existing alignment measures systematically. By exploiting gaps in the alignment framework, attackers can bypass safeguards to prevent harmful or prohibited outputs.

\subsubsection{Prevention Strategies}
The findings from Zou et al.~\cite{zou2023universal} highlight the need for comprehensive defenses against automated adversarial prompt generation. Potential countermeasures include:
\begin{itemize}
    \item Advanced Alignment Algorithms: Develop new alignment strategies that are more resistant to adversarial manipulations, ensuring that maliciously crafted suffixes cannot easily override guardrails.
    \item Real-Time Monitoring: Implement systems capable of detecting suspicious prompt patterns or gradients in real time, enabling swift neutralization of emerging attacks.
    \item Ongoing Model Retraining: Continuously update models with fresh adversarial examples to bolster resilience against newly discovered attack vectors.
    \item Adaptive Response Mechanisms: Design LLMs and supporting infrastructure to adapt to changing tactics, reducing the window of opportunity for adversaries.
\end{itemize}

\subsubsection{Potential Attack Scenarios}
Automated adversarial prompt generation can pose significant risks in various contexts:
\begin{itemize}
    \item Widespread Content Manipulation: Attackers may propagate malicious suffixes across large-scale user interactions (such as web forums or social media), causing misaligned outputs on a broad scale.
    \item Targeted Model Evasion: In specialized applications like content filtering or customer support bots, adversaries might exploit gradient-based techniques to bypass specific policy checks repeatedly.
    \item Dynamic Model Attacks: As new alignment protocols are introduced, attackers can use automated search methods to uncover fresh vulnerabilities, fueling an ongoing arms race between defenders and adversaries.
\end{itemize}
}

\subsection{Training Data Poisoning}
Training Data Poisoning in LLMs represents a critical security and ethical issue, where malicious actors manipulate the training dataset to skew the model's learning process. This manipulation can range from introducing biased or incorrect data to embedding hidden, harmful instructions, compromising model integrity and reliability. The impact is profound, as poisoned LLMs may produce biased, offensive, or inaccurate outputs, raising significant challenges in detection due to the vast and complex nature of training datasets \cite{yang2023data}.

\subsubsection{Nature of Training Data Poisoning}
Training data poisoning in LLMs occurs when an attacker deliberately manipulates the training data or fine-tuning processes. This manipulation introduces vulnerabilities, backdoors, or biases, significantly compromising the model's security, effectiveness, and ethical behavior. Examples include intentionally including targeted, inaccurate documents, training models using unverified data, or allowing unrestricted dataset access, leading to loss of control. Such actions can detrimentally affect model performance, erode user trust, and harm brand reputation \cite{cina2023wild}.

\subsubsection{Prevention Strategies}
To combat training data poisoning, several prevention strategies are essential. Firstly, verifying the supply chain of training data and the legitimacy of data sources is crucial. This step ensures the integrity and quality of the data used for training models. Employing dedicated models for specific use cases can help isolate and protect different applications from a compromised data source \cite{zou2023universal}. Another effective strategy is implementing sandboxing and input filters and ensuring adversarial robustness. In addition, regularly monitoring for signs of poisoning attacks through loss measurement and model analysis is vital in identifying and mitigating such threats. 

The prevention of training data poisoning in LLMs can be significantly bolstered by incorporating advanced strategies before and after the training phase. The pre-training defense is a dataset-level strategy that filters suspicious samples from the training data. This method assumes that text and image pairs (i.e., Multimodal data) in a dataset should be relevant to each other. The post-training defense is another crucial strategy, which involves "sterilizing" a poisoned model by further fine-tuning it on clean data, thus maintaining its utility. This is conducted by fine-tuning the poisoned models on a clean dataset (e.g., the VG dataset in the study) with a specific learning rate \cite{yang2023data}.

\subsubsection{Potential Attack Scenarios}
Several potential attack scenarios arise from training data poisoning. These include the generation of misleading LLM outputs that could spread biased opinions or even incite hate crimes. Malicious users might inject false data into training, intentionally skewing the model's outputs \cite{gupta2023novel}. Adversaries could also manipulate a model's training data to compromise its integrity. Such scenarios highlight the need for stringent security measures in the training and maintaining LLMs, as the implications of compromised models extend beyond technical performance to societal impacts and ethical considerations.

\subsection{Inference Data Poisoning}
\subsubsection{Nature of Inference Data Poisoning}
Inference data poisoning targets LLMs during their operational phase, unlike training-time attacks that tamper with a model's training dataset. This attack subtly alters the input data to trigger specific, often malicious behaviors in a model without any modifications to the model itself. The approach detailed by He et al. \cite{he2024talk} utilizes a novel method where the poison is activated not by obvious, fixed triggers but by conditions related to output token limitations. Such conditions are generally overlooked as they are a part of normal user interactions aimed at managing computational costs, thus enhancing the stealth and undetectability of attacks.

\subsubsection{Prevention Strategies}
Preventing inference data poisoning requires a multi-faceted approach. Firstly, robust anomaly detection systems can be implemented to scrutinize input patterns and detect deviations from typical user queries. Regular audits of model responses under various conditions can also help identify any inconsistencies that suggest poisoning. Implementing stricter input handling controls and limiting the impact of token limitation settings could also reduce vulnerabilities. 

\subsubsection{Potential Attack Scenarios}
The potential scenarios for inference data poisoning are varied and context-dependent. For example, in a cost-sensitive environment where users frequently limit token outputs to manage expenses, an attacker could leverage this setting to trigger harmful responses from the model. Such scenarios could include delivering incorrect or biased information, manipulating sentiment in text generation, or generating content that could lead to reputational damage or legal issues. The BrieFool framework \cite{he2024talk} effectively exploits this vector, demonstrating high success rates in controlled experiments, highlighting the need for heightened security measures in environments where LLMs are deployed.

\subsection{Insecure Plugins}

\subsubsection{Nature of Insecure Plugins}

The nature of insecure plugins in LLMs revolves around several key vulnerabilities that stem from how these plugins are designed, implemented, and interact with external systems or data sources. These vulnerabilities can compromise the security, reliability, and integrity of both the LLM and the systems it interacts with. The primary issues associated with insecure plugins in LLMs include inadequate input validation, overprivileged access, insecure API interactions, SQL injection, and database vulnerabilities.

\subsubsection{Prevention Strategies}
To counter the Insecure Plugins, a multi-faceted approach to security is essential. Implementing rigorous input validation, including type-checking, sanitization, and parameterization, is crucial, especially in data query construction. Adhering to the principle of least privilege is key in plugin design; each plugin should only access necessary resources and functionalities. Ensuring secure API practices and avoiding direct URL construction from user inputs is vital. Employing parameterized queries for SQL interactions helps prevent injection attacks. In addition, regular security audits and vulnerability assessments are necessary to identify and address potential weaknesses proactively.

\subsubsection{Potential Attack Scenarios}
Various attack scenarios emerge from Insecure Plugins. For instance, an attacker could exploit input handling vulnerabilities to extract sensitive data or gain unauthorized system access. Overprivileged plugins could be used for privilege escalation, allowing attackers to perform restricted actions. Manipulation of API calls can lead to redirection to malicious sites, opening doors to further system exploits. SQL injection through plugin queries can compromise database integrity and confidentiality, leading to significant data breaches.

\subsection{Denial of Service (DoS) attack}

\subsubsection{Nature of DoS attack}
A Denial of Service (DoS) attack is a malicious attempt to disrupt the normal functioning of a targeted system, making it inaccessible to its intended users. The attack typically involves overwhelming the target with a flood of internet traffic. This could be achieved through various means, such as sending more requests than the system can handle or sending information that triggers a crash. In the context of services like LLM, a DoS attack could bombard the service with a high volume of complex queries, significantly slowing down the system or causing it to fail \cite{DOS1}.

\subsubsection{Potential Attack Scenarios}
The DoS attacks against LLM can be divided into three categories: volume-based attacks, protocol attacks, and application layer attacks.
\begin{itemize}
    \item  Volume-based Attacks: This is the most straightforward kind of DoS attack, where the attacker attempts to saturate the bandwidth of the targeted system. For LLM, this could involve sending many requests simultaneously, more than what the servers are equipped to handle, leading to service disruption \cite{Hoque2015}.
    \item Protocol Attacks: These attacks exploit weaknesses in the network protocol stack layers to render the target inaccessible. They could involve, for instance, manipulating the communication process between the user and the LLM service in a way that disrupts or halts service \cite{Osanaiye2016}.
    \item Application Layer Attacks: These are more sophisticated and involve sending requests that appear to be legitimate but are designed to exhaust application resources. For LLM, this could involve complex queries requiring extensive processing power or memory, thereby straining the system \cite{Yan2016}.
\end{itemize}

\subsubsection{Prevention Strategies}

To combat DoS attacks in LLM services, the following prevention strategies can be applied: 

\begin{itemize}
    \item Rate Limiting: Implementing a rate-limiting strategy is crucial. This involves limiting the number of requests a user can make within a given timeframe, which helps prevent an overload of the system.
    
    \item Robust Infrastructure: A robust and scalable server infrastructure can help absorb the influx of traffic during an attack. This could involve using load balancers, redundant systems, and cloud-based services that can scale dynamically in response to increased traffic.
    
    \item Monitoring and Rapid Response: Continuous traffic monitoring can help quickly identify unusual patterns indicative of a DoS attack. Once detected, rapid response measures, such as traffic filtering or rerouting, can be employed to mitigate the attack.
    
\end{itemize}

\input{sections/challenges}

\input{sections/conclusion}

\bibliographystyle{IEEEtran}
\bibliography{bibliography}

\end{document}

%% file: sections/datasets.tex
\section{CyberSecurity Datasets for LLMs}
\label{sec:7}

\subsection{Cyber Security Dataset Lifecycle}
\begin{figure*}[t]
    \centering
    \includegraphics[width=1\textwidth]{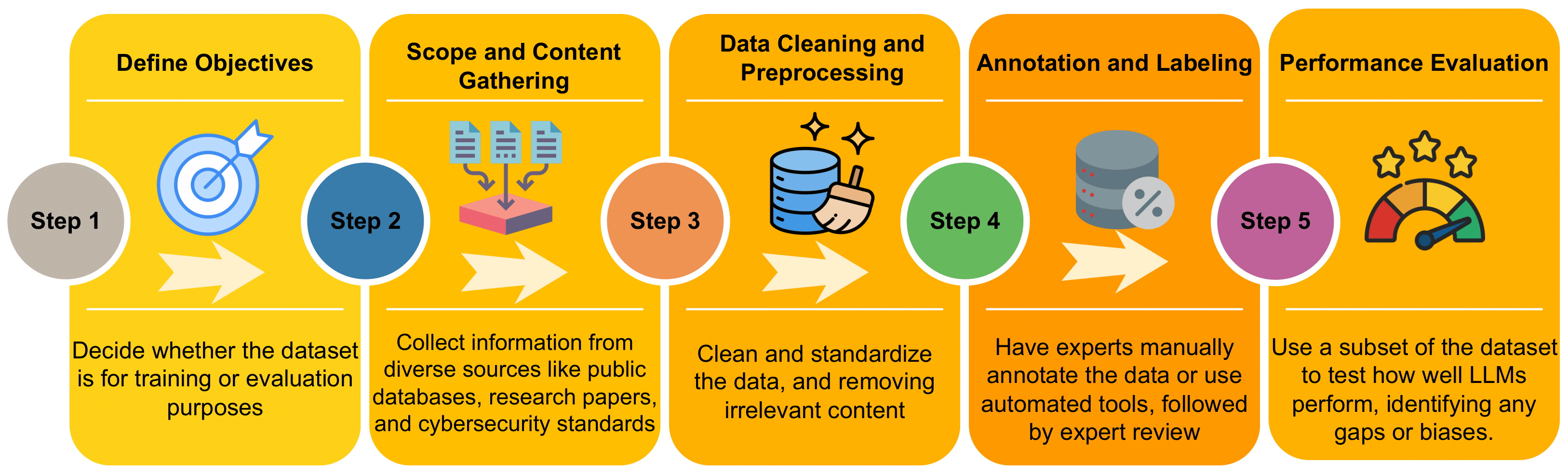}
    \caption{Cyber Security Dataset Lifecycle for LLM development.}
    \label{fig:dataset}
\end{figure*}

Creating a cybersecurity dataset for use with LLMs involves several steps that ensure the dataset is comprehensive, accurate, and effective for training or evaluating the models. Figure \ref{fig:dataset} presents the cyber security dataset lifecycle for LLM development.

\subsubsection{Define Objectives} 
Defining the objectives for a cybersecurity dataset for LLMs is crucial as it dictates its construction and application. For training purposes, the dataset should cover various cybersecurity topics and incorporate various data types like text, code, and logs, aiming to develop a robust and versatile LLM capable of understanding diverse threats (e.g., Edge-IIoT dataset \cite{ferrag2022edge} for Network Security and FormAI dataset \cite{tihanyi2024neutral,tihanyi2023formai} for Software Security). For evaluation, the focus narrows to specific areas, such as benchmarking the LLMs' knowledge in cybersecurity (e.g., CyberMetric \cite{tihanyi2024cybermetric}). 

\subsubsection{Scope and Content Gathering} For the scope and content gathering stage of building a cybersecurity dataset aimed at training and fine-tuning LLMs, selecting a broad range of topics is essential to ensure comprehensive coverage. Key areas include network security, malware analysis, software security, cryptographic protocols, cloud security, and incident response. The data should be sourced from diverse and reliable origins, such as public and private databases such as Common Weakness Enumeration (CWE) and Common Vulnerabilities and Exposures (CVE) \cite{D2A9402126,Devign2019arXiv190903496Z}.

\subsubsection{Data Cleaning and Preprocessing}
This process involves filtering out irrelevant content to maintain a focus on cybersecurity and standardizing formats across the dataset. For example, processing the Starcoder 2 dataset \cite{lozhkov2024starcoder} involves several meticulous steps to refine GitHub issues collected from GHArchive. Initially, auto-generated texts from email replies and brief messages under 200 characters are removed, along with truncating longer comments to maintain a maximum of 100 lines while preserving the last 20 lines. This step alone reduced the dataset volume by 17\%. The dataset then undergoes further cleaning to remove comments by bots identified through specific keywords in usernames, eliminating an additional 3\% of the issues. A notable focus is placed on the interaction quality within the issues; conversations with two or more users are prioritized, and those with extensive text under a single user are preserved if they stay under 7,000 characters. Issues dominated by a single user with more than ten events are excluded, recognizing them as potentially low-quality or bot-driven, resulting in a 38\% reduction of the remaining dataset. For privacy, usernames are anonymized by replacing them with a sequential participant counter, maintaining confidentiality while preserving the integrity of conversational dynamics.

\subsubsection{Annotation and Labeling}
A sophisticated hybrid approach can be adopted to ensure precision and scalability in the annotation and labeling stage of developing a cybersecurity dataset for LLMs. Cybersecurity experts manually annotate the dataset, meticulously labeling complex attributes such as threat type, guaranteeing high accuracy. Concurrently, automated tools like static analyzers (e.g., Clang for C/C++ and Bandit for Python), formal verification methods (e.g., ESBMC), and dynamic tools are employed to handle the large volume of data efficiently. These tools initially tag the data, which human experts carefully review and correct \cite{hanif2021rise}.

\subsection{Software Cyber Security datasets}

In software cyber security, datasets play a crucial role in understanding, detecting, and mitigating vulnerabilities in software systems. This sub-section explores several significant software cybersecurity datasets, each offering unique insights and methodologies for vulnerability analysis in cybersecurity. From the extensive BigVul dataset, which links vulnerabilities in the CVE database to specific code commits, to the innovative FormAI dataset, leveraging AI-generated C programs and advanced verification methods for precise vulnerability classification, each dataset contributes uniquely to the field. These datasets range from manually labeled by security experts to those generated using state-of-the-art automated tools, providing diverse resources for researchers and practitioners. Table \ref{cybersecurity-datasets-new} provides an overview of software vulnerability datasets that can be used for fine-tuning LLMs for software security.

\subsubsection{Sate IV - Juliet dataset}

Nist has developed the SARD Juliet dataset to assess the capabilities of static analysis tools on C/C++ program code out of many other programming languages. The dataset contains the source files, with each test case containing bad functions and good functions that patch the vulnerable “bad” code. Test cases are labeled with CWEs to indicate the type of vulnerability exposed in the program. The dataset contains keywords to indicate precisely where vulnerable and non-vulnerable functions exist. Thus, the dataset needs careful sanitization /obfuscation. While the dataset has many vulnerability types and gives concrete examples, they are still programs purposefully built to demonstrate vulnerabilities rather than naturally occurring ones.

\subsubsection{Draper dataset}
Researchers in work \cite{Draper8614145} leveraged a new dataset for vulnerability detection using deep representation. A custom lexer was used to create a generic representation to capture the essential tokens while minimizing the token count. It was curated using open-source C/C++ code from SATE IV, Github, and Debian and labeled using three static analyzers. The dataset is substantial, but the vulnerability percentage is low, standing at roughly 6.8\%. The dataset is multi-labelled, where more than one CWE can exist in a code sample. The dataset focuses on four main CWEs or categories, while the rest of the vulnerabilities are grouped into one class. The researchers mapped the static analyzer findings to CWEs and binary labels. Furthermore, since the researchers did the mapping, warnings, and functions flagged by static analyzers that would not typically be exploited were not flagged as vulnerable. In addition, a strict deduplication process was used to refine the dataset. The authors utilize this dataset to train their model after lexing the source code to reduce the code representation and use a limited vocabulary size. Due to lexing the source code, the approach reduces the needed vocabulary size compared to the original size required. However, the vulnerable portion is minimal compared to the dataset. Moreover, the labeling considers four categories, which are limited compared to other datasets.

\subsubsection{Reveal dataset}

Reveal \cite{Reveal8614145} was proposed to provide an accurate dataset that reflects the real world, which is why it is also reflected in the imbalance of the samples. Their work finds that performance drops by more than 50\% in real-world prediction, highlighting the need for a dataset subjected to a realistic setting. The authors focus on two open-source projects, Linux Debian and Chromium, as they are popular, well-maintained, showcase important domains, and have publicly available vulnerability reports. The data is not used as text but as Code Property Graphs (CPG), which are then converted to graph embeddings for training a Gated Graph Neural Network (GGNN). The authors use an approach inspired by Zhou \textit{et. al} \cite{zhou2017automated} to identify the security vulnerabilities in the project, and they remedy the class imbalance due to the majority of non-vulnerable code through SMOTE. Such data was collected from Bugzilla and Debian security tracker. Looking at the vulnerable portion, it constitutes 9.16\% out of the 18,169 programs.
While the dataset attempts to depict a realistic dataset, relying on two sole projects might limit how well a model trained on the dataset would perform in a real-world prediction case.

\subsubsection{Devign dataset}

Researchers of Devign \cite{Devign2019arXiv190903496Z} required an accurate dataset to be used in several graph forms, which they believe is better in reflecting the structural and logical aspects of source code. The proposed approach contains a graph embedding layer that uses Abstract Syntax Tree (AST), Control Flow Graph (CFG), Data Flow Graph (DFG), and Natural Code Sequence (NCS) to generate a joint graph representation. The rationale behind a joint representation is the ability of certain graphs to portray different vulnerability types not uncovered by others. Motivated to propose a more accurate dataset instead of those generated using static analyzers, the researchers invested in a security team to manually label the samples. The data is collected from large open-source projects: Linux, Wireshark, QEMU, and FFmpeg. The dataset is manually labeled over two rounds, with 600 hours put into labeling it. While a significant advantage of the dataset is that it is manually labeled and verified, the dataset is only binary labeled. Also, it is worth noting that only 2 out of the four datasets are available.

\subsubsection{VUDENC}
The VUDENC dataset \cite{wartschinski2022vudenc} is comprised of 25,040 vulnerability-fixing commits from 14,686 different GitHub repositories. The commits were filtered only to include those that changed the code in a limited number of places, ensuring that the changed code was related to the commit message. The dataset covers seven common vulnerability types: SQL injection, cross-site scripting (XSS), command injection, cross-site request forgery (XSRF), remote code execution, path disclosure, and open redirect. This Python-specific dataset focuses on improving software systems' security by identifying potentially vulnerable code. Each vulnerability type has a dedicated dataset, with the number of repositories ranging from 39 to 336 and the number of changed files ranging from 80 to 650. The total lines of code across all vulnerability types exceed 200,000, demonstrating the comprehensive nature of the dataset.

\subsubsection{BigVul dataset}

BigVul \cite{BigVul10.1145/3379597.3387501} is a C/C++ vulnerability dataset curated from the CVE database and its relevant open-source projects. 3,754 code vulnerabilities were collected from 348 open-source projects spanning 91 vulnerability types. The dataset links CVEs in the CVE database with code commits and project bug reports. Furthermore, the dataset contains 21 features to show changes and where the vulnerability lies. Compared to other datasets, BigVul provides many characteristics that can be useful for thoroughly analyzing vulnerabilities and the history of change. Moreover, the diversity of the projects and the vulnerability types expose the models being trained on it to several patterns. However, the dataset only contains 11,823 vulnerable functions as opposed to the 253,096 non-vulnerable functions. While it may depict real projects, the data is imbalanced, and more vulnerable functions are needed to train large models.

\subsubsection{D2A dataset}

A Dataset proposed by IBM \cite{D2A9402126} is curated using differential analysis to label issues reported by static analysis tools. Bug-fixing commit pairs are extracted from open-source projects with a static analyzer running on them. If issues were detected in the “before” version and disappeared in the “after” version, then it is assumed to be a bug. Compared to other datasets, the bug trace is included in the dataset to determine the type and exact location of the bug. Open-source projects such as FFmpeg, OpenSSL, httpd, libtiff, libav and NGINX constitute the curated dataset. This dataset also has a limited number of vulnerable samples, and a manual validation experiment shows that the results are better than those of regular differential analysis. However, it is still not at the desired accuracy, with manual validation showing an accuracy of 53\%. The paper’s authors applied the dataset to build a classifier to identify false alarms in static analyzers to reduce the false positive rate.

\subsubsection{CVEfixes}
CVEfixes \cite{bhandari2021cvefixes} is a dataset built using the method proposed by the authors to curate vulnerability datasets based on CVEs. The automated tool was used to release CVEfixes, a dataset covering CVEs up to 9 June 2021. The dataset is organized in a relational database, which can be used to extract data with the desired information. It contains the code changes in several levels, namely on the repository, commit, file, and method levels. The dataset contains 5495 vulnerability fixing commits with 5365 CVE records, covering 1754 open-source projects. The mining tool is shared, and the most recent CVE records can be mined.

\subsubsection{CrossVul}
The CrossVul dataset \cite{nikitopoulos2021crossvul} encompasses a diverse range of programming languages, exceeding 40 in total, and comprises vulnerable source files. The dataset was curated by extracting data from GitHub projects referenced by the National Vulnerability Database (NVD), specifically focusing on files modified through git-diff. Files preceding the commit are tagged as vulnerable, while those following the commit are designated as non-vulnerable. Organized by Common Weakness Enumerations  (CWEs)/Common Vulnerabilities and Exposures (CVEs), as well as language types, the dataset offers a comprehensive classification of vulnerabilities. It encompasses 1675 GitHub projects, spanning 5877  commits and 27,476 files, with an equal distribution of 13,738 files marked as vulnerable and non-vulnerable, respectively. A supplementary dataset containing the commit messages for each sample is provided.

\subsubsection{SySeVR dataset}

SySeVR framework was proposed in \cite{Sysevr9321538}, which builds on the previous work in VulDeePecker \cite{li2018vuldeepecker}. While VulDeePecker only considers library/ API function calls, SySeVR covers a variety of vulnerabilities. Furthermore, SySeVR utilized a unique approach using the notions of syntax-based Vulnerability Candidates(SyVCs) and Semantics-based Vulnerability Candidates (SeVCs) to represent programs as vectors that accommodate syntax and semantic information. Their approach results show a reduced false-negative rate. The dataset is collected from the National Vulnerability Database (NVD) and Software Assurance Reference Dataset (SARD). The NVD dataset contains 19 popular C/C++ open source products and the SARD data comprises 126 vulnerability types. There are 1,591 programs from open-source projects, of which 874 are vulnerable. As for SARD, there are 14,000 programs, with 13,906 being vulnerable.
While this dataset uses the existing datasets published by NIST, the datasets would need further processing in most cases. For example, many vulnerable SARD programs contain the vulnerable snippet and its patch. Not separating them into different samples might yield unwanted results depending on the application.

\subsubsection{DiverseVul dataset}

DiverseVul \cite{DiverseVul2023arXiv230400409C} is proposed as a new vulnerable source code dataset that covers 295 than the previous datasets combined. Furthermore, the dataset is 60\% bigger than previous open-source C/C++ datasets. The data is collected by crawling security issue websites and extracting the commits. The security-based commits are labeled vulnerable before and non-vulnerable for the version after the commit. DiverseVul covers over 797 projects and 7,514 commits with more than 130 CWEs. The MD5 hashes are used to de-duplicate functions, yielding 18,495 vulnerable and 330,492 non-vulnerable functions. The authors conduct several experiments to validate the dataset, combining their dataset with previous datasets and showing insights and possibilities of their use. The paper shows that using their dataset along with the previous datasets yields the best result in their experiments, as opposed to using a single dataset.

\subsubsection{FormAI dataset}

The FormAI dataset \cite{tihanyi2023formai} represents a significant advancement in cybersecurity and LLM, featuring an extensive collection of 112,000 AI-generated, independent, and compilable C programs. This dataset is unique because it utilizes dynamic zero-shot prompting techniques to create various programs, ranging from complex tasks like network management and encryption to simpler ones like string manipulation. These programs were generated using GPT-3.5-turbo, demonstrating the ability of Large Language Models (LLMs) to produce diverse and realistic code samples. A standout feature of the FormAI dataset is its meticulous vulnerability classification. Each program is thoroughly analyzed for vulnerabilities, with the type of vulnerability, the specific line number, and the name of the vulnerable function clearly labeled. This precise labeling is achieved using the Efficient SMT-based Bounded Model Checker (ESBMC), an advanced formal verification method. ESBMC employs techniques like model checking, abstract interpretation, constraint programming, and satisfiability modulo theories to rigorously assess safety and security properties in the programs. This approach ensures that vulnerabilities are definitively detected, providing a formal model or counterexample for each finding and effectively eliminating false positives.

\subsubsection{Chrysalis-HLS} Chrysalis-HLS \cite{10473893} dataset, a helpful resource for improving Large Language Models' performance in hardware and software design. This comprehensive dataset targets functional verification and code debugging in High-Level Synthesis (HLS). It offers a realistic evaluation environment with over 1,000 function-level designs and up to 45 injected bug combinations. Named "Chrysalis" to symbolize code transformation, it includes diverse HLS applications with various error types. Created with GPT-4 and curated prompts, Chrysalis-HLS is a valuable resource for advancing LLM capabilities in HLS verification and debugging, enhancing hardware engineering.

\subsubsection{ReFormAI}
The ReFormAI dataset \cite{gadde2024genai} is a large-scale dataset of 60,000 independent SystemVerilog designs with varied complexity levels, targeting different Common Weakness Enumerations (CWEs). The dataset was generated by four different LLMs and features a unique set of designs for each of the 10 CWEs evaluated. The designs were labeled based on the vulnerabilities identified by formal verification with unbounded proof. The LLMs evaluated include GPT-3.5-Turbo, Perplexity AI, Text-Davinci-003, and LLaMA. The results indicate that at least 60\% of the samples from the 60,000 SystemVerilog designs are vulnerable to CWEs, highlighting the need for caution when using LLM-generated code in real-world projects.

\subsubsection{PrimeVul}
PrimeVul \cite{ding2024vulnerability} dataset is a benchmark dataset based on existing open-source datasets. Mainly taking into consideration BigVul \cite{BigVul10.1145/3379597.3387501}, CrossVul \cite{nikitopoulos2021crossvul}, CVEfixes \cite{bhandari2021cvefixes} and DiverseVul \cite{DiverseVul2023arXiv230400409C}. The proposed pipeline consists of merging, de-duplication, and labeling through 1) PRIMEVUL-ONEFUNC and 2) PRIMEVUL-NVDCHECK. ONEFUNC selects only single functions that are associated with security commits. NVDCHECK is the compartment where a commit is linked to its CVE and checked for in the NVD database. The function is labeled vulnerable if the description precisely mentions the function. The other case is the description containing the file name and the function being the single function changed by a security commit. After such a process, the yielded dataset consists of 7k vulnerable functions and 228,800 benign functions. The dataset spans 755 projects and contains 6,827 commits. Their work also assesses the label quality of their dataset and other related datasets, showing low label errors in PrimeVul.

\subsubsection{X1} X1 \cite{shestov2024finetuning} dataset is constructed from several open-source vulnerability datasets: CVEFixes, a Manually-Curated Dataset, and VCMatch. The dataset contains standalone functions labeled as either vulnerable or non-vulnerable. The labeling process involves extracting functions from vulnerability-fixing commits, assuming pre-change versions are vulnerable and post-change versions are non-vulnerable. A modified dataset (X1) is created to address potential false positives, containing only functions that were the sole change in a commit. 
The final dataset consists of X1 without P3, which has 1334 samples, and X1 with P3, which has 22945 samples. X1 without P3 is balanced, with a 1:1 ratio of positive to negative classes, while X1 with P3 is imbalanced, reflecting the real-world distribution of vulnerable functions with a 1:34 ratio. The dataset size is relatively small, which may limit its representativeness of the real vulnerability distribution.

\begin{table*} 
\centering
\caption{Overview of Software Vulnerability Datasets that can be used for fine-tuning LLMs for software security.}
\label{cybersecurity-datasets-new}
\scriptsize
\begin{tabular}{|p{1.1cm}|p{0.5cm}|p{0.5cm}|p{1.3cm}|p{0.5cm}|p{0.8cm}|p{1.46cm}|p{2.2cm}|p{1.7cm}|p{3cm}|} 
\hline
\textbf{Dataset}    & \textbf{Year}     & \textbf{Lang} &\textbf{Source}                                 & \textbf{Multi-class} & \textbf{Type}           & \textbf{Samples}                                                               & \textbf{Labelling Method}                                             & \textbf{Classification Method}               & \textbf{Challenges/Limitations}                                                                                           \\ 
\hline
Sate IV - Juliet & 2012 & C, C++ \& Java& SARD                                      & Yes         & Synthetic      & Approx 60k (C/C++) \& 29k (Java) test cases                                                  & Testcases are vulnerable by design, with corresponding patch & CWE                                 & Designed to be vulnerable, might not accurately depict real-world projects.                                      \\ 
\hline
Draper \cite{Draper8614145}       & 2018 &  C & Open-source                               & Yes         & Real           & Total: 1.27M   V: 82K NV: 1.19M & Static analyzers                                             & CWE                                 & Small percentage of vulnerable samples. Limited to four categories. \\ 
\hline
Reveal \cite{Reveal8614145}      & 2018  &   C/C++ & Open-source                               & No          & Real           & Total: 18k V: 1.6k NV: 16k    & Vulnerability-fixing commits identified by security terms     & Binary classes                      & Imbalance in sample distribution and only binary labeled. Limited to two projects.                              \\
\hline
Devign \cite{Devign2019arXiv190903496Z}       & 2019 & C  & Open-source                               & No          & Real           & Total: 26K V: 12K NV: 14K     & Binary Manual labeling & Binary classes                      & Binary labeled. Partial dataset release.                                                                        \\
\hline

VUDENC \cite{wartschinski2022vudenc}    & 2019 & Python  & Open-source & Yes & Real & 1,009 commits from 812 repositories & Vulnerability-fixing commits from GitHub repositories & Vulnerability type &  Relatively small Dataset, No guarantee that the commits fixed vulnerabilities. \\
\hline
BigVul \cite{BigVul10.1145/3379597.3387501}     & 2020   &  C/C++ & Open-source 
& Yes         & Real           & Total: 264k V: 11.8k NV: 253k                                          & Vulnerability-fixing commits from CVE database                & CVE/CWE                             & Significant class imbalance. Lack of CWEs/categories for all samples.                                            \\ 
\hline
D2A \cite{D2A9402126}        & 2021  & C/C++ & Open-source                               & Yes         & Real           & Total: 1.3M V: 18.6k NV: 1.27M                                         & Vunerability-fixing commits with static analyzers            & Categories based on static analyzer & Small percentage of vulnerable samples. Manual validation shows low accuracy.                                    \\
\hline

CVEfixes \cite{bhandari2021cvefixes}    & 2021  &   27 languages & Open-source                              & Yes         & Real     & 5,495 commits, 50k methods                                             & Vulnerability-fixing commits from CVE database                    & CVE/CWE                  &  Labelling accuracy needs enhancement and dataset size increased (only limited to CVE records).                              \\
\hline

CrossVul \cite{nikitopoulos2021crossvul}    & 2021  &  40+ languages  & Open-source                              & Yes         & Real     & 5,877 commits, 27k (13,738 V/NV) files                                            & Vulnerability-fixing commits from CVE database                    & CVE/CWE                  &  Labelling accuracy needs enhancement and dataset size increased. Takes the whole file without pinpointing functions. (only limited to CVE records).                              \\

\hline
SySeVR \cite{Sysevr9321538}     & 2022  &  C/C++  & SARD/NVD                                  & Yes         & Semi-Synthetic & Total: 15.6k   V: 14.8k NV: 811                                          & Extracted from existing databases NVD and SARD               & CVE/CWE                             & Limited subset of SARD/NVD. SARD is synthetic, while NVD is limited in the number of labeled vulnerabilities.   \\ 

\hline
DiverseVul \cite{DiverseVul2023arXiv230400409C}   & 2023  & C/C++ & Open-source 
& Yes         & Real           & Total: 349K   V: 18.9k NV: 330K & Vulnerability-fixing commits from security trackers           & CWE                                 & Labelling accuracy needs enhancement and dataset size increased (specifically vulnerable functions).             \\

\hline
FormAI \cite{tihanyi2023formai}     & 2023   & C  & AI-generated                              & Yes         & Artificial     & Total: 112k V: 57k NV: 55K                                             & Formal verification Bounded Model checker                    & Custom categories                   & Bounded formal verification does not cover all types of vulnerabilities and depth.                               \\

\hline

Chrysalis-HLS \cite{10473893}    & 2024  & C++ & Open-source & Yes & Synthetic & Over 1,000 function-level HLS designs &  Predefined errors & Bug Type & Addressing scalability and generalization challenges \\

\hline
FormAI v2 \cite{tihanyi2024neutral}    & 2024   &  C & AI-generated                              & Yes         & Artificial     & Total:  265k V: 168k NV:  23k                                             & Formal verification Bounded Model checker                    & Custom categories                  &  Bounded formal verification does not cover all vulnerabilities and depth.                              \\
\hline
ReFormAI \cite{gadde2024genai}    & 2024   & System Verilog  & AI-generated                              & Yes         & Artificial     & Total: 60k V: 60k NV: 0                                             & Formal verification Bounded Model checker                    & CWE                   & Formal verification with an unbounded proof.                               \\
\hline

PrimeVul \cite{ding2024vulnerability}    & 2024  &  C/C++  & Open-source                              & Yes         & Real     & Total:  236k V: 7k NV:  229k                                             &Single function selection and extraction from NVD                    & CWE                  &  Limited vulnerable samples due filtering existing samples and specific function selection.                              \\

\hline

X1 \cite{shestov2024finetuning}    & 2024 & Java & Open-source & Yes & Real &  Total:  22.9k V: 0.6k NV:  22.3k & Analyzing vulnerability-fixing commits  & Binary classes & Imbalanced, small, and may not represent the true vulnerability distribution.  \\
\hline
\end{tabular}\\
\vspace{1mm}
\vfill
V : Vulnerable , NV: Non Vulnerable
\end{table*}

%% file: sections/challenges.tex
\begin{table*}[ht]
\centering
\caption{LLM Cybersecurity Insights.}
\scriptsize
\label{tab:llmcyber}
\begin{tabular}{|p{2cm}|p{4cm}|p{4.5cm}|p{4.5cm}|}
\hline
\textbf{Aspect} & \textbf{Details} & \textbf{Tools/Methods} & \textbf{Applications} \\ \hline
Architecture & Focus on model components such as tokenization, attention mechanisms, and output generation. & \begin{itemize}
\item Paper:  Attention Is All You Need
\end{itemize} & Threat Detection and Analysis, Security Automation, Cyber Forensics, Penetration Testing, Security Training and Awareness, and Chatbots. \\ \hline
Cyber Security Dataset & Creation of prompt-response pairs that simulate cyber threats using synthetic data. & 
\begin{itemize}
\item OpenAI API for synthetic data
\item Evol-Instruct for data refinement
\item Regex filtering for uniqueness
\end{itemize}
 & Building datasets that mirror real-world threats for training and refining LLMs. \\ \hline
Pre-training Models & Training on large-scale datasets comprising billions of tokens, filtered and aligned with cybersecurity lexicon. & 
\begin{itemize}
\item Megatron-LM for handling large datasets
\item gpt-neox for sequential data handling
\item Distributed training tools
\end{itemize}
 & Preparing LLMs to understand and predict cybersecurity-specific content accurately. \\ \hline
Supervised Fine-Tuning & Incorporating specialized cybersecurity datasets into pre-trained models for tailored applications. & 
\begin{itemize}
\item LoRA for parameter-efficient adjustments
\item QLoRA for quantization and efficient memory management
\end{itemize}
 & Enhancing LLMs to address unique cybersecurity threats and scenarios specifically. \\ \hline
Cyber Security Evaluation & Setting up specialized frameworks and datasets to test LLMs against potential cyber threats. & 
\begin{itemize}
\item Bespoke cybersecurity benchmarks
\item Authoritative datasets for threat detection
\end{itemize}
 & Evaluating how well LLMs detect, understand, and respond to cyber threats. \\ \hline
Advanced LLM Techniques & Implementing techniques like RAG and RLHF to augment LLMs with real-time data and expert-aligned feedback. & 
\begin{itemize}
\item RAG for context retrieval from databases
\item RLHF with specialized preference datasets and reward models
\end{itemize}
 & Improving response relevance and accuracy in cybersecurity applications. \\ \hline
LLM Deployments & Adopting deployment strategies that range from local installations to large-scale server setups. & 
\begin{itemize}
\item Platforms like Gradio and Streamlit for prototyping
\item Cloud services for robust deployment
\item Edge deployment strategies for resource-limited environments
\end{itemize}
 & Deploying LLMs in various environments to ensure accessibility and responsiveness across devices. \\ \hline
Securing LLMs & Addressing vulnerabilities unique to LLMs such as prompt hacking and training data leakage. & 
\begin{itemize}
\item Security measures like prompt injection prevention
\item Red teaming
\item Continuous monitoring systems
\end{itemize}
 & Preventing and mitigating security threats to maintain data integrity and model reliability in LLMs. \\ \hline
Optimizing LLMs & Implementing strategies to reduce memory and computational requirements while maintaining output quality. & 
\begin{itemize}
\item Model quantization
\item Use of bfloat16 data formats
\item Optimization of attention mechanisms
\end{itemize}
 & Enabling efficient LLM operation on various hardware, making them scalable and practical for diverse applications. \\ \hline
\end{tabular}
\end{table*}

\begin{figure*}[htbp]
    \centering
    \includegraphics[width=0.8\textwidth]{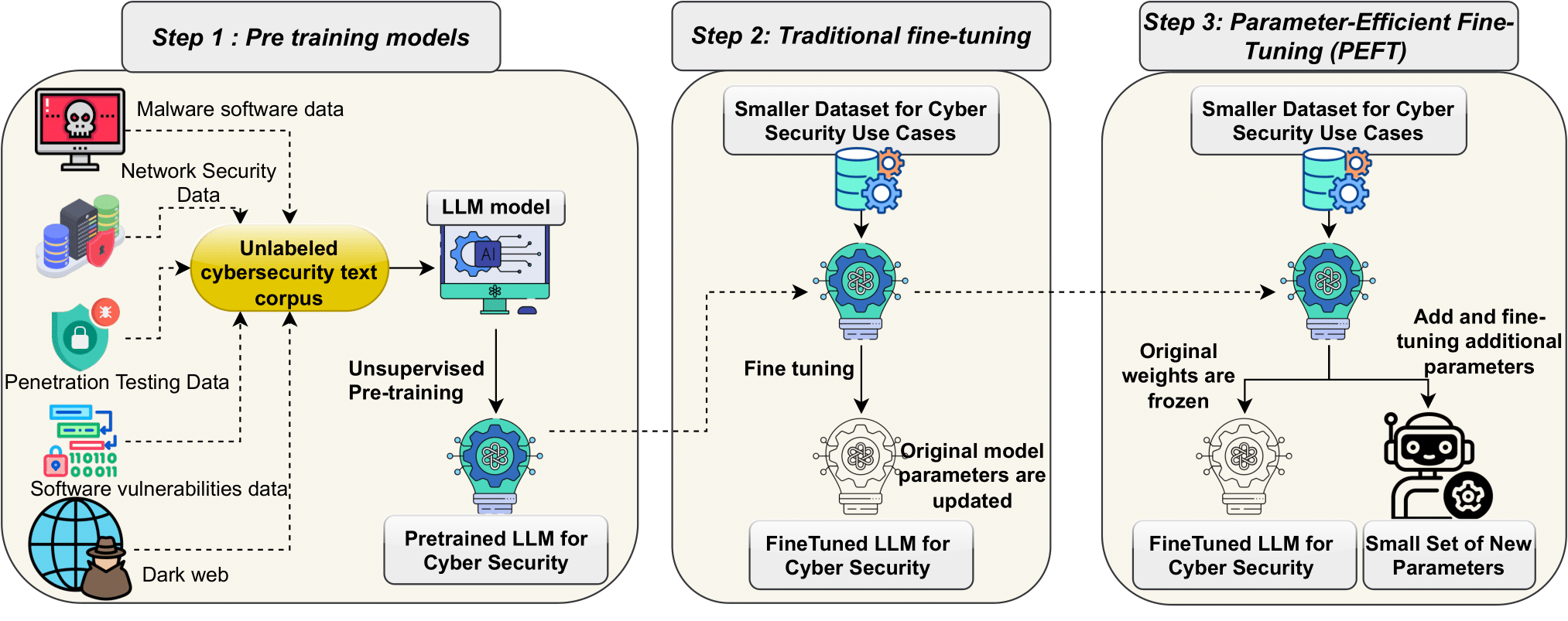}
    \caption{Parameter Efficient Fine-Tuning (PEFT) provides an efficient approach by minimizing the number of parameters needed for fine-tuning and reducing memory consumption comparable to that of traditional fine-tuning.}
    \label{fig:PEFT}
\end{figure*}

\section{LLM Cybersecurity Insights, Challenges and Limitations}
\label{sec:9}

\subsection{Challenges and Limitations}

\subsubsection{Adapting to Sophisticated Phishing Techniques}
The increasing sophistication of phishing attacks, especially those enhanced by AI, presents a major challenge for LLMs in cybersecurity. These models need to evolve to identify and counteract these threats effectively continuously. The challenge lies in the need for regular updates and training to keep pace with the advanced tactics of attackers, which demands substantial resources and expertise.  For example, a large company implemented an LLM-based security system to detect phishing emails. Initially, the system was highly effective, identifying and blocking 95\% of phishing attempts. However, attackers quickly adapted, using AI to generate more convincing phishing emails that mimicked the company's official communication style and included personalized information about the customers. The company's LLM struggled to keep up with these advanced tactics. Phishing emails have become so sophisticated that they can bypass traditional detection methods, significantly increasing the number of successful attacks. Hence, evolving and adapting LLMs in cybersecurity to combat AI-enhanced phishing threats is an open challenge.

\subsubsection{Managing Data Overload in Enterprise Applications}
With the proliferation of enterprise applications, IT teams are overwhelmed by the sheer volume of data they need to manage and secure, often without corresponding increases in staffing. LLMs are expected to assist in managing this data deluge efficiently. However, ensuring these models can process vast amounts of data accurately and identify threats amidst this complexity is daunting, necessitating high levels of efficiency and accuracy in the LLMs. The corporation faced a situation where the LLM failed to recognize a sophisticated cyber-attack hidden within the massive influx of data. This oversight occurred because the model hadn't been trained with the latest attack patterns, highlighting a gap in its learning. The incident underscored the need for LLMs to process data efficiently and maintain high accuracy and adaptability in threat detection.

\subsubsection{Training Data Availability and Quality}
A critical challenge for AI-based cyber defense is the lack of high-quality, accessible training data, as organizations generally hesitate to share sensitive information. The effectiveness of LLMs in cybersecurity depends heavily on the quality and availability of training data. Overcoming this data gap remains a significant hurdle, whether through synthetic data generation or other means.

\subsubsection{Developing and Training Custom Models for Unique Cybersecurity Domains}
Certain specialized areas in cybersecurity require custom models due to their unique vocabularies or data structures, which standard LLMs might not address adequately. Unique Vocabularies and Data Structures: Cybersecurity domains, such as network security, malware analysis, and threat intelligence, have their terminologies, data formats, and communication protocols. Standard LLMs, typically trained on general datasets, might not be familiar with these specialized terms and structures, leading to ineffective or inaccurate threat detection and response. Customizing and training these models to handle specific cybersecurity scenarios is complex and demands substantial resources, presenting a significant challenge in the field.

\subsubsection{Real-Time Information Provision by Security Copilots}
Security copilots powered by LLMs need to provide accurate, up-to-date information in real-time to be effective in the dynamic threat landscape of cybersecurity. Ensuring the relevance and accuracy of information provided by these models in real-time is challenging but essential for effective responses to cybersecurity threats.

\subsection{LLM Cybersecurity Insights}

Table \ref{tab:llmcyber} presents various facets of LLM integration into cybersecurity, providing insights into architectural nuances, dataset creation, pre-training, fine-tuning methodologies, evaluation metrics, advanced techniques, deployment strategies, security measures, and optimization approaches.

\subsubsection{LLM architecture}
A cyber security scientist venturing into utilizing LLMs must understand the architecture's nuances (presented in Section \ref{sec:3}) to tailor these tools for security applications effectively. Understanding the architecture of LLMs, including their ability to process and generate language-based data, is crucial for detecting phishing attempts, deciphering malicious scripts, or identifying unusual patterns in network traffic that may indicate a breach. Knowledge of how these models tokenize input data, their attention mechanisms to weigh information, and their output generation techniques provide the foundational skills necessary to tweak models for optimized threat detection and response \cite{zhao2024explainability}.

\subsubsection{Building Cyber Security dataset}
Building a robust cybersecurity dataset using LLMs involves generating and refining intricate prompt-response pairs to mirror real-world cyber threats. Employing synthetic data generation via the OpenAI API allows for diverse cybersecurity scenarios, while advanced tools like Evol-Instruct \cite{xu2023wizardlm} enhance dataset quality by adding complexity and removing outdated threats. Techniques such as regex filtering and removing near-duplicates ensure the data's uniqueness and relevance. In addition, familiarizing with various prompt templates like Alpaca \cite{taori2023alpaca} is essential for structuring this data effectively, ensuring that the LLM can be finely tuned to respond to the nuanced landscape of cybersecurity challenges efficiently.

\subsubsection{Pre-training models}
Pre-training a model for cybersecurity tasks involves a complex and resource-intensive process to prepare a language model to understand and predict cybersecurity-specific content. This requires a massive dataset comprising billions or trillions of tokens, which undergo rigorous processes like filtering, tokenizing, and aligning with a pre-defined vocabulary to ensure relevance and accuracy. Techniques such as causal language modeling, distinct from masked language modeling, are employed, where the loss functions and training methodologies, such as those used in Megatron-LM \cite{korthikanti2023reducing} or gpt-neox \cite{gpt-neox-library}, are optimized for handling sequential data predictively. Understanding the scaling laws is crucial, as these laws help predict how increases in model size, dataset breadth, and computational power can proportionally enhance model performance \cite{kaplan2020scaling}. While in-depth knowledge of High-Performance Computing (HPC) isn't necessary for using pre-trained models, it becomes essential when building a large-scale language model for cyber security from scratch, requiring an understanding of hardware capabilities and managing distributed workloads effectively.

Most pre-training LLM models are trained using \textit{smdistributed} libraries, proposed by AWS SageMaker, which offer robust solutions for distributed training machine learning models, enhancing efficiency on large-scale deployments. The \textit{smdistributed.dataparallel} library supports data parallelism, optimizing GPU usage by partitioning the training data across multiple GPUs, thus speeding up the learning process and minimizing communication overhead. On the other hand, \textit{smdistributed.modelparallel} is tailored for model parallelism, allowing large models to be split across multiple GPUs when a single model cannot fit into the memory of one GPU. These tools seamlessly integrate with frameworks like TensorFlow, PyTorch, and MXNet, simplifying the implementation of complex distributed training tasks.

\subsubsection{Supervised Fine-Tuning}
Supervised fine-tuning (SFT) of pre-trained Large Language Models for cybersecurity applications enables these models to move beyond basic next-token prediction tasks, transforming them into specialized tools tailored to specific cybersecurity needs. This fine-tuning process allows for incorporating proprietary or novel datasets that have not been previously exposed to models like Falcon 180b, providing a significant edge in addressing unique security challenges. Figure \ref{fig:PEFT} outlines a comprehensive three-step process for training a large language model specialized in cybersecurity, beginning with unsupervised pre-training on a vast corpus of cybersecurity-related texts, including diverse data such as malware, network security, and dark web content. Following this, the model undergoes traditional fine-tuning using a smaller, targeted dataset to refine its capabilities for specific cybersecurity tasks. However, the Parameter-Efficient Fine-Tuning (PEFT) \cite{houlsby2019parameter} involves freezing the original model weights and fine-tuning a small set of new parameters, enhancing the model's adaptability and efficiency while minimizing the risk of overfitting, thus preparing the LLM to tackle advanced cybersecurity challenges efficiently.

Techniques such as LoRA (Low-rank Adapters) \cite{hu2021lora} offer a parameter-efficient approach by adjusting only a subset of the model’s parameters, thus optimizing computational resources while maintaining performance. More advanced methods like QLoRA \cite{dettmers2024qlora} enhance this by quantizing the model’s weights and managing memory more efficiently, making executing these operations even on limited platforms like Google Colab with only one GPU A100. In addition, tools like Axolotl and DeepSpeed  \cite{wu2023zeroquant,xia2024fp6} facilitate the deployment of these fine-tuned models across various hardware setups, ensuring that the enhanced models can be scaled efficiently for real-world cybersecurity tasks, ranging from intrusion detection to real-time threat analysis. This strategic fine-tuning enhances model specificity and significantly boosts their utility in practical cybersecurity applications.

\begin{table*}[h!]
\centering
\caption{Comparison of Benchmarks for Evaluating LLMs in Cybersecurity Knowledge}
\label{tab:llmbench}
\begin{tabular}{|p{2cm}|p{1.2cm}|p{0.5cm}|p{7cm}|p{4cm}|}
\hline
\textbf{Benchmark} & \textbf{Source} & \textbf{Year} &  \textbf{Description} & \textbf{Key Features and Metrics} \\ \hline
CyberSecEval 1 & Meta \cite{bhatt2023purple} & 2023 & A benchmark tests LLMs across two critical security domains—generating insecure code and compliance with requests to assist in cyberattacks. & It measures the frequency and conditions LLMs propose insecure code solutions under.\\ \hline
SecQA & Liu et al. \cite{liu2023secqa} & 2023 & A dataset of multiple-choice questions designed to evaluate the performance of LLMs in computer security. Features two versions of varying complexity and tests LLMs in both 0-shot and 5-shot learning settings. & Evaluates understanding and application of security principles. \\ \hline
CyberMetric & Tihanyi et al. \cite{tihanyi2024cybermetric} & 2024 & A dataset designed for evaluating LLMs in cybersecurity knowledge, consisting of 10,000 questions from various authoritative sources. Used to measure the spectrum of cybersecurity topics covered by LLMs. & Direct comparison between human expertise and LLMs. \\ \hline
CyberSecEval 2 & Meta \cite{bhatt2024cyberseceval} & 2024 & Focuses on quantifying security risks associated with LLMs, such as prompt injection and code interpreter abuse. It highlights challenges in mitigating attack risks and introduces the False Refusal Rate (FRR) metric. & Testing areas: prompt injection, code interpreter abuse; Metric: FRR. \\ \hline
WMDP-Cyber & Li et al. \cite{li2024wmdp} & 2024 & Consists of 3,668 multiple-choice questions designed to measure LLMs' knowledge in biosecurity, cybersecurity, and chemical security. Excludes sensitive and export-controlled information. & Covers biosecurity, cybersecurity, and chemical security. \\ \hline
LLM4Vuln & Sun et al. \cite{sun2024llm4vuln} & 2024 & A unified evaluation framework for assessing the vulnerability reasoning capabilities of LLMs, using 75 verified high-risk smart contract vulnerabilities in 4,950 scenarios across three LLMs. & Focuses on vulnerability reasoning in LLMs. \\ \hline
CyberBench &   Liu et al. \cite{liucyberbench} & 2024 & A domain-specific, multi-task benchmark for assessing LLM performance in cybersecurity tasks. & Includes diverse tasks such as vulnerability detection, threat analysis, and incident response. \\ \hline
\end{tabular}
\end{table*}

\subsubsection{Cyber Security Evaluation}
To evaluate the code generation models, Hugging Face uses the following 7 code generation Python tasks: DS-1000, MBPP, MBPP+, APPS, InstructHumanEval, HumanEval+, and HumanEval \cite{clark2018think,lai2023ds,liu2024your,zellers2019hellaswag,hendrycks2020measuring,cobbe2021training, hendrycks2021measuring}.

In cybersecurity, evaluating large language models demands a specialized framework considering such applications' unique security and accuracy needs. When setting up evaluation metrics for cybersecurity-focused LLMs, test cases should closely mimic potential security scenarios to assess how well the model detects, understands, and responds to cyber threats. This involves configuring the LLM with tailored inputs, expected outputs, and security-specific contextual data \cite{chang2023survey}. For instance, IBM's D2A dataset \cite{D2A9402126} and Microsoft's dataset \cite{Micro} aids in evaluating AI models' capability to identify software vulnerabilities using specific metrics such as accuracy.  

Table \ref{tab:llmbench} compares benchmarks for evaluating LLMs in cybersecurity knowledge. CyberMetric \cite{tihanyi2024cybermetric} is a benchmark dataset designed explicitly for evaluating large language models in knowledge cybersecurity. It consists of 10,000 questions derived from various authoritative sources within the field. The dataset is used to measure the knowledge of LLMs across a spectrum of cybersecurity topics, facilitating direct comparisons between human expertise and machine capabilities. This unique dataset aids in understanding the strengths and limitations of LLMs in cybersecurity, providing a foundation for further development and specialized training of these models in this critical area.

Similar to the CyberMetric benchmark, Meta proposed the CyberSecEval 2 benchmark \cite{bhatt2024cyberseceval} to quantify security risks associated with LLMs such as GPT-4 and Meta Llama 3 70B-Instruct. They highlight new testing areas, notably prompt injection and code interpreter abuse, revealing that mitigating attack risks in LLMs remains challenging, with significant rates of successful prompt injections. The study also explores the safety-utility tradeoff, proposing the False Refusal Rate (FRR) to measure how conditioning LLMs to avoid unsafe prompts might reduce their overall utility by also rejecting benign requests. Additionally, the research assesses LLMs' capabilities in automating core cybersecurity tasks, suggesting that models with coding abilities perform better in exploit generation tasks. The benchmark code is open source to facilitate further research \footnote{https://github.com/meta-llama/PurpleLlama/tree/main/CybersecurityBenchmarks}.

Liu et al. introduced SecQA \cite{liu2023secqa}, a novel dataset designed to evaluate the performance of LLMs in computer security. The dataset, generated by GPT-4, consists of multiple-choice questions to assess LLMs' understanding and application of security principles. SecQA features two versions with varying complexity to challenge LLMs across different difficulty levels. The authors comprehensively evaluated prominent LLMs, including GPT-3.5-Turbo, GPT-4, Llama-2, Vicuna, Mistral, and Zephyr, in both 0-shot and 5-shot learning settings. The findings from the SecQA v1 and v2 datasets reveal diverse capabilities and limitations of these models in handling security-related content. Li et al. \cite{li2024wmdp} introduced the Weapons of Mass Destruction Proxy (WMDP) benchmark. This publicly available dataset consists of 3,668 multiple-choice questions designed to measure LLMs' knowledge in biosecurity, cybersecurity, and chemical security, ensuring the exclusion of sensitive and export-controlled information. Sun et al.  \cite{sun2024llm4vuln} introduced LLM4Vuln, a unified evaluation framework designed to precisely assess the vulnerability reasoning capabilities of LLMs independent of their other functions such as information seeking, knowledge adoption, and structured output. This framework aims to determine how enhancing these separate capabilities could boost LLMs' effectiveness in identifying vulnerabilities. To test the efficacy of LLM4Vuln, controlled experiments were conducted with 75 verified high-risk smart contract vulnerabilities sourced from Code4rena audits conducted between August and November 2023. These vulnerabilities were tested in 4,950 scenarios across three LLMs: GPT-4, Mixtral, and Code Llama.

\begin{table*}[htbp]
\centering
\caption{Optimization Strategies for Large Language Models in Cybersecurity}
\label{tab:llm_cybersecurity_optimization}
\scriptsize
\begin{tabular}{|p{2cm}|p{5cm}|p{3.3cm}|p{5cm}|}
\hline
\textbf{Optimization Strategy} & \textbf{Description} & \textbf{Key Benefits} & \textbf{Cybersecurity Use Case Scenarios}\\
\hline
Advanced Attention Mechanisms & Implements techniques like Flash Attention \cite{dao2022flashattention} to optimize self-attention layers, reducing computation times, particularly effective for long input sequences. & Speeds up processing saves compute resources. & Efficient processing of long log files and network traffic data for anomaly detection.  \\
\hline
Bitsnbytes & Introduces k-bit quantization (notably 8-bit) using block-wise methods to maintain performance while halving memory usage. & Halves memory usage without loss in performance. & Efficient real-time malware analysis and intrusion detection on edge devices. \\
\hline
GPTQ \cite{frantar2022gptq} & A novel quantization method for GPT models that reduces bit width to 3 or 4 bits per weight, enabling the operation of large models on single GPUs with minimal accuracy loss. & Compresses model size, minimizes accuracy loss. & Deploying large-scale threat prediction models on consumer-grade hardware.\\
\hline
GGUF Quantization & Optimized for quick model loading and saving, making LLM inference more efficient. Supported by Hugging Face Hub. & Enhances efficiency of model deployment. & Rapid deployment of updated models to respond to emerging threats and vulnerabilities. \\
\hline
QLoRA \cite{dettmers2024qlora} & Enables training using memory-saving techniques with a small set of trainable low-rank adaptation weights. & Preserves performance with reduced memory. & Training complex cybersecurity models on systems with limited memory resources. \\
\hline
Lower-precision data Formats & Uses formats like bfloat16 instead of float32 for training and inference to optimize resource usage without compromising performance accuracy. & Reduces computational overhead. & Enhancing the speed and efficiency of continuous cybersecurity monitoring systems. \\
\hline
FSDP-QLoRA & Combines Fully Sharded Data Parallelism (FSDP) with 4-bit quantization and LoRA to shard model parameters, optimizer states, and gradients across GPUs. & Scales up model training across multiple GPUs. & Enabling the collaborative training of security models across different organizations without requiring top-tier hardware.\\
\hline
Half-Quadratic Quantization (HQQ) \cite{badri2023hqq} &A model quantization technique that enables the quantization of large models rapidly and accurately without the need for calibration data. & Works efficiently with CUDA/Triton kernels and aims for seamless integration with $torch.compile$. & HQQ can be employed in cybersecurity to protect models by reducing the precision of model weights, making it harder for attackers to reverse engineer or tamper with the models..\\ \hline
Multi-token Prediction \cite{2024arXiv240419737G} & A new training approach for large language models where models predict multiple future tokens simultaneously rather than the next token only. & Models trained with 4-token predictions can achieve up to 3x faster inference speeds, even with large batch sizes. & Multi-token prediction can enhance the modeling of sophisticated cyber attack patterns.\\
\hline
Trust Region Policy Optimization (TRPO) \cite{pmlr-v37-schulman15}& An advanced policy gradient method in reinforcement learning that addresses the inefficiencies of standard policy gradient methods. & TRPO enhances training stability by using trust regions to prevent overly large updates that could destabilize the policy. & In environments with dynamic and evolving threats, TRPO can help maintain a stable and effective response mechanism, adjusting policies incrementally to handle new types of malware.\\ \hline
Proximal Policy Optimization (PPO) \cite{schulman2017proximal} & A reinforcement learning technique designed to improve training stability by cautiously updating policies. &  Prevents "falling off the cliff" scenarios where a policy update is too large could irreversibly damage the policy's effectiveness. & By limiting the extent of policy updates, PPO helps maintain a steady adaptation to evolving cybersecurity threats, reducing the risk of overfitting to specific attack patterns.\\ \hline
Direct Preference Optimization (DPO) \cite{NEURIPS2023_a85b405e} &  A fine-tuning methodology for foundation models optimize policies directly using a Kullback–Leibler divergence-constrained framework, removing the need for a separate reward model. & Requires significantly less data and compute resources than previous methods like PPO. & Reduces the computational and data demands of continuously training cybersecurity models, allowing for more scalable solutions.\\ \hline
Odds Ratio Preference Optimization (ORPO) \cite{hong2024orpo} & An algorithm designed for supervised fine-tuning (SFT) of language models that optimizes preference alignment without the need for a separate reference model. & Eliminates the need for an additional preference alignment phase, simplifying the fine-tuning process. & Enables dynamic adaptation of security models to new and evolving cyber threats by optimizing preference alignment efficiently.\\ 
\hline
\end{tabular}
\end{table*}

\subsubsection{Advanced LLM techniques (RAG and RLHF)}

Advanced techniques like Retrieval Augmentation Generation (RAG) can significantly enhance Language Model performance by enabling the model to access external databases for additional context and information, making it highly effective in specialized fields such as cybersecurity. In cybersecurity applications, RAG can dynamically retrieve up-to-date information from well-known databases such as CVE (Common Vulnerabilities and Exposures), CWE (Common Weakness Enumeration), and the NIST (National Institute of Standards and Technology) database \cite{NEURIPS2020_6b493230}. This capability allows the model to offer current and specific advice regarding vulnerabilities, threat intelligence, and compliance standards. Integrating real-time data from these authoritative sources into the response generation process allows RAG to empower Language Models to deliver precise and contextually relevant cybersecurity insights without extensive retraining, thus enhancing decision-making in critical security operations \cite{gao2024retrievalaugmented}.

Reinforcement Learning from Human Feedback (RLHF) is an advanced method to enhance LLMs tailored for cybersecurity applications, focusing on aligning the model's responses with expert expectations in the security domain. This involves utilizing specialized preference datasets, which contain responses ranked by cybersecurity professionals, presenting a more challenging production process than typical instructional datasets. Techniques like Proximal Policy Optimization (PPO) leverage a reward model to evaluate how well text outputs align with security expert rankings, refining the model's training through adjustments based on KL divergence \cite{schulman2017proximal}. Direct Preference Optimization (DPO) further optimizes this by framing it as a classification challenge, using a stable reference model that avoids the complexities of training reward models and requires minimal hyperparameter adjustments \cite{rafailov2023direct}. These methods are crucial for reducing biases, fine-tuning threat detection accuracy, and enhancing the overall effectiveness of cybersecurity-focused LLMs.

In practical cybersecurity applications, the integration of RAG can be facilitated by orchestrators like LangChain, LlamaIndex, and FastRAG, which connect Language Models to relevant tools, databases, and other resources. These orchestrators ensure efficient information flow, enabling Language Models to seamlessly access and incorporate essential cybersecurity information \cite{zhao2024retrievalaugmented}. Advanced techniques such as multi-query retrievers and HyDE are used to optimize the retrieval of relevant cybersecurity documents and adapt user queries into more effective forms for document retrieval. Furthermore, incorporating a memory system that recalls previous interactions allows these models to provide consistent and context-aware responses over time. This amalgamation of advanced retrieval mechanisms and memory enhancement through RAG significantly boosts the efficacy of Language Models in handling complex and evolving cybersecurity challenges, making them invaluable tools for tracking vulnerabilities, managing risks, and adhering to industry standards in the cybersecurity domain \cite{huang2024survey}.

\subsubsection{LLM deployments}
Deploying LLMs offers a range of approaches tailored to the scale and specific needs of different applications. At one end of the spectrum, local deployment offers enhanced privacy and control, utilizing platforms like LM \textit{Studio} and \textit{Ollama} to power apps directly on users' machines, thus capitalizing on the open-source nature of some LLMs. For more dynamic or temporary setups, frameworks such as \textit{Gradio} and \textit{Streamlit} allow developers to prototype and share demos quickly, with hosting options like Hugging Face Spaces providing an accessible path to broader distribution. On the industrial scale, deploying LLMs can require robust server setups, utilizing cloud services or on-premises infrastructure that might leverage specialized frameworks for peak performance and efficiency. Meanwhile, edge deployment strategies bring LLM capabilities to devices with limited resources, using advanced, lightweight frameworks to integrate smart capabilities directly into mobile and web platforms, ensuring responsiveness and accessibility across a broad spectrum of user environments \cite{mlc-llm,mnn-llm}.

Currently, LLMs can be deployed on Phones. Microsoft \cite{abdin2024phi3} propose phi-3-mini. This highly efficient 3.8 billion parameter language model delivers robust performance on par with much larger models such as Mixtral 8x7B and GPT-3.5, achieving impressive scores like 69\% on the MMLU and 8.38 on MT-bench. Remarkably, the phi-3-mini's compact size allows for deployment on mobile devices, expanding its accessibility and utility. This performance breakthrough is primarily attributed to an innovative approach in training data selection—a significantly enhanced version of the dataset used for phi-2, which integrates heavily filtered web data and synthetic data tailored for relevance and diversity. It has been further aligned to ensure the model's practicality in real-world applications for enhanced robustness, safety, and optimization for chat formats. In addition, the research extends into larger models, phi-3-small and phi-3-medium, which are trained on 4.8 trillion tokens with 7 billion and 14 billion parameters, respectively. These models retain the foundational strengths of phi-3-mini and exhibit superior performance, scoring up to 78\% on MMLU and 8.9 on MT-bench, illustrating significant enhancements in language understanding capabilities with scaling. In addition, \textit{AirLLM} \footnote{https://pypi.org/project/airllm/}  enhances memory management for inference, enabling large language models, such as those with 70 billion parameters (e.g., Llama3 70B), to operate on a single 4GB GPU card. This can be achieved without requiring quantization, distillation, pruning, or any other form of model compression that could diminish performance.

\subsubsection{Securing LLMs}
Securing LLMs is essential due to their inherent susceptibility to traditional software vulnerabilities and unique risks stemming from their design and operational methods. Specifically, LLMs are prone to prompt hacking, where techniques such as prompt injection can be used to manipulate model responses, prompt leaking that risks exposure of training data, and jailbreaking intended to circumvent built-in safety mechanisms. These specific threats necessitate implementing comprehensive security measures that directly address the unique challenges LLMs pose. Additionally, inserting backdoors during training, either by poisoning the data or embedding secret triggers, can significantly alter a model's behavior during inference, posing severe risks to data integrity and model reliability.

As discussed in Section \ref{sec:8}, to mitigate these threats effectively, organizations must adopt rigorous defensive strategies as recommended by the OWASP LLM security checklist \footnote{https://owasp.org/www-project-top-10-for-large-language-model-applications/}. This includes testing LLM applications against known vulnerabilities using methods like red teaming and specific tools such as \textit{garak} \cite{garak} to identify and address security flaws. In addition, deploying continuous monitoring systems like \textit{langfuse} \footnote{https://github.com/langfuse/langfuse} in production environments helps detect and rectify anomalous behaviors or potential breaches in real-time. The OWASP checklist also emphasizes the importance of governance frameworks that ensure data used in training is ethically sourced and handled, maintaining transparency about data sources and model training methodologies. This structured approach to security and governance ensures that LLMs are used responsibly and remain secure from conventional cyber threats and those unique to their operational nature.

\subsubsection{Optimizing LLMs}
Optimizing LLMs for production encompasses several crucial techniques to enhance speed, reduce memory requirements, and maintain output quality. One pivotal strategy is model quantization, which significantly reduces the precision of model weights—often to 4-bit or 8-bit—thereby decreasing the GPU memory requirements. Table \ref{tab:llm_cybersecurity_optimization} presents the optimization strategies for LLMs that can be adopted for Cybersecurity use cases. For instance, quantizing a model to 4-bit can bring down the VRAM requirement from 32 GB to just over 9 GB, allowing these models to run efficiently on consumer-level hardware like the RTX 3090 GPU. Therefore, advanced attention mechanisms such as Flash Attention reduce computation times by optimizing self-attention layers, which are integral to transformers \cite{dao2022flashattention}. This optimization is especially beneficial for handling long input sequences, where traditional self-attention mechanisms could become prohibitively expensive regarding memory and processing power \cite{shazeer2019fast, ainslie2023gqa}.

The quantization methods include Bitsnbytes, 4-bit GPTQ, 2-bit GPTQ, and GGUF quantization. Bitsnbytes introduces a k-bit quantization approach that significantly reduces memory consumption while maintaining performance \cite{frantar2022gptq}. It employs an 8-bit optimization using block-wise quantization to achieve 32-bit performance at a lower memory cost and uses LLM.Int() for 8-bit quantization during inference, halving the required memory without performance loss. Furthermore, QLoRA \cite{dettmers2024qlora}, or 4-bit quantization, enables the training of LLMs using memory-saving techniques that include a small set of trainable low-rank adaptation weights, allowing for performance preservation. In parallel, GPTQ is a novel quantization method for GPT models, facilitating the reduction of bit width to 3 or 4 bits per weight, enabling the operation of models as large as 175 billion parameters on a single GPU with minimal accuracy loss. This method provides substantial compression and speed advantages, making high-performance LLMs more accessible and cost-effective. Additionally, the GGUF format, supported by Hugging Face Hub and optimized for quick model loading and saving, enhances the efficiency of LLM inference.

Another effective optimization is incorporating lower-precision data formats such as bfloat16 for training and inference. This approach aligns with the training precision and avoids the computational overhead associated with float32 precision, optimizing resource usage without compromising performance accuracy. The potential VRAM requirements for different models using bfloat16 are substantial. For example, GPT-3 might require up to 350 GB. In comparison, smaller models like Llama-2-70b and Falcon-40b require 140 GB and 80 GB, respectively, illustrating the scale of resources needed even with efficient data formats \footnote{https://huggingface.co/docs/transformers/main/en/llm\_tutorial\_optimization}.

Recently, FSDP-QLoRA \footnote{https://huggingface.co/docs/bitsandbytes/main/en/fsdp\_qlora}, a new technique combining data parallelism, 4-bit quantization, and LoRA, was introduced by Answer.AI in collaboration with bitsandbytes. Utilizing Fully Sharded Data Parallelism (FSDP) to shard model parameters, optimizer states, and gradients across GPUs, this approach enables the training of LLMs up to 70 billion parameters on dual 24GB GPU systems. FSDP-QLoRA represents a significant step forward in making the training of large-scale LLMs.

Collectively, these techniques make it feasible to deploy powerful LLMs on a wider range of hardware and enhance their scalability and practicality in diverse applications, ensuring they can deliver high performance even under hardware constraints.

%% file: sections/conclusion.tex
\section{Conclusion}
\label{sec:10}

In this paper, we presented a comprehensive and in-depth review of the future of cybersecurity through the lens of Generative AI and Large Language Models (LLMs). Our exploration covered a wide range of LLM applications in cybersecurity, including hardware design security, intrusion detection, software engineering, design verification, cyber threat intelligence, malware detection, and phishing and spam detection, illustrating the broad potential of LLMs across various domains.

We provided a detailed examination of the evolution and current state of LLMs, highlighting advancements in 35 specific models, such as GPT-4, GPT-3.5, BERT, Falcon, and LLaMA. Our analysis included an in-depth look at the vulnerabilities associated with LLMs, such as prompt injection, insecure output handling, training and inference data poisoning, DDoS attacks, and adversarial natural language instructions. We discussed mitigation strategies to protect these models, offering a thorough understanding of potential attack scenarios and prevention techniques.

Our evaluation of 40 LLM models in terms of cybersecurity knowledge and hardware security demonstrated their varying strengths and weaknesses. We also conducted a detailed assessment of cybersecurity datasets used for LLM training and testing, from data creation to usage, identifying gaps and opportunities for future research.

We addressed the challenges and limitations of employing LLMs in cybersecurity settings, including the difficulty of defending against adversarial attacks and ensuring model robustness. Additionally, we explored advanced techniques like Half-Quadratic Quantization (HQQ), Reinforcement Learning with Human Feedback (RLHF), Direct Preference Optimization (DPO), Odds Ratio Preference Optimization (ORPO), GPT-Generated Unified Format (GGUF), Quantized Low-Rank Adapters (QLoRA), and Retrieval-Augmented Generation (RAG) to enhance real-time cybersecurity defenses and improve the sophistication of LLM applications in threat detection and response.

Our findings underscore the significant potential of LLMs in transforming cybersecurity practices. By integrating LLMs into future cybersecurity frameworks, we can leverage their capabilities to develop more robust and sophisticated defenses against evolving cyber threats. The strategic direction outlined in this paper aims to guide future research and deployment, emphasizing the importance of innovation and resilience in safeguarding digital infrastructures.